\documentclass[twocolumn,secnumarabic,amssymb, nobibnotes, nofootinbib, aps, prd]{revtex4-2}

\newcommand{\abc}{{a+b\rightarrow c}}
\newcommand{\C}{{\cal C}} % for Cab
\usepackage{bigdelim}
\usepackage{amsmath}
\usepackage{booktabs}
\usepackage{natbib}
\usepackage{epsfig}
\usepackage{natbib}
\usepackage{color}
\usepackage{array}
\usepackage{longtable}
\usepackage{multirow}
\usepackage{hyperref}
\usepackage{etoolbox}
\usepackage[normalem]{ulem}
\usepackage{color,soul}
\usepackage{xcolor}
\usepackage{pifont}% http://ctan.org/pkg/pifont
\setulcolor{blue}
\setstcolor{red}
\sethlcolor{yellow}

\definecolor{rosso}{cmyk}{0,.88,.77,.40}
\definecolor{darkblue}{rgb}{0, 0, 0.8}
\definecolor{darkgreen}{rgb}{0,0.4,0}
\definecolor{dgreen}{rgb}{0.1,0.6,0.7}
\definecolor{violet}{rgb}{0.8, 0.0, 1.0}
\definecolor{bole}{rgb}{0.47, 0.27, 0.23}
\definecolor{airforceblue}{rgb}{0.36, 0.54, 0.66}
\definecolor{amber}{rgb}{1.0, 0.49, 0.0}
\definecolor{darkred}{rgb}{0.55, 0.0, 0.0}
\definecolor{modelref}{rgb}{0.247,0.3647,0.491}
\definecolor{firebrick}{rgb}{0.698,0.1333,0.1333}
\definecolor{seagreen}{rgb}{0.180,0.545,0.341}
\colorlet{notgreen}{orange!40!white}
\sethlcolor{notgreen}
\newcommand{\cmark}{\ding{51}}%

\defcitealias{2018PhRvC..98c4611G}{Paper~I}

\hypersetup{
  bookmarksopen=true, bookmarksopenlevel=1,
  unicode=false, pdftoolbar=true, pdfmenubar=true,
  colorlinks=true, linkcolor=darkblue, citecolor=darkblue,
  filecolor=darkblue, urlcolor=darkblue
}
\def\aap{A\&A}%
\def\adsr{Adv. in Space Res.}
\def\aipcs{Am.~Inst.~Phys.~Conf.~Series}
\def\anrps{Ann. Rev. Nucl. Part. Sci.}
\def\ap{Astropart.~Phys.}%
\def\apj{ApJ}%
\def\apjl{ApJ}%
\def\apjs{ApJS}%
\def\icrc{ICRC}%
\def\jcap{J. Cosmology Astropart. Phys.}%
\def\mnras{MNRAS}%
\def\nar{New A Rev.}%
\def\nat{Nature}%
\def\nimb{Nucl.~Instr.~Meth.~Phys.~Res.~B}%
\def\nphysa{Nucl.~Phys.~A}%
\def\physrep{Phys.~Rep.}%
\def\prc{Phys.~Rev.~C}%
\def\prd{Phys.~Rev.~D}%
\def\prl{Phys.~Rev.~Lett.}%
\def\ssr{Space~Sci.~Rev.}%
\begin{document}
%\AtEndEnvironment{thebibliography}{\input{biblio_SigTot.tex}}

\title{Current status and desired accuracy of the isotopic production cross sections relevant to astrophysics of cosmic rays II. Fluorine to Silicon, and updated results for Li, Be, and B}

\author{Yoann G\'enolini}
\email[]{yoann.genolini@lapth.cnrs.fr}
\affiliation{LAPTh, Universit\'e Savoie Mont Blanc \& CNRS, Chemin de Bellevue, 74941 Annecy Cedex, France}

\author{David Maurin}
\email[]{dmaurin@lpsc.in2p3.fr}
\affiliation{LPSC, Universit\'e Grenoble-Alpes, CNRS/IN2P3, 53 avenue des Martyrs, 38026 Grenoble, France}

\author{Igor V. Moskalenko}
\email[]{imos@stanford.edu}
\affiliation{W. W. Hansen Experimental Physics Laboratory and Kavli Institute for Particle Astrophysics and Cosmology, Stanford University, Stanford, CA 94305, USA}

\author{Michael Unger}
\email[]{michael.unger@kit.edu}
\affiliation{Institute for Astroparticle Physics (IAP),
  Karlsruhe Institute of Technology (KIT), Karlsruhe, Germany}
\affiliation{Institutt for fysikk, Norwegian University of Science and Technology (NTNU), Trondheim, Norway}

\date{\today}

\begin{abstract}
High-precision cosmic-ray data from ongoing and recent past experiments (Voyager, ACE-CRIS, PAMELA, ATIC, CREAM, NUCLEON, AMS-02, CALET, DAMPE) are being released in the tens of MeV/n to multi-TeV/n energy range. Astrophysical and dark matter interpretations of these data are limited by the precision of nuclear production cross sections. In Paper~I, Phys.~Rev.~C 98, 034611 (2018), we set up a procedure to rank nuclear reactions whose desired measurements will enable us to fully exploit currently available data on CR Li to N ($Z=3-7$) species.
Here we extend these rankings to O up to Si nuclei ($Z=8-14$), also updating our results on the LiBeB species. We also highlight how comprehensive new high precision nuclear data, that could e.g.\ be obtained at the  Super Proton Synchrotron at CERN, would be a game-changer for the determination of key astrophysical quantities (diffusion coefficient, halo size of the Galaxy) and indirect searches for dark matter signatures.
\end{abstract}

\maketitle
%N.B.: revtex does not rely on "\setcounter{tocdepth}{3}"
\makeatletter
\def\l@subsubsection#1#2{}
\makeatother
% \tableofcontents
%\clearpage

\section{Introduction} \label{sec:intro}
%%%%%%%%%%%%%%%%%%%%%%%%%%%%%%%%%%%%%%%%%%%%%%%%%%%%%%
%%%%%%%%%%%%%%%%%%%%%%%%%%%%%%%%%%%%%%%%%%%%%%%%%%%%%%

In the last two decades, Galactic cosmic-ray (GCR) physics has entered a precision era with the PAMELA \cite{2007APh....27..296P, 2017NCimR..40..473P} and AMS-02 \cite{2021PhR...894....1A} experiments. The energy domain of GCR direct high-precision measurements now starts from a few MeV/n with Voyager 1, 2~\cite{2013Sci...341..150S} and ACE-CRIS \cite{2018ApJ...865...69I}, and goes up to hundreds of TeV/n with CALET \cite{2019PhRvL.122r1102A}, DAMPE~\cite{2017APh....95....6C}, CREAM \cite{2017ApJ...839....5Y}, and NUCLEON \cite{2019AdSpR..64.2546G}. Many GCR datasets are now dominated by systematic uncertainties, with AMS-02 data reaching an unprecedented few-percentage precision. On the other hand, the modeling of the GCR nuclear component (compared to these data) involves a large network of production reactions whose precision only reaches $10-20\%$ level on average (see for instance \cite{2018PhRvC..98c4611G}, hereafter \citetalias{2018PhRvC..98c4611G}). This is a severe limitation to provide answers to important questions in this field of research, such as the origin of the CRs, transport in the Galaxy and unveiling astrophysical dark matter (DM).

The aim of this series of paper is to rank the most crucial nuclear isotopic production cross sections, whose new measurements will enable us to fully exploit the present and future GCR data. We refer the reader to \citetalias{2018PhRvC..98c4611G} for a more extensive introduction, motivations, and references, in particular for the historical context of nuclear data and models relevant for GCR physics.
While \citetalias{2018PhRvC..98c4611G} dealt with isotopic cross section reactions to produce Li, Be, B, C, and N elements (i.e.\ $Z=3-7$), this second paper extends the study to O--Si ($Z=8-14$) species. In particular, overwhelmingly secondary F and mostly primary Si and the F/Si ratio can be used as a heavier analog \cite{2022ApJ...925..108B,2022arXiv220801337F,2023PhRvD.107f3020Z} to the B/C ratio, which is widely used to deduce the CR propagation parameters. Given that various CR species can have different origin, a set of such secondary to primary ratios with different atomic mass numbers and properties can be utilized to probe interstellar propagation on different spatial scales \cite{2003A&A...402..971T,2016ApJ...824...16J}.

Our results allow the experimental teams to concentrate on a small number of crucial reactions rather than blindly measure thousands of relevant reactions. The implications are broad and can potentially lead to major breakthroughs.

The number and breadth of the newly discovered features are remarkable. Perhaps the first we should mention is the discovery of a rise of the positron fraction $e^+/(e^+$$+$$e^-)$ up to 100 GeV by PAMELA \cite{2009Natur.458..607A}, contrary to the expectations of a monotonic decrease with energy \cite{1982ApJ...254..391P, 1998ApJ...493..694M}. The rise was confirmed by {\it Fermi}-LAT \cite{2012PhRvL.108a1103A}, and with higher precision and up to $\approx$500 GeV by AMS-02 \cite{2014PhRvL.113l1101A}. The latest AMS-02 data indicate a cutoff in the $e^+$ spectrum at $\approx$350 GeV \cite{2021PhR...894....1A} that may hint at the origin of the excess.  The all-electron ($e^-$$+$$e^+$) spectrum up to 1 TeV was measured by {\it Fermi}-LAT \cite{2009PhRvL.102r1101A, 2010PhRvD..82i2004A}. It appears to be too flat, contrary to the expectations of a steep decrease. A sharp cutoff above $\approx$1 TeV was reported by H.E.S.S.\  \cite{2008PhRvL.101z1104A, 2009A&A...508..561A} and confirmed by VERITAS \cite{2018PhRvD..98f2004A}. Subsequent measurements by PAMELA \cite{2017RNCim.40...473A},  AMS-02 \cite{2014PhRvL.113v1102A, 2019PhRvL.122j1101A, 2021PhR...894....1A}, CALET \cite{2018PhRvL.120z1102A}, DAMPE \cite{2017Natur.552...63D}, and {\it Fermi}-LAT \cite{2017PhRvD..95h2007A}, though do not fully agree in detail, confirmed these results and reveal some features in the 100 GeV--TeV region. The all-electron spectrum includes the $e^+$ excess (a so-called ``signal'') and may include the identical $e^-$ ``signal'' if the source is charge-sign symmetric (e.g., pulsars or DM) \cite{2018ApJ...854...94B, 2019PhRvL.122j1101A}. Fast energy losses due to the inverse Compton and synchrotron emission at TeV energies ensure that TeV electrons could only come from nearby DM clumps or local sources \cite{2004ApJ...601..340K, 2019ApJ...887..250P, 2019JCAP...04..024M, 2021PhRvD.103h3010E} that would appear as spectral bumps (yet to be detected).

Analysis of the {\it Fermi}-LAT observations yielded a weak extended residual component at a few GeV peaked at the Galactic Center (GC) \cite{2011PhLB..697..412H, 2016ApJ...819...44A}, consistent with minimal supersymmetric model (MSSM) and limits from direct detection experiments \cite{2017JCAP...12..040A, 2020ARNPS..70..455M}.
A recent detection of the $\gamma$-ray emission from the extensive 400-kpc-across halo of M31 at 3-20 GeV \cite{2019ApJ...880...95K} hints at the similarity with the GC and $\bar{p}$ excesses
\cite{2021PhRvD.103b3027K, 2021PhRvD.103f3023B}.
The excess of 10-20 GeV $\bar{p}$ \cite{2019PhRvD..99j3026C, 2020JHEP...07..163H}, which is clearly seen in a high-precision spectrum by AMS-02 \cite{2016PhRvL.117i1103A, 2021PhR...894....1A}, may be an artifact due to uncertainties in the $\bar{p}$ production cross section, but surprisingly this excess appears in the same energy range as the excesses in $\gamma$-ray emission from the GC and the halo of the M31 galaxy. Meanwhile, outside of the excess range the $\gamma$-ray emission and $\bar{p}$ data agree well with conventional predictions \cite{2017ApJ...840..115B, 2019ApJ...880...95K}. The absence of a corresponding signal in direct detection experiments may point to the hidden sector DM \cite{2020JHEP...07..163H}.

The most striking is the AMS-02 preliminary claim of the detection of six $^3\overline{\rm He}$ and two $^4\overline{\rm He}$ events \cite{Ting2018}. The number of detected events (8) and the ratio $^4\overline{\rm He}/^3\overline{\rm He}$\,$=$$1/3$ is too high for them to be produced through coalescence \cite{2018PhR...760....1C, 2018PhRvD..97j3011K, 2019PhRvD..99b3016P, 2020JCAP...08..048K}.

Besides, new data yield a deeper understanding of the internal works of our Galaxy. Observations of the breaks (hardening) in the spectra of (mostly) primary ($p$, He, C, O, Ne, Mg, Si, Fe), secondary (Li, Be, B, F), and intermediate (N, Na, Al) nuclei at the same magnetic rigidity $R$$\approx$300 GV by PAMELA \cite{2011Sci...332...69A}, {\it Fermi}-LAT \cite{2014PhRvL.112o1103A}, AMS-02 \cite{2015PhRvL.114q1103A, 2015PhRvL.115u1101A, 2017PhRvL.119y1101A, 2018PhRvL.120b1101A, 2018PhRvL.121e1103A, 2020PhRvL.124u1102A, 2021PhR...894....1A, 2021PhRvL.126d1104A, 2021PhRvL.126h1102A, 2021PhRvL.127b1101A}, while the spectral slopes of these groups of nuclei are different. This may imply a change in the spectrum of the interstellar turbulence \cite{2012ApJ...752...68V, 2012PhRvL.109f1101B, 2020ApJS..250...27B} or an influence of a passing star (such as $\epsilon$ Eri) \cite{2021ApJ...911..151M, 2022ApJ...933...78M}; for a summary of current models see \cite{2023FrPhy..1844301M}. There is also an evidence of some fraction of primary Li in CRs \cite{2020ApJ...889..167B}, the unexpected low-energy excess in CR Fe \cite{2021ApJ...913....5B} and Al \cite{2022ApJ...933..147B}, and a hint at the presence of primary F in CRs \cite{2022ApJ...925..108B}, all discovered using the combined GalProp--HelMod framework; alternative interpretations of the Li and F data invoke spatially-dependent diffusion model \cite{2023PhRvD.107f3020Z} or, what concerns us particularly in this paper, uncertainties on the production cross sections \citep{2020A&A...639A.131W,2021JCAP...07..010D,2021PhRvD.103j3016K,2022A&A...668A...7M,2022arXiv220801337F}.

Anomalously high CR isotopic ratios $^{12}$C/$^{16}$O, $^{22}$Ne/$^{20}$Ne, and $^{58}$Fe/$^{56}$Fe \cite{2001SSRv...99...15W, 2006NewAR..50..516B} and a recent detection of radioactive $^{60}$Fe in CRs by the ACE-CRIS \cite{2016Sci...352..677B} indicate a relatively recent supernovae (SNe) activity in the solar neighborhood and align well with the discovered low-energy excess in CR Fe \cite{2021ApJ...913....5B}.
The discovery of the {\it Fermi} bubbles in the {\it Fermi}-LAT data \cite{2010ApJ...724.1044S,2014ApJ...793...64A}, huge $\approx$10-kpc-across structures emanating from the GC, and even larger bubbles by eROSITA \cite{2020Natur.588..227P} add to our understanding of the activity of the central supermassive black hole and its ability to accelerate particles.

The listed discoveries imply that new physics may be involved, however, the significance of these findings critically depends on the accuracy of the underlying datasets, especially isotopic production cross sections. The latter are used for evaluation of the CR propagation parameters, which, in turn, is the cornerstone of all calculations involving CR particles, their emissions, and related backgrounds.

The paper is organized as follows: in Section~\ref{sec:setup}, we present the propagation setup and reference isotopic production cross-section parametrizations used for our calculations. We also highlight the contributing fractions to CR fluxes (CR source, fragmentation on the interstellar medium, or radioactive decay of parent nucleus). In Section~\ref{sec:definitions}, we gather all relevant definitions and formulas used for the ranking of the individual isotopic production cross sections (in terms of their contribution to the production of a given CR isotope or element), and provide formulas to propagate the cross-section uncertainties to the calculated GCR fluxes. We also provide additional coefficients (built from the ranking production cross section coefficients) enabling the calculation of the beam time and number of event estimates to improve GCR flux modeling to a user-desired precision. Of interest for the broader CR community, we also define additional rankings in terms of the dominant reactions for the direct production (from GCR fluxes as measured), the most important production channels (defined to be the ensemble of 1-step or multistep production paths starting from primary fluxes), and the most important progenitors (defined to be the sum of all contributions starting from a primary flux). In Section~\ref{sec:results}, we show a subset of our results, illustrating the various ranking coefficients at 10~GeV/n in different graphical views, the desired reactions to improve GCR flux predictions, and the number of events necessary to measure these reactions at a precision that would be on par to model GCR fluxes at an accuracy below the $3\%$ precision reached by current GCR data. In Section~\ref{sec:forecasts}, we consider mock nuclear data obtained from the above beam-time properties to make forecasts on the improvement brought on key CR quantities for different possible scenarios of new nuclear measurements (for instance at NA61/SHINE).

For readability, we report a lot of our figures and tables in the Appendices: Appendix~\ref{app:plot_Edep_rankings} shows the energy-dependence of (i) the primary/secondary content of GCR elemental fluxes and the relative importance of 1-step vs.\ multistep production, (ii) the most important production channels, (iii) the most important progenitors. Appendix~\ref{app:fig_Dij_Pij} provides a graphical view at 10~GeV/n of the most important direct production reactions and most important progenitors at the isotopic level. Appendix~\ref{app:tablesxs_plot2DfaHc} provides tables of the $f_{abc}$ coefficients (also at 10~GeV/n), i.e.\ the ranking of the most important production reactions $a+b\to c$ for Li to Si GCR fluxes; we also show graphical views of the $f_{aHc}$ (i.e.\ restricting our consideration by reactions on H, the dominant component of the intersellar medium (ISM)).

We also provide a Supplemental Material \cite{supp..mat..arxiv} (including references [155-253]) with plots of the most important reaction channels (parameterizations and available data) discussed in the paper.

\section{Calculation setup and flux origin} \label{sec:setup}
%%%%%%%%%%%%%%%%%%%%%%%%%%%%%%%%%%%%%%%%%%%%%%%%%%%%%%
%%%%%%%%%%%%%%%%%%%%%%%%%%%%%%%%%%%%%%%%%%%%%%%%%%%%%%

The propagation of CR nuclei in the Galaxy is described by a steady-state system of second-order differential equations (e.g., \cite{2007ARNPS..57..285S}), which include CR sources and transport (diffusion, convection), energy gains and losses, and production of secondary isotopes and particles and destruction of nuclei species in the ISM, assumed to be made of $90\%$ of H and $10\%$ of He in number. {In order to highlight the role of the network of nuclear cross sections (and the impact of their uncertainties) at the core of our study, we write a compact version of the propagation equation for the differential density $N_k$ of CR isotope $k$ (energy dependencies are implicit):}
\begin{equation}
\begin{split}
  {\cal L}N_k &= Q_k^p + Q_k^s,\\
  {\rm with}\quad Q_k^s &= \sum_{i\,\in\rm ISM} \sum_s n_i v_s\,\!\sigma_{s+i\to k}^{\rm cumul}N_s\,.
\end{split}
\label{eq:1D}
\end{equation}
{In this equation, the left-hand side corresponds to a generic operator ${\cal L}$ that includes diffusion and convection, continuous energy gains and losses, and the nuclear destruction rate of $k$ on ISM species of density $n_i$, $\sum_{i\,\in\rm ISM} n_i v_k\sigma_{k+i}^{\rm inel}$: the uncertainties on the inelastic cross sections on H targets is at the level $(\Delta\sigma^{\rm inel})/\sigma^{\rm inel}\approx 5\%$  (see discussion in \citetalias{2018PhRvC..98c4611G}), which is subdominant in current flux calculations.
The right-hand side of Eq.~(\ref{eq:1D}) describes a generic astrophysical or DM primary source term, $Q^k_p$, and the secondary source term, $Q_k^s$. The latter is given by the sum over all nuclear interactions of CR isotope $s$ (with velocity $v_s$) on the ISM species $i$ (of density $n_i$) ending up into species $k$ (directly or via short-lived intermediate steps), with a cumulative cross section $\sigma_{s+i\to k}^{\rm cumul}$ detailed in Sect.~\ref{sec:ghosts}---the straight-ahead approximation, whereby CR fragments carry the same energy per nucleon as their CR progenitor, is assumed.
At high energy (starting above a few GeV/n), the operator ${\cal L}$ becomes dominated by diffusion, so that $N_k\propto (Q^p+Q^s_k)/D$, with $D$ the diffusion coefficient whose energy dependence is discussed below. This means that, for a secondary species (i.e. $Q_k^p=0$), the relative uncertainty on the calculated fluxes, $(\Delta N_k)/N_k$, is directly proportional to $(\Delta\sigma^{\rm cumul})/\sigma^{\rm cumul}$, estimated to be at the $20\%$ level on individual reactions. In practice, the calculation of the flux of CR isotopes can be written in a matrix form. Omitting for readability the target index $i$, and with $N_n$ the heavier species (always a pure primary) and $N_0$ the lighter one, we have
\begin{equation}
\nonumber
  {\cal L}
  \begin{bmatrix}
   N_n\\
   \vdots\\
   N_1\\
   N_0
  \end{bmatrix}
  \!=\!
  \begin{bmatrix}
   Q^p_n\\
   \vdots\\
   Q^p_1\\
   Q^p_0
  \end{bmatrix}
   \!+\! n v
    \begin{bmatrix}
     0  & 0  &  \dots  & 0 \\
     \sigma_{n\to (n-1)}^{\rm cumul}& 0  &  \dots  & 0\\
     \vdots&  \ddots &  \ddots  & \vdots\\
     \sigma_{n\to 0}^{\rm cumul} & \dots  &  \sigma_{1\to 0}^{\rm cumul} &  0\\
  \end{bmatrix}
  \begin{bmatrix}
   N_n\\
   \vdots\\
   N_1\\
   N_0
  \end{bmatrix}.
\end{equation}
The matrix of production cross sections is triangular, as fragments of a CR can only be produced by heavier ones (the fusion channel is always negligible at the energies considered, except for the fusion of protons into deuterons). We thus have a network of roughly a thousand production reactions (including short-lived nuclei) to consider to go up to $Z=30$. This network needs to be carefully inspected to rank the most important reactions for the flux calculation.}

In order to produce the observed amount of secondary species, GCRs must cross a specific amount of matter in the ISM, called the grammage. In this respect, the discussion and the results presented here are largely independent of the specific GCR model implementation, as long as the model reproduces standard observables of GCR physics \cite{2001ApJ...547..264J}.

\subsection{Propagation setup} \label{sec:setup_propag}
%%%%%%%%%%%%%%%%%%%%%%%%%%%%%%%%%%%%%%%%%%%%%%%%%%%%%%

As in \citetalias{2018PhRvC..98c4611G}, we rely on the 1D semianalytical propagation model implemented in the USINE package\footnote{\url{https://dmaurin.gitlab.io/USINE/}} \cite{2020CoPhC.24706942M}. Since \citetalias{2018PhRvC..98c4611G}, several high-precision elemental fluxes have been released by the AMS-collaboration \cite{2018PhRvL.120b1101A, 2018PhRvL.121e1103A, 2020PhRvL.124u1102A, 2021PhRvL.126d1104A, 2021PhRvL.126h1102A, 2021PhRvL.127b1101A}, in particular F to Si fluxes and also the Fe flux. These data have been used to update the values of the transport parameters employed in different models \cite{2020ApJS..250...27B, 2021ApJ...913....5B, 2022ApJ...925..108B, 2023FrPhy..1844301M}, including studies done with the USINE code \cite{2019PhRvD..99l3028G,2020A&A...639A.131W,2020A&A...639A..74W,2022A&A...668A...7M}. We use here the so-called \textit{SLIM} configuration introduced in \cite{2019PhRvD..99l3028G}, that is a purely diffusive configuration without convection and distributed reacceleration. The diffusion coefficient $D(R)$ depends on the rigidity $R=pc/Ze$ and is taken to be a broken-power law with breaks both at low (around a few GV) and high (at a few hundreds of GV) rigidities:
\begin{equation}
  \label{eq:def_DR}
  \!\!\!D(R) \!=\! {\beta} D_{0}
  \!\!\left[ \frac{R}{\!1\,{\rm GV}\!} \right]^{\delta}
  \!{\left[ 1 \!+\! \left(\!\frac{R_l}{R} \right)^{\!\!\!\frac{\delta-\delta_l}{s_l}}\!\right]^{\!s_l}}
  \!{\left[ 1 \!+\! \left(\!\frac{R}{R_h} \right)^{\!\!\!\frac{\delta-\delta_h}{s_h}}
    \!\right]^{\!-s_h}}\!\!\!\!\!\!\!\!\!.\!\!
\end{equation}
For practical purposes, the values of the parameters in the above equation are taken from \cite{2022arXiv220801337F}. It is important to stress that for these parameters, $D$ is a decreasing function of the energy below the low-energy break, and a growing function above this break.

At variance with \citetalias{2018PhRvC..98c4611G}, we assume species-dependent broken power laws for the CR source spectra instead of a universal power-law for all species: this ensures a better match of the existing GCR data, but it has in practice a very minor impact on the results \cite[see, e.g.,][]{2022arXiv220801337F}. The isotopic fractions and relative elemental abundances at source are initialized to their solar system values \cite{2003ApJ...591.1220L} and iteratively adjusted  until they match both the elemental fluxes and isotopic ratios: for elemental fluxes, we use AMS-02 data at 50 GV for H to Si and Fe, and HEAO-3 CR data \cite{1990A&A...233...96E} at 10.6~GeV/n for other elements up to Zn; for isotopic ratios, we use low-energy data points mostly from ACE-CRIS \cite{2001AdSpR..27..773W,2005ApJ...634..351B,2009ApJ...695..666O} but also from ULYSSES \cite{1997ICRC....3..381C}.

We stress that other hypotheses for the transport setup or the source spectra would lead to slightly different energy dependences of the various rankings presented in this study, respectively at low and high-energy.
First, considering reacceleration and convection (instead of, or in addition to, the low-energy diffusion break) would change the residence time in the Galaxy below a few GeV/n, hence the grammage crossed and the amount of secondary CR produced. However, in this nonrelativistic regime, the interactions rate also decreases by a $\beta$ factor, and eventually goes to zero when the threshold for the cross section is reached. Furthermore, below a few hundreds of MeV/n, energy losses in the ISM become the dominant process that shapes the fluxes, so that all these details become somewhat irrelevant. Actually, our results are shown for solar-modulated fluxes (using the Force-Field approximation \cite{1967ApJ...149L.115G,1968ApJ...154.1011G}), and the so-called top-of-atmosphere (TOA) fluxes considered here correspond to calculation made for interstellar energies always above a few hundreds of MeV/n. All in all, changing the transport setup should only slightly change the relative importance of secondary production with regard to the total flux (for mixed species) in the low-energy regime; more importantly, the relative importance of the various cross sections (and hence the rankings) should not. Second, assuming a universal spectral index for all species or allowing for different low- and high-energy spectral indices for different elements \cite{2020ApJS..250...27B, 2021ApJ...913....5B, 2022ApJ...925..108B, 2022arXiv220801337F} would change the ranking of the most-important cross sections above several tens of TeV, where high-energy data are not constraining enough, and in particular AMS-02 data are dominated by statistical uncertainties. To mitigate this potential effect, we provide the ranking of the production cross sections at 10 GeV/n: this is an energy around which the overall production level (or grammage) is perhaps best-constrained (thanks to the best precision reached for GCR data there).

\subsection{Nuclear cross-section models} \label{sec:setup_XS}
%%%%%%%%%%%%%%%%%%%%%%%%%%%%%%%%%%%%%%%%%%%%%%%%%%%%%%

Both types of the cross sections, the total fragmentation and isotopic production cross sections, are necessary for the GCR modeling, the former being measured and modeled at a better precision than the latter. For flux calculations, uncertainties from production cross sections dominate compared to uncertainties from inelastic cross sections, and we fix the parameterization of the latter to that of \cite{1996NIMPB.117..347T}; we refer readers to \cite{2022ApJS..262...30P} for a recent discussion of the inelastic cross-sections models\footnote{Note that there are many typos in the original paper \cite{1996NIMPB.117..347T}, which are fixed in \cite{2022ApJS..262...30P}.} used in GCR physics.

The existing models for the isotopic production cross sections were discussed at length in \citetalias{2018PhRvC..98c4611G} \footnote{Note that, among the approaches described, the accuracy of Monte Carlo transport codes is still questionable and we prefer to rely on parameterization based both on data and dedicated nuclear codes.}. To encompass the typical uncertainties on these models, as in \citetalias{2018PhRvC..98c4611G}, we consider two parameterizations available from the GALPROP code\footnote{\url{http://galprop.stanford.edu}} \cite{2022ApJS..262...30P}, namely its options OPT12 and OPT22. Both are based on a careful inspection of the quality and systematics of all available nuclear datasets for each reaction in MeV/n--GeV/n range, and the use of specifically designed approximations, semiempirical formulas, and nuclear codes normalized to data when available \cite{1998nucl.th..12069M, 2001ICRC....5.1836M, 2003ApJ...586.1050M, 2003ICRC....4.1969M, 2004AdSpR..34.1288M, 2005AIPC..769.1612M}: \emph{for marginally important reactions,} OPT12 uses the WNEW code \cite{1990PhRvC..41..566W, 1998PhRvC..58.3539W, 1998ApJ...508..940W, 1998ApJ...508..949W} and OPT22--the YIELDX code \cite{1998ApJ...501..911S, 1998ApJ...501..920T}, providing a representative dispersion for cross sections where no data are available. For more details on the nuclear package in GALPROP see \cite{2020ApJS..250...27B} and references therein. For most of our rankings, we actually use a third parameterization, based on a mixture of GALPROP-OPT12 and GALPROP-OPT22 \cite{2022A&A...668A...7M,2022arXiv220801337F}. This parameterization is based on GALPROP-OPT12 and renormalizes, for the most important reaction involved in the production of LiBeB \cite{2022A&A...668A...7M} and F \cite{2022arXiv220801337F}, GALPROP-OPT22 cross sections to nuclear data, which became available after the GALPROP nuclear cross-section package was developed. Indeed, with these new nuclear data, the fragmentation of Fe appears to be very important \cite{2022A&A...668A...7M}, compared to how it was ranked in \citetalias{2018PhRvC..98c4611G}. This parameterization can be seen as a partial update of the original GALPROP routines, which is used here to update the ranking for the LiBeB elements provided in \citetalias{2018PhRvC..98c4611G}.

A full update of the GALPROP parameterizations, based on a comprehensive survey of the literature for nuclear data relevant for the production of all elements considered here would be ideal. However, this is an extremely time-consuming task yet to be completed. While working to achieve this goal, we provide the best rankings available at this time. As new data becomes available, our results will be updated in the follow-up papers.

\subsection{Primary vs.\ secondary origin of F to Si} \label{sec:origin}
%%%%%%%%%%%%%%%%%%%%%%%%%%%%%%%%%%%%%%%%%%%%%%%%%%%%%%

Before defining which rankings are performed and what they are, it is important to briefly introduce the most salient properties, production-wise, of the CR elements under scrutiny. For an advanced discussion on the origin of these elements in GCRs, see \cite{2021MNRAS.508.1321T}.

GCRs are broadly separated in two categories, namely {\em primary} species, accelerated in sources (e.g., $^1$H, $^4$He, O, Si, Fe), and {\em secondary} species, resulting from the nuclear fragmentation of heavier species in the ISM (e.g., $^2$H, $^3$He, LiBeB, F, Sc, Ti, and V). Most species are formed from an energy-dependent mixture of these two categories. Schematically, for each secondary component, its spectrum appears to be steeper than the spectrum of its progenitor by the value of the index of the diffusion coefficient.
Given the shape of $D(R)$, see Eq.~(\ref{eq:def_DR}), which is minimum at a few GeV/n and increases otherwise, the maximum of secondary content in CRs peaks at a few GeV/n. Moreover, in case of consequent nuclear interactions, the energy dependence of the spectrum of the daughter nucleus will suffer such steepening each time, resulting in an even steeper spectrum (see also \citetalias{2018PhRvC..98c4611G}). Obviously, secondary-dominated species will be much more sensitive to uncertainties in the cross sections than primary-dominated ones, but the maximum of uncertainties will arise at this peak, where the direct and multistep contributions maximized. Showing calculations at a few GeV/n thus highlights the role of nuclear uncertainties on CR fluxes for the most unfavorable case.

We show in Table~\ref{tab:origin} the different origins of the elemental fluxes (and their isotopes) calculated at 10~GeV/n. To avoid any possible ambiguity, we denote the contributing fractions shown as $f_{\rm prim}$ (for primary), $f_{\rm sec}$ (for secondary, including direct and multistep production), and $f_{\rm rad}$ (for radioactive, a rare case of a CR species receiving a contribution from the $\beta$ or EC capture decay of another CR progenitor\footnote{Relevant unstable species for GCR studies are those whose decay lifetime is not too small or too large compared to the propagation time in the Galaxy, see Section~\ref{sec:ghosts}.}), whose sum makes up $100\%$ of the total flux. We further separate $f_{\rm sec}$ into direct, two-step, and more than two-step contributions (whose total makes up $100\%$ of the fragmentation flux).

\begingroup
\begin{table}[t]
\caption{Fractions of primary, fragmentation, and radioactive origin (with regard to total flux), and contributions of 1-step, 2-step, and more-than-2-step channels (with regard to total fragmentation production) for fluxes at 10 GeV/n modulated with force-field modulation potential $\phi_{\rm FF}=700$~MV.\label{tab:origin}}
\begin{tabular}{rrp{4pt}rrrp{6pt}rrr}
\hline\hline
\multicolumn{2}{c}{CR element}&&\multicolumn{3}{c}{\% of} &&\multicolumn{3}{c}{\% of multi-step}\\[-2pt]
\multicolumn{2}{c}{or isotope (\%)}&&\multicolumn{3}{c}{total flux} &&\multicolumn{3}{c}{secondaries}\\
\cmidrule(l{10pt}r{10pt}){4-6}\cmidrule(l{10pt}r{10pt}){8-10}
\multicolumn{2}{r}{} &&$f_{\rm prim}$&$f_{\rm sec}$&$f_{\rm rad}$&&~~~~1&~~~~~2&$>\!2$\\[2pt] \hline
 {\bf Li  }      &        && {\bf   0} & {\bf 100} & {\bf   0}&&{\bf  74} & {\bf  20} & {\bf   6}\\[-3pt]
  $^{6}$Li& (61\%)  &&   0 & 100 &   0&& 75 &  19 &   6\\[-3pt]
  $^{7}$Li& (39\%)  &&   0 & 100 &   0&& 72 &  22 &   6\\[3pt]
 {\bf Be  }      &        && {\bf   0} & {\bf 100} & {\bf   0}&&{\bf  75} & {\bf  19} & {\bf   6}\\[-3pt]
  $^{7}$Be& (58\%)  &&   0 & 100 &   0&& 80 &  15 &   5\\[-3pt]
  $^{9}$Be& (33\%)  &&   0 & 100 &   0&& 69 &  24 &   7\\[-3pt]
 $^{10}$Be& (9\%)  &&   0 & 100 &   0&& 70 &  24 &   6\\[3pt]
 {\bf B   }      &        && {\bf   0} & {\bf  96} & {\bf   4}&&{\bf  77} & {\bf  18} & {\bf   5}\\[-3pt]
  $^{10}$B& (34\%)  &&   0 &  90 &  10&& 71 &  23 &   6\\[-3pt]
  $^{11}$B& (66\%)  &&   0 & 100 &   0&& 80 &  15 &   4\\[3pt]
 {\bf C   }      &        && {\bf  80} & {\bf  20} & {\bf   0}&&{\bf  78} & {\bf  17} & {\bf   5}\\[-3pt]
  $^{12}$C& (90\%)  &&  88 &  12 &   0&& 72 &  22 &   6\\[-3pt]
  $^{13}$C& (10\%)  &&   7 &  93 &   0&& 84 &  12 &   3\\[-3pt]
  $^{14}$C& (0.03\%)  &&   0 & 100 &   0&& 60 &  33 &   6\\[3pt]
 {\bf N   }      &        && {\bf  22} & {\bf  76} & {\bf   2}&&{\bf  88} & {\bf   8} & {\bf   3}\\[-3pt]
  $^{14}$N& (51\%)  &&  43 &  52 &   5&& 86 &  11 &   3\\[-3pt]
  $^{15}$N& (49\%)  &&   0 & 100 &   0&& 90 &   7 &   3\\[3pt]
 {\bf O   }      &        && {\bf  94} & {\bf   6} & {\bf   0}&&{\bf  65} & {\bf  24} & {\bf  10}\\[-3pt]
  $^{16}$O& (97\%)  &&  97 &   3 &   0&& 64 &  25 &  11\\[-3pt]
  $^{17}$O& (1\%)  &&   2 &  98 &   0&& 65 &  24 &  10\\[-3pt]
  $^{18}$O& (1\%)  &&   9 &  91 &   0&& 69 &  22 &   9\\[3pt]
 {\bf F   }      &        && {\bf   0} & {\bf 100} & {\bf   0}&&{\bf  74} & {\bf  18} & {\bf   8}\\[-3pt]
  $^{19}$F& (100\%)  &&   0 & 100 &   0&& 74 &  18 &   8\\[3pt]
 {\bf Ne  }      &        && {\bf  69} & {\bf  31} & {\bf   0}&&{\bf  71} & {\bf  20} & {\bf   9}\\[-3pt]
 $^{20}$Ne& (61\%)  &&  85 &  15 &   0&& 66 &  24 &  10\\[-3pt]
 $^{21}$Ne& (10\%)  &&   0 & 100 &   0&& 70 &  21 &   9\\[-3pt]
 $^{22}$Ne& (29\%)  &&  62 &  38 &   0&& 76 &  17 &   7\\[3pt]
 {\bf Na  }      &        && {\bf  13} & {\bf  87} & {\bf   0}&&{\bf  80} & {\bf  13} & {\bf   7}\\[-3pt]
 $^{23}$Na& (100\%)  &&  13 &  87 &   0&& 80 &  13 &   7\\[3pt]
 {\bf Mg  }      &        && {\bf  83} & {\bf  16} & {\bf   1}&&{\bf  70} & {\bf  19} & {\bf  10}\\[-3pt]
 $^{24}$Mg& (74\%)  &&  90 &  10 &   0&& 72 &  18 &  10\\[-3pt]
 $^{25}$Mg& (13\%)  &&  53 &  47 &   0&& 70 &  20 &  11\\[-3pt]
 $^{26}$Mg& (13\%)  &&  69 &  22 &   9&& 68 &  22 &  10\\[3pt]
 {\bf Al  }      &        && {\bf  35} & {\bf  65} & {\bf   0}&&{\bf  85} & {\bf   9} & {\bf   7}\\[-3pt]
 $^{26}$Al& (9\%)  &&   0 & 100 &   0&& 77 &  15 &   8\\[-3pt]
 $^{27}$Al& (91\%)  &&  38 &  62 &   0&& 86 &   8 &   6\\[3pt]
 {\bf Si  }      &        && {\bf  91} & {\bf   9} & {\bf   0}&&{\bf  50} & {\bf  29} & {\bf  21}\\[-3pt]
 $^{28}$Si& (90\%)  &&  96 &   4 &   0&& 52 &  26 &  21\\[-3pt]
 $^{29}$Si& (6\%)  &&  50 &  50 &   0&& 48 &  31 &  22\\[-3pt]
 $^{30}$Si& (4\%)  &&  40 &  60 &   0&& 50 &  29 &  20\\[3pt]
\hline\hline
\end{tabular}
\end{table}
\endgroup

%%%%%%%%%%%%%%%%%%%%%

In Table~\ref{tab:origin}, we first see that between O and Si, Mg is the only element with a radioactive contribution (Be, C, N were already discussed in \citetalias{2018PhRvC..98c4611G}), owing to a non-negligible amount of $^{26}$Al decaying into $^{26}$Mg, weighted down in Mg by the subdominant isotopic content of $^{26}$Mg. We then see that F is a quasipure secondary species, as discussed for instance in \cite{2022arXiv220801337F}. We then have Ne, Mg, and Si predominantly of primary origin, and finally Na and Al as prototypes of mixed species. The last three columns show that direct production (1-step) is always the dominant fraction with regard to the total fragmentation production, but with a significant amount of 2-step production (up to $20\%$), and a non-negligible amount of $>2$-step production (up to $10\%$);  these numbers must be compared to the $\lesssim3\%$ precision we have nowadays on CR data, a precision we wish to match from the modeling side. We stress that the odds of having multistep reactions decrease with the number of steps $n$: the timescale for $n$ successive interactions to happen is $n$ times that of a single interaction, which makes it $n$ times less likely than escape from the Galaxy to happen. As a consequence, as we will see later on, this significant $10\%$ level of multistep production seen in Table~\ref{tab:origin} is actually made up of a huge number of tiny contributions.

\section{Definitions and formulas} \label{sec:definitions}
%%%%%%%%%%%%%%%%%%%%%%%%%%%%%%%%%%%%%%%%%%%%%%%%%%%%%%
%%%%%%%%%%%%%%%%%%%%%%%%%%%%%%%%%%%%%%%%%%%%%%%%%%%%%%

In \citetalias{2018PhRvC..98c4611G}, definitions and formulas used in our ranking analyses were scattered in different sections. Here, we gather them all in one section to facilitate reading. We also expand the formulas to new cases and configurations not considered in \citetalias{2018PhRvC..98c4611G}.

We start with the definition of {\em ghost} nuclei, highlighting the scope of cumulative cross-sections in a GCR context in Section~\ref{sec:ghosts}. We then define the $f_{abc}$ coefficients to rank the production cross-sections to measure with high priority in Section~\ref{sec:coeffs_fabc}. From these coefficients, we define in Section~\ref{sec:error_propag} formulas for the error propagation of nuclear cross-section uncertainties to the GCR fluxes (under several plausible assumptions on nuclear data). In Section~\ref{sec:Cab_beamtime}, we consider the case where all fragments from many progenitors can be measured at once, for which we can provide formulas to calculate the beam time properties (reactions and associated number of events) necessary to reach a desired precision on the predicted GCR fluxes. To do so, we introduce the $\C_{ab}$ coefficients, formed from the $f_{abc}$ coefficients and the reaction cross sections.

In order to rank other quantities of interest for CR physicists, other families of coefficients are introduced in Section~\ref{sec:coeffs_others}: the $D_{i \to j}$ coefficients to rank the most important direct production channels calculated from GCR fluxes {\em as measured} (not in \citetalias{2018PhRvC..98c4611G}); the $P^{\rm 1-step}_{ij}$, $P^{\rm 2-step}_{ikj}$\dots\ to rank the production {\em channels}, i.e.\ reaction paths linking a primary flux to a fragment, accounting for direct and multistep production; the $P_{i\to j}$ and $P_{Z_i\to j}$ coefficients (not in \citetalias{2018PhRvC..98c4611G}) to rank the most important progenitors of a given CR species.

\subsection{Ghost nuclei in cumulative cross sections} \label{sec:ghosts}
%%%%%%%%%%%%%%%%%%%%%%%%%%%%%%%%%%%%%%%%%%%%%%%%%%%%%%

Unstable nuclei created by interactions of CRs (on the ISM) whose half-life are $\lesssim$~kyr play a specific role in the GCR modeling. These short-lived nuclei decay before having another interaction and are thus dubbed {\em ghost} nuclei: they are not detected in CR experiments, but they increase the production of their stable daughter nuclei (detected in CR experiments), which is reflected in the cumulative cross sections $\sigma^{\rm cumul}$. In nuclear physics experiments, the production cross section of these ghosts can be measured if their half-life is longer than the time of flight between the target and the detector. If this is not the case, then this implies that the measured cross sections are themselves cumulative of some sort. Hence, going from measured or modeled production cross sections to cumulative cross sections for CR applications cannot be ignored for most CR species.

For a CR projectile $X$ into a stable CR nucleus $Y$ (for any given target $T$), we define $\sigma^{\rm cumul}_{X\rightarrow Y}$ to be the direct production of $Y$, plus the production of all related ghost nuclei $g$ weighted by their branching-ratio decay ${\cal B}r_g$ in $Y$:
\begin{equation}
  \sigma^{\rm cumul}_{X+T\rightarrow {\rm Y}} = \sigma_{X+T\rightarrow {\rm Y}} + \sum_g {\cal B}r_g\cdot \sigma_{X+T\rightarrow g}\;.
\label{eq:cumul}
\end{equation}
We use in the following the ghost and branching ratios\footnote{Note that GALPROP nuc\_package also includes the nuclear reaction network with branching ratios. These include naked and H-like atoms, which may have different decay channels.} gathered in \cite{1984ApJS...56..369L} and expanded in \cite{2001ApJ...555..585M}. {In practice, this nuclear reaction network is built from
Nuclear Data Sheets \cite{2018iii}, which follows multistage chains
of p, n, d, t, ${}^3$He, $\alpha$, $\beta$-decays, and electron K-capture, and,
in many cases, more complicated reactions.} We list in Table~\ref{tab:ghosts} the properties of the ghost nuclei whose production for F to Si fluxes is large enough to make them appear in the ranking tables shown in App.~\ref{sec:coeffs_fabc}.

For the rankings presented in the next section, it is also useful to define $\sigma^{\rm cumul}_{i\rightarrow j}$, the cumulative cross section of CR species $Y$ from $X$ over all ISM targets $T$:
\begin{equation}
  \sigma^{\rm cumul}_{X\rightarrow Y}\equiv \frac{\sum_T n_{\rm ISM}^T \sigma^{\rm cumul}_{X+T\rightarrow Y}}{\sum_T n_{\rm ISM}^T}\,,
  \label{eq:cumulISM}
\end{equation}
with $n_{\rm ISM}^T$ the density of the ISM target $T$.

%\begingroup
%\squeezetable
\begin{table}[!th]%The best place to locate the table environment is directly after its first reference in text
\caption{List of ghost nuclei with significant contributions to CR fluxes $Z=3-14$, i.e.\ appearing in Tables \ref{tab:sortedxs_Be} to \ref{tab:sortedxs_Si}. The different columns show their half-life, decay channel, and branching ratio, taken from \cite{2017ChPhC..41c0001A}.\label{tab:ghosts}}
\begin{tabular}{crcl}
\hline\hline
Nucleus &$T_{1/2}$~~~~~~~~~& Daughter & Decay mode\\
\hline
$^{6}$He  & 806.92~ms & $^{6}$Li  & $\beta^-$  ($100\%$) \\
$^{9}$Li  & 178.3~ms  & $^{9}$Be  & $\beta^-$  ($49.2\%$) \\
$^{11}$Be & 13.76~s   & $^{11}$B  & $\beta^+$  ($97.1\%$) \\
$^{10}$C  & 19.3009~s & $^{10}$B  & $\beta^+$  ($100\%$) \\
$^{11}$C  & 20.364~min  & $^{11}$B  & $\beta^+$  ($100\%$) \\
%$^{15}$C  & 2.449~s   & $^{15}$N  & $\beta^-$  ($100\%$) \\
%$^{16}$C  & 747~ms    & $^{15}$N  & $\beta^-n$  ($97.9\%$) \\
$^{12}$B  & 20.20~ms  & $^{12}$C  & $\beta^-$  ($98.4\%$)\\
$^{13}$N  & 9.965~min   & $^{13}$C  & $\beta^+$  ($100\%$) \\
$^{16}$N  & 7.13~s    & $^{16}$O  & $\beta^-$  ($99.99855\%$) \\
$^{17}$N  & 4.173~s   & $^{16}$O  & $\beta^-n$ ($95\%$)\\
\multirow{2}{*}{$^{13}$O}  & \multirow{2}{*}{8.58~ms\;}\rdelim\{{2}{3mm}\!\!\!\!\!\!\!& $^{13}$C & $\beta^+$ ($89.1\%$) \\
                           &                          & $^{12}$C & $\beta^+p$ ($10.9\%$)\\
$^{14}$O  & 70.620~s  & $^{14}$N  & $\beta^+$  ($100\%$) \\
$^{15}$O  & 122.24~s  & $^{15}$N  & $\beta^+$  ($100\%$) \\
$^{19}$O  & 26.470~s  & $^{19}$F  & $\beta^-$  ($100\%$) \\
$^{17}$F  & 64.370~s  & $^{17}$O  & $\beta^+$  ($100\%$) \\
$^{18}$F  & 109.739~min & $^{18}$O  & $\beta^+$  ($100\%$) \\
$^{20}$F  & 11.163~s  & $^{20}$Ne & $\beta^-$  ($100\%$) \\
$^{21}$F  & 4.158~s   & $^{21}$Ne & $\beta^-$  ($100\%$) \\
$^{19}$Ne & 17.274~s  & $^{19}$F  & $\beta^+$  ($100\%$) \\
$^{23}$Ne & 37.140~s  & $^{23}$Na & $\beta^-$  ($100\%$) \\
$^{21}$Na & 22.422~s  & $^{21}$Ne & $\beta^+$  ($100\%$) \\
$^{22}$Na & 2.60~yr   & $^{22}$Ne & $\beta^+$  ($100\%$) \\
$^{24}$Na & 14.957~h  & $^{24}$Mg & $\beta^-$  ($100\%$) \\
$^{25}$Na & 59.1~s    & $^{25}$Mg & $\beta^-$  ($100\%$) \\
$^{22}$Mg & 3.8755~s  & $^{22}$Ne & $\beta^+$  ($100\%$) \\
$^{23}$Mg & 11.317~s  & $^{23}$Na & $\beta^+$  ($100\%$) \\
$^{27}$Mg & 9.435~min   & $^{27}$Al & $\beta^-$  ($100\%$) \\
$^{25}$Al & 7.183~s   & $^{25}$Mg & $\beta^+$  ($100\%$) \\
$^{28}$Al & 2.245~min   & $^{28}$Si & $\beta^-$  ($100\%$) \\
$^{29}$Al & 6.56~min    & $^{29}$Si & $\beta^-$  ($100\%$) \\
$^{27}$Si & 4.15~s    & $^{27}$Al & $\beta^+$  ($100\%$) \\
$^{30}$P  & 2.498~min   & $^{30}$Si & $\beta^+$  ($100\%$) \\
$^{32}$P  & 14.268~d  & $^{32}$S  & $\beta^-$  ($100\%$) \\
$^{44}$Sc & 4.0420~h  & $^{44}$Ca & $\beta^+$  ($100\%$) \\
\hline\hline
\end{tabular}
\end{table}
%\endgroup

\subsection{Ranking of cross sections: $f_{abc}$ coefficients} \label{sec:coeffs_fabc}
%%%%%%%%%%%%%%%%%%%%%%%%%%%%%%%%%%%%%%%%%%%%%%%%%%%%%%

In the context of GCRs, we wish to establish a priority list of nuclear production cross sections to measure and improve, in order to better model F to Si CR fluxes. Each element has its own ranking (of the most important nuclear reaction), and both the production cross sections and CR fluxes (progenitors) have different energy dependences, so that the various contributions are also energy dependent. Nevertheless, for legibility, throughout Section~\ref{sec:definitions}, these dependences on the specific element considered and energy is left implicit in the various definitions and formulas.

Let us consider the production cross section $\sigma_{\abc}$, where $a$ is a CR projectile, $b$ an ISM target (H or He), and $c$ a fragment (CR or ghost). The impact of this cross section on the total secondary flux $\psi_{\rm sec}$ of a given CR isotope or element is obtained from the relative difference between the standard (or reference) flux calculation, $\psi_{\rm sec}^{\rm ref}$, and the calculation in which we set this cross section to zero, $\psi_{\rm sec}^{\sigma_{\abc}=0}$. In \citetalias{2018PhRvC..98c4611G}, we defined the $f_{abc}$ coefficients to be exactly this difference (i.e.\ impact on the flux):
   \begin{equation}
   \label{eq:coeffs_fabc}
      f_{abc} = \frac{\psi_{\rm sec}^{\rm ref}-\psi_{\rm sec}^{\sigma_{\abc}=\,0}}{\psi_{\rm sec}^{\rm ref}}\;.
   \end{equation}

With this definition, the ranking of the most important production cross sections amounts to the ranking of the $f_{abc}$ coefficients; we stress that this ranking remains unchanged if we consider the impact on the total flux $\psi_{\rm tot}$ instead of $\psi_{\rm sec}$ in Eq.~(\ref{eq:coeffs_fabc}). As underlined in \citetalias{2018PhRvC..98c4611G}, the $f_{abc}$ can be roughly interpreted as fractional contributions, although
\begin{equation}
  \label{eq:sum_fabc}
  \sum_{\forall (a,b,c)}f_{abc}\gtrsim 1\,,
\end{equation}
i.e.\ the sum of all cross-section contributions (for a CR species) is usually larger than one. Indeed, most $\sigma_{a+b\to c}$ are involved in both direct and multistep production, leading to a double counting of some sort\footnote{In a multistep reaction, we kill the associated production when we set to zero the cross section of any of the steps involved: when we sum the associated $f_{abc}$, we add multiple times the same production, hence the double (or multiple) counting.}: the more important these multistep contributions are (which peaks at a few GeV and then decrease with energy, see \citetalias{2018PhRvC..98c4611G}), the larger (than one) the sum of the $f_{abc}$ is.

\subsection{Impact of nuclear uncertainties on fluxes} \label{sec:error_propag}
%%%%%%%%%%%%%%%%%%%%%%%%%%%%%%%%%%%%%%%%%%%%%%%%%%%%%%

The uncertainty on any calculated CR flux $\psi_{\rm tot}$ depends on the individual cross-section uncertainties $\Delta\sigma_{\abc}$. As discussed in \citetalias{2018PhRvC..98c4611G}, this uncertainty can be related to the $f_{abc}$ coefficients (introduced in Section~\ref{sec:coeffs_fabc}) and $f_{\rm sec}$ fraction (shown in Table~\ref{tab:origin}).

We consider here the case of nuclear data coming from patchy measurements of subsets of reactions at different facilities; this is the current status of nuclear cross-section data. Different plausible assumptions on the presence or absence of correlations between these datasets (and in practice, on the modeling of the cross sections based on these data) lead to different error propagation formulas. We discussed three cases in \citetalias{2018PhRvC..98c4611G}, namely:
\begin{itemize}
   \item fully correlated uncertainties (or `corr' for short)
\begin{eqnarray}
\left(\frac{\Delta  \psi_{\rm tot}}{\psi_{\rm tot}}\right)^{\rm corr}\!\! \approx f_{\rm sec}\, \sum_{a,b,c} f_{abc} \frac{\Delta\sigma_{\abc}}{\sigma_{\abc}}\,;
\label{eq:uncertainty_sumCorr}
\end{eqnarray}
   \item fully uncorrelated uncertainties (`uncorr')
\begin{eqnarray}
\left(\frac{\Delta  \psi_{\rm tot}}{\psi_{\rm tot}}\right)^{\rm uncorr}\!\!\!\!\!\!\! &\approx& f_{\rm sec}\, \sqrt{\sum_{a,b,c} \left(f_{abc} \frac{\Delta\sigma_{\abc}}{\sigma_{\abc}}\right)^2}\,;
\label{eq:uncertainty_sumUncorr}
\end{eqnarray}
   \item mixed case  (hereafter `mix'), with uncorrelated uncertainties for fragments of the same projectile, but correlated uncertainties for different projectiles
\begin{eqnarray}
\left(\frac{\Delta  \psi_{\rm tot}}{\psi_{\rm tot}}\right)^{\rm mix} \!\!\!\!&\approx&\!\! f_{\rm sec} \sum_a
\sqrt{\sum_{b,c} \left(f_{abc} \frac{\Delta\sigma_{\abc}}{\sigma_{\abc}}\right)^2}\!.
\label{eq:uncertainty_sumMix}
\end{eqnarray}
\end{itemize}

The first and second cases correspond, respectively, to an optimistic and pessimistic view on the existing status of nuclear data, and consequently on the estimated uncertainties on the CR fluxes. The third case is probably the most realistic, though this is difficult to assess in practice (see discussion in \citetalias{2018PhRvC..98c4611G}).
These formulas were used in \citetalias{2018PhRvC..98c4611G} to illustrate how the flux uncertainties decrease (below the target precision of AMS-02 data), when a growing number of important reactions are perfectly measured. We use these formulas for the same purpose here in Section~\ref{sec:res_errevol}.

\subsection{Beam time calculation: $\C_{ab}$ coefficients} \label{sec:Cab_beamtime}

In our previous paper, we introduced a simple method to use the ranked
production reactions as a guidance for laboratory measurements on how
to estimate the required number of interactions to achieve a specific
flux uncertainty. Since in an experiment dedicated to the measurements
of fragmentation cross sections, many fragments associated with a
single projectile are measured at once, we can estimate the
statistical uncertainty that such an experiment can achieve depending
on the amount of recorded interactions.

\paragraph*{Flux uncertainty from one measured reaction.}
In the reaction $a+b$, fragments of type $c$ are produced with
probability
\begin{equation}
  p_c = \frac{\sigma_{\abc}}{\sigma_{a+b}},
\end{equation}
where $\sigma_{a+b}$ is the total inelastic cross section for $a+b$
reaction and $\sigma_{\abc}$ is the fragmentation cross section $\abc$
to produce fragment $c$. If $N$ interactions are recorded, then
the measured number of fragments, $n_c = p_c\, N$, is multinomially
distributed. If the number of interactions itself is a Poissonian
random variable, then the number of fragments is independent and
follows a Poissonian distribution with variance $n_c$, see
e.g.~\cite{1997sda..book.....C}, leading to an experimental variance
of the partial cross section of
\begin{equation}
  V(\sigma_\abc)  = \frac{1}{N} \, \sigma_{a+b}^2 \, p_c = \frac{1}{N} \,\sigma_{a+b} \,\sigma_\abc.
  \label{eq:Var}
\end{equation}

The combined relative uncertainty on the flux (total flux $\psi_{\rm tot}$ denoted $\psi$ in the rest of the section) from all reactions in $a+b$ interactions follows
from error propagation,
\begin{equation}
   \left(\frac{\Delta \psi}{\psi}\right)_{a+b}^2 = \frac{f_{\rm sec}^2}{\psi^2}
   \sum_{c} \left(\frac{\partial \psi}{\partial \sigma_{\abc}}\right)^2 V(\sigma_\abc)\,,
   \label{eq:errProp}
\end{equation}
where $f_{\rm sec}$ denotes the fraction of secondaries contributing to the
flux (cf.\ Table~\ref{tab:origin}). Using
$\partial \psi/\partial \sigma_\abc = (\psi/\sigma_\abc) f_{abc}$ and Eq.~(\ref{eq:Var}), this can be written as
\begin{equation}
  \left(\frac{\Delta \psi}{\psi}\right)_{a+b}^2 = \frac{ f_{\rm sec}^2}{N} \C_{ab}^2,
\end{equation}
where we introduced the $\C_{ab}$ coefficients
\begin{equation}
   \C_{ab}^2 \equiv \sum_{c} f_{abc}^2 \frac{\sigma_{a+b}}{\sigma_\abc}.
   \label{eq:Cab}
\end{equation}
Note that the $\C_{ab}$ defined here is mathematically equivalent to
our previous definition after expanding the squared sum in Eq.~(18) of
\citetalias{2018PhRvC..98c4611G}:
$$\left(\,\sum x_i\right)^2 = \sum_i x_i^2 + 2\sum_{i=1}^{m-1}\sum_{j=i+1}^{m} x_i x_j,$$
and canceling the common terms.

\paragraph*{Flux uncertainty from all contributing reactions.}
The above uncertainty is only for one contributing reaction.
The total flux uncertainty is obtained from the quadratic sum over all
contributing reactions (i.e.\ reactions $a+b$, $d+e$, $f+g$, etc.).
Labelling these $n_r$ reactions with the index $k$, we can write
$\{\C_{ab}, \C_{de}, \C_{fg},\dots\}$ as $\{\C_k\}_{k=1\dots n_r}$,
and assuming $N_k$ interactions recorded for each reaction $k$,
we get
\begin{equation}
    \left(\frac{\Delta \psi}{\psi}\right)^2 = f_{\rm sec}^2 \sum_{k=1}^{n_r} \frac{1}{N_k}\, \C_{k}^2.
\end{equation}

\paragraph*{Optimizing the total number of reactions to measure.}
The requirement of having the total relative flux uncertainty to be smaller than a certain value
$\xi$,
\begin{equation}
  f_{\rm sec}^2 \sum_{k=1}^{n_r} \frac{1}{N_k}\, \C_{k}^2 \equiv \sum_{k=1}^{n_r} {V}_k \leq \xi^2\,,
\end{equation}
has no unique solution for the $N_{k=1\dots n_r}$ number of interactions (for each reaction $k$) to be
measured. We try the ansatz
\begin{equation}
{V}_k = \xi^2 \frac{\C_{k}^\beta}{\sum_{k=1}^{n_r} \C_{k}^\beta}
\end{equation}
for the partial contribution ${V}_k$ of reaction $k$ to the overall
variance, which corresponds to
\begin{equation}
  N_k =  \frac{f_{\rm sec}^2}{\xi^2}\,  \,  \C_{k}^{2-\beta} \sum_{k=1}^{n_r} \C_{k}^\beta
  \label{eq:nk}
\end{equation}
interactions to be measured for reaction $k$, and in total
\begin{align}
  N_\text{tot} =&  \sum_{k=1}^{n_r} N_k = \frac{f_{\rm sec}^2}{\xi^2} \, \left(\sum_{k=1}^{n_r} \C_{k}^{2-\beta}\right) \left(\sum_{k=1}^{n_r} \C_{k}^\beta\right)\,.
  \label{eq:ntot}
\end{align}

In \citetalias{2018PhRvC..98c4611G} we investigated equal partial variances, $\beta=0$ and ${V}_k=\xi^2/n_r$, which leads to $N_k = n_r\, (f_{\rm sec}\C_{k}/\xi)^2$. This scheme has
the undesired effect
that adding reactions with low flux impact (small $\C_{k}^2$) will artificially increase $N_k$ of all the other reactions via the number of considered reactions $n_r$.

The scheme that \emph{minimizes} the total number of interactions to be recorded is given by $\beta=1$, since for this value the derivative of Eq.~(\ref{eq:ntot}),
\begin{align}
  \frac{\partial N_\text{tot}}{\partial \beta} =
  \frac{f_{\rm sec}^2}{\xi^2} \Biggl[&\;\quad
    \left(\sum_{k=1}^{n_r}  \C_{k}^\beta \log_{10} \C_k\right)
    \left(\sum_{k=1}^{n_r} \C_{k}^{2-\beta}\right)  \nonumber \\
    &- \left(\sum_{k=1}^{n_r} \C_{k}^{2-\beta}\log_{10} \C_k \right)\left(\sum_{k=1}^{n_r} \C_{k}^\beta\right)\Biggr],
\end{align}
is zero. Such an optimization allows significant work effort to be conserved and yet fulfills our primary goal.

We show in Section~\ref{sec:res_beamtime} the $\C_{ab}$ coefficients we obtain for Li to Si, and the desired number of events to reach a specified accuracy in the modeled GCR fluxes. These numbers are then be used to provide forecasts in Section~\ref{sec:forecasts}, in order to highlight how such measurements would be the game changer for GCR physics.

\subsection{Other rankings of astrophysical interest} \label{sec:coeffs_others}
%%%%%%%%%%%%%%%%%%%%%%%%%%%%%%%%%%%%%%%%%%%%%%%%%%%%%%

We stress that nuclear or particle physicists, whose goal is to make new measurements, should focus on the ranking of reactions obtained via the $f_{abc}$ (and associated $\C_{ab}$) coefficients above. However, other coefficients, corresponding to different groupings of the production cross-sections, can be formed. These coefficients enable the ranking of other quantities of interest for CR physicists, as described below.

\subsubsection{Direct contributions from measured fluxes: $D_{i\to j}$ coefficients} \label{sec:coeffs_Dij}
%%%%%%%%%%%%%%%%%%%%%%%%%%%%%%%%%%%%%%%%%%%%%%%%%%%%%%

At first order, the most important reactions for the production of CR species $j$ are those responsible from the direct production of $j$ from the measured CR isotopic fluxes $i$, as provided for instance in \cite{2013ICRC...33..803M}. Indeed, as some of the measured fluxes are partly or fully of secondary origin, accounting for the direct production only misses cross sections involved in two-step reactions (reactions allowing to go from a primary species into this secondary flux).

Denoting $\sigma^{\rm cumul}_{i\rightarrow j}$ the cumulative cross section of CR species $j$ from $i$ over all ISM targets $t$, see Eq.~(\ref{eq:cumulISM}), these direct contributions are calculated from
\begin{equation}
  \label{eq:coeffs_Dij}
  D_{i\to j} = \frac{\psi_{\rm sec}^{j,\,\rm ref}-\psi_{\rm sec}^{j,\,\sigma^{\rm cumul}_{i\to j} =\,0}}{\psi_{\rm sec}^{j,\,\rm ref}}\,,
\end{equation}
and by construction we have
\begin{equation}
  \sum_i D_{i\to j} = 1.
\end{equation}
The ranking of the above coefficients is directly the ranking of most important production reactions (not separating the various targets and ghost contributions), as opposed to the $f_{abc}$ corresponding to the ranking of all involved cross sections (Section \ref{sec:coeffs_fabc}).

We did not consider these coefficients in \citetalias{2018PhRvC..98c4611G}, but we do in this paper (see next section). We also consider the case of an element $Z_i$ ($n_i$ isotopes) into $Z_j$ ($n_j$ isotopes), given by the average
\begin{equation}
  \displaystyle D_{Z_i\to Z_j} = \frac{1}{n_{Z_i}n_{Z_j}} \times \sum_{i\in Z_i} \sum_{j\in Z_j} D_{i\to j}\,;
  \label{eq:D_Zi_Zj}
\end{equation}
these coefficients also satisfy $\sum_{Z_i} D_{Z_i\to Z_j}=1$.

\subsubsection{Channels from primary fluxes: $P^{\rm 1-step}_{ij}$ and $P^{\rm 2-step}_{ikj}$ coefficients} \label{sec:coeffs_Pij_steps}
%%%%%%%%%%%%%%%%%%%%%%%%%%%%%%%%%%%%%%%%%%%%%%%%%%%%%%

In \citetalias{2018PhRvC..98c4611G}, a {\em channel} was defined to be a unique production path linking a CR progenitor $i$ to a CR fragment~$j$. This production path can be of length one (direct production) or larger than one (multistep production), associated in this paper to the $P^{\rm 1-step}_{ij}$, $P^{\rm 2-step}_{ikj}$, etc. coefficients defined below. What we did not stress enough in \citetalias{2018PhRvC..98c4611G} is that a channel can only start from a CR having a primary source term, hence the notation $P$ here (at variance with the notation chosen in \citetalias{2018PhRvC..98c4611G}). Compared to the $D_{ij}$ coefficients above (Section~\ref{sec:coeffs_Dij}), the $P^{\rm 1-step}_{ij}$ coefficients only start from species $i$ having a non-zero primary component. However, the $P^{\rm 2-step}_{ij}$ coefficients describe the 2-step contributions not accounted for in the $D_{ij}$. In that respect, the full list of the production cross sections that matter most is directly the reactions appearing in the ranked channels.

These coefficients are obtained from the relative difference between the secondary flux (of a CR species $j$) calculated from the case in which only a specific channel is open (i.e.\ all cross sections set to zero except for those involved in this channel) and the reference calculation $\psi^{j,\,\rm ref}_{\rm sec}$ (where all cross sections are set to their fiducial values) define the $P^{\rm 1-step}_{ij}$ (and $P^{\rm 2-step}_{ikj}$, \dots) coefficients associated with 1-step (2-step, \dots) channels:
\begin{equation}
   \begin{cases}
   \label{eq:coeffs_Pij_Pikj}
      \displaystyle P^{\rm 1-step}_{ij} = \frac{\psi_{\rm sec}^{j,\,\sigma^{\rm cumul}_{m\rightarrow n} =\,0 \;\,\forall (m,n)\neq (i,j)}}{\psi_{\rm sec}^{j,\,\rm ref}}\;,\\
      \displaystyle P^{\rm 2-step}_{ikj} = \frac{\psi_{\rm sec}^{j,\,\sigma^{\rm cumul}_{m\rightarrow n}=\,0 \;\,\forall (m,n)\neq \{(i,k),(k,j)\}}}{\psi_{\rm sec}^{j,\,\rm ref}}\,,\\[0.4cm]
      \displaystyle P^{\rm 3-step}_{ikpj}=\dots\\[0.2cm]
      \dots
   \end{cases}
\end{equation}

As underlined in \citetalias{2018PhRvC..98c4611G}, the list of channels for all possible progenitors $i$ of a CR fragment $j$ (under scrutiny) defines all its possible and independent contributions. Hence, contrarily to the $f_{abc}$ coefficients (see Eq.~[\ref{eq:sum_fabc}]), but similarly to the $D_{ij}$ coefficients, we do have, for the sum of these coefficients,
\begin{equation}
   \sum_{i}^{i>j} \Bigg( P^{\rm 1-step}_{ij} + \sum_{k}^{i>k>j} \bigg(P^{\rm 2-step}_{ikj} + \sum_{k}^{i>k>j} \big(\dots \big)\bigg)\Bigg)= 1.
   \label{eq:sum_Pij_Pijk}
\end{equation}
In practice, we only calculate the $P^{\rm 1-step}_{ij}$ and $P^{\rm 2-step}_{ikj}$ coefficients, and regroup the remaining coefficients under
\begin{equation}
  P_j^{\rm >2-step} =  1 - \sum_{i}^{i>j} \Bigg( P^{\rm 1-step}_{ij} + \sum_{k}^{i>k>j} P^{\rm 2-step}_{ikj} \Bigg).
  \label{eq:sum_sup2step}
\end{equation}
This is sufficient to rank the most important direct and 2-step channels, and to estimate the relative importance of the remaining multistep contributions. Although individual multistep reactions are generally subdominant, their sum can contribute to up to $25\%$ at 10~GeV/n (see Table~\ref{tab:origin}). Actually, multistep contributions decrease with energy, and the ratio of contributing fractions $P^{(n+1)-\rm step}/P^{n-\rm step}\propto R^{-\delta}$ (with $\delta\approx 0.5$ the slope of the diffusion coefficient), as discussed in detail in \citetalias{2018PhRvC..98c4611G}.

Finally, if one is interested in the ranking of the production channels of a CR element $Z_j$ instead of a CR isotope $j$, then the above formulas can still be used by substituting $\psi_{\rm sec}^{j\,,\rm ref} \to \psi_{\rm sec}^{Z_j\,,\rm ref} $ into Eq.~(\ref{eq:coeffs_Pij_Pikj}), and summing also over all $j \in Z_j$ in Eqs.~(\ref{eq:sum_Pij_Pijk})-(\ref{eq:sum_sup2step}), with the meaning that $Z_j$ is the element under scrutiny and $j$ is any isotope of the element $Z_j$.

\subsubsection{Progenitors from cumulative of channels: $P_{i\to j}$ coefficients} \label{sec:coeffs_Pij}
%%%%%%%%%%%%%%%%%%%%%%%%%%%%%%%%%%%%%%%%%%%%%%%%%%%%%%

As introduced in \cite{2022A&A...668A...7M}, we can further group the above coefficients in order to specifically highlight the relative importance of different CR primary progenitors $i$ in the production of a CR isotope $j$. By summing over all possible intermediate fragments between $i$ and $j$, we form
\begin{equation}
   \label{eq:coeffs_Pij}
    P_{i\to j} \!=\! P^{\rm 1-step}_{ij} + \!\!\!\sum_{k}^{i>k>j}\!\!\! \Bigg(P^{\rm 2-step}_{ikj} + \!\!\!\sum_{p}^{k>p>j}\!\!\! \bigg(P^{\rm 3-step}_{ikpj} + \dots\bigg)\!\Bigg).
\end{equation}
As for the channel coefficient, see Eq.~(\ref{eq:sum_Pij_Pijk}), the sum of these new coefficients is also one, i.e.
\begin{equation}
   \label{eq:sum_fitoj}
   \sum_i P_{i\to j} = 1\,.
\end{equation}
We stress that the $P_{i\to j}$ and $D_{i\to j}$ (see Section~\ref{sec:coeffs_Dij}) correspond to different quantities: while the $P_{i\to j}$ coefficients quantify the importance of the primary species from which a CR $j$ under scrutiny comes from (accounting from multistep fragmentation starting from this primary progenitor), the $D_{i\to j}$ coefficients quantify the importance of the direct production of CR $j$ from the measured CR fluxes (independently of their primary and secondary origin).

In practice, because it is time consuming to calculate all orders of the $P_{i\dots j}^{n-\rm step}$ coefficients, we directly calculate the $P_{i\to j}$ via
\begin{equation}
  \label{eq:coeffs_fitoj_num}
  P_{i\to j} = \frac{\psi_{\rm sec}^{j,\;\sigma_{k\to p}=\,0\,\;\forall k>i,\,\forall p}\;-\;\psi_{\rm sec}^{j,\;\sigma_{i\to k}=\,0\,\;\forall k}}{\psi_{\rm sec}^{j,\,\rm ref}}\;,
\end{equation}
that is the ratio of the difference between the calculation `all contributions from CRs heavier than $i$ set to zero' and `all contributions starting from $i$ set to zero', to the reference calculation $\psi^{j,\,\rm ref}_{\rm sec}$; by construction and as also checked directly from the numerical calculation of these coefficients, their sum is one.

Instead of the production of an isotope $j$, we can extend these coefficients to the production of an element $Z_j$, i.e.\ $P_{i\to Z_j}$, by substituting $\psi_{\rm sec}^{j} \to \psi_{\rm sec}^{Z_j} $ in Eq.~(\ref{eq:coeffs_fitoj_num}) and summing over all isotopes $j$ (of this element $Z_j$) in the right-hand side of Eqs.~(\ref{eq:coeffs_Pij}) and (\ref{eq:sum_fitoj}). We can also extend the formula to the production from an element $Z_i$ (instead of an isotope $i$), with $n_i$ the number of isotopes of element $Z_i$,
\begin{equation}
  \begin{cases}
  \displaystyle P_{Z_i\to j} =  \frac{1}{n_{Z_i}}\times \sum_{i\in Z_i} P_{i\to j}\,,\\
  \displaystyle P_{Z_i\to Z_j} =  \frac{1}{n_{Z_j}}\times \sum_{i\in Z_i} P_{i\to Z_j}\,,
  \end{cases}
  \label{eq:P_Zi_Zj}
\end{equation}
which also satisfies $\sum_{Z_i} P_{Z_i\to j}=1$ and $\sum_{Z_i} P_{Z_i\to Z_j}=1$.
The $P_{Z_i\to Z_j}$ coefficients are shown in Sect.~\ref{sec:res_rankings} and their isotopic counterparts, the $P_{i\to j}$ coefficients, are shown in App.~\ref{app:fig_Dij_Pij}.

The coefficients introduced in this section are directly the contributing fractions of CR $i$ (or element $Z_i$) into CR $j$ (or element $Z_j$), they weight the importance of both the various CR fluxes and production cross sections in a single energy-dependent number. As a result, ranking these coefficients amounts to ranking the various contributors (or progenitors) of $j$. We stress that in some cases, a progenitor may rank quite high while the associated channels (i.e.\ $P_{i\dots j}$ coefficient discussed in Section~\ref{sec:coeffs_Pij_steps}) all rank low: the contribution of numerous subpercentage multistep production channels sometimes combine into a significant overall contribution. Finally, because these coefficients include direct and multistep contributions in variable amounts, their energy dependence is a weighted-combination of the single and multistep associated energy dependence, and thus less trivial than their $P_{i\dots j}$ counterparts.

\section{Results at 10~GeV/n (on TOA fluxes)} \label{sec:results}
%%%%%%%%%%%%%%%%%%%%%%%%%%%%%%%%%%%%%%%%%%%%%%%%%%%%%%
%%%%%%%%%%%%%%%%%%%%%%%%%%%%%%%%%%%%%%%%%%%%%%%%%%%%%%

We discuss here our main results in a few summary figures. We provide more information and details in the Appendix, where additional plots and tables are gathered.

\subsection{Graphical views of the various rankings} \label{sec:res_rankings}
%%%%%%%%%%%%%%%%%%%%%%%%%%%%%%%%%%%%%%%%%%%%%%%%%%%%%%

In \citetalias{2018PhRvC..98c4611G}, we provided tables for the rankings of the $f_{abc}$ coefficients (\ref{sec:coeffs_fabc}) and the channels (\ref{sec:coeffs_Pij_steps}). Here, we highlight via graphics the different information contained in the $D_{Z_i\to Z_j}$ (\ref{sec:coeffs_Dij}), $P_{Z_i\to Z_j}$, and $f_{abc}$ coefficients, and what they tell us about GCRs and more importantly, about the production cross sections.

\begin{figure}[t]
\includegraphics[width=1.02\columnwidth]{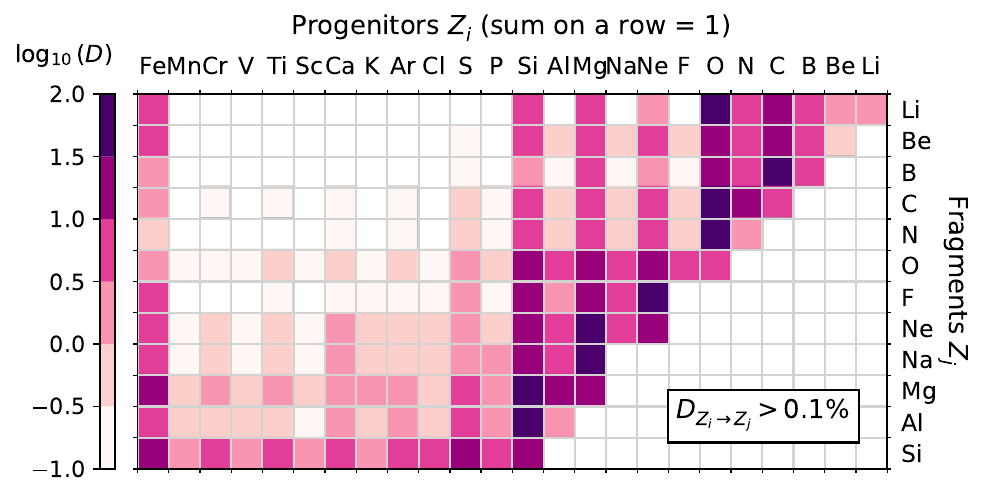}
\vspace{0.05cm}\\
\includegraphics[width=1.02\columnwidth]{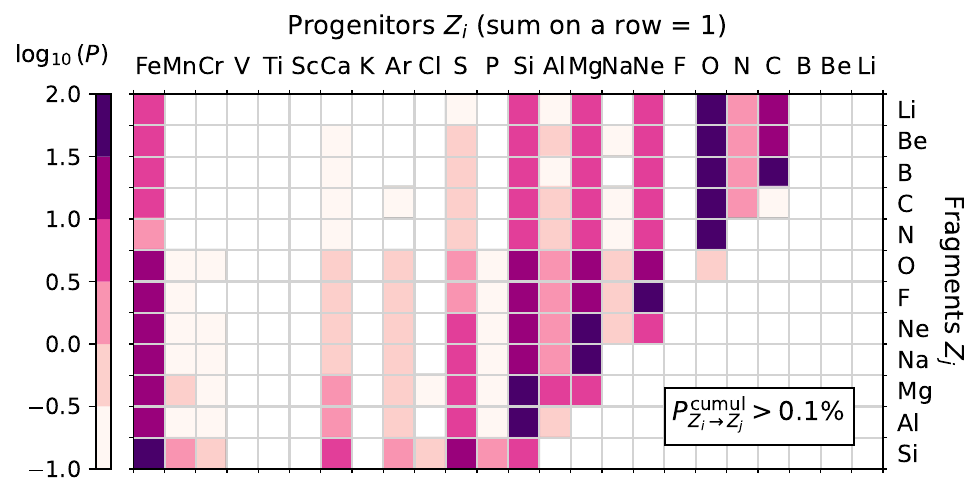}
\caption{Secondary production of GCR elements $Z_j$ (right-hand side labels) from GCR elements $Z_i$ at 10~GeV/n. The color bar encodes $\log_{10}$ of the contributing fractions for this production, from 2.0 ($100\%$) down to -1.0 ($0.1\%$); only contributions larger than $0.1\%$ are shown. The top panel corresponds to the $D_{Z_i\to Z_j}$ coefficients (\ref{sec:coeffs_Dij}) and the bottom panel to the $P_{Z_i\to Z_j}$ ones (\ref{sec:coeffs_Pij}). For both cases, the sum of coefficients on one row is 100\%, i.e.\ all the secondary production, but we recall that the latter ($f_{\rm sec}$) is sometimes only a small fraction of the total flux (see Table~\ref{tab:origin}). See text for discussion.
\label{fig:Zi_Zj}
}
\end{figure}

Figure~\ref{fig:Zi_Zj} shows the contributing production fractions (larger than $0.1\%$) of elements $Z_i$ into $Z_j$ at 10~GeV/n; for each $Z_j$, the sum of the coefficients on a row, by definition, adds up to $100\%$ of the secondary production.
The top panel shows the $D_{Z_i\to Z_j}$ coefficients (not considered in \citetalias{2018PhRvC..98c4611G}), defined in Eq.~(\ref{eq:D_Zi_Zj}), corresponding to the direct production of the elements under scrutiny ($y$-axis on the right) from GCR elemental fluxes ($x$-axis on top) as measured. This plot highlights the standard results, for instance, that most of LiBeB comes from the fragmentation of C and O, with some contributions from other primary species (eg., Si, Fe). For the mostly secondary species F, the main progenitors are Ne, Mg, and Al. Moving to heavier elements, the main progenitors shift to Mg, Si, and Fe. Going back to Li, this plot also shows that some of its fraction comes from the fragmentation of Be and B (and even its own isotopes, like $^7$Li), i.e.\ secondary species. What this plot does not show is the origin and importance of the reactions contributing to the production of these secondary species.

This information is actually encoded into the ranking of channels, via the associated coefficients defined in Eq.~(\ref{eq:coeffs_Pij_Pikj}). A full discussion of the ranking and importance of these channels for Li to Si elements, and their energy dependences, are shown in Appendix~\ref{app:plot_Edep_rankings}. Here, we only show in the bottom panel of Fig.~\ref{fig:Zi_Zj} the $P_{Z_i\to Z_j}$ coefficients defined in Eq.~(\ref{eq:P_Zi_Zj}): these coefficients are built from regrouping the 1-step and multistep channels, and correspond to the cumulated production of $Z_j$ from the primary fluxes $Z_i$. Compared to the top panel, as illustrated on the Li case, we stress that: (i) there is no production of Li from B, Be, and Li, because the latter have no primary component; (ii) the production of Li or F from Fe is more important in the bottom panel, because not only does it include direct fragmentation of Fe into Li (as seen in the top panel) but it also includes multistep fragmentation of Fe into Li. In other words, these coefficients provide a direct view of which primary elements contribute to the production of all GCR fluxes. We stress that we focused the discussion on elements, but the behavior and conclusions we just drew are similar if isotopes are considered instead; the associated $D_{i\to j}$ and $P_{i\to j}$ coefficients are shown in Appendix~\ref{app:fig_Dij_Pij}.

\begin{figure*}[t]
\includegraphics[width=0.53\textwidth]{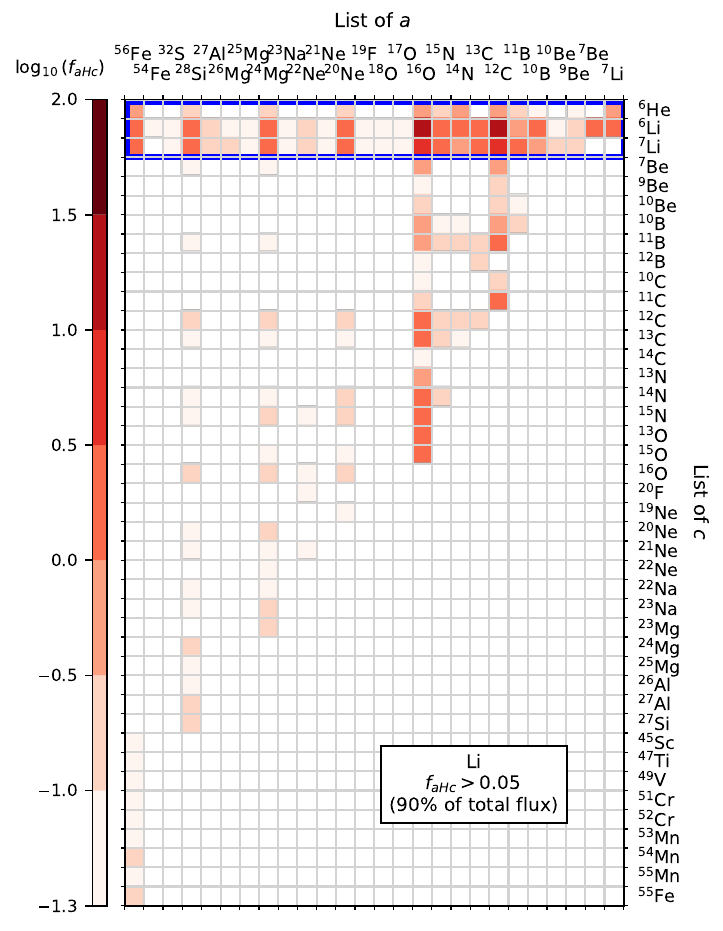}
\includegraphics[width=0.445\textwidth]{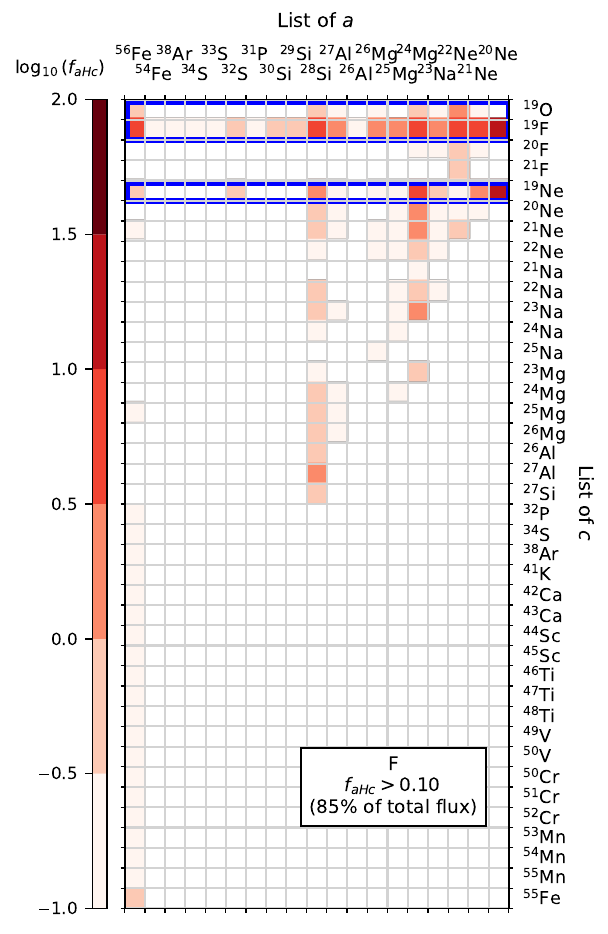}  
\caption{Graphical view of the ranked reactions to measure for the production of GCR Li (left) and F (right) at 10~GeV/n; similar plots for other elements are gathered in Appendix~\ref{app:tablesxs_plot2DfaHc}. The reactions shown are those making up $90\%$ (respectively $85\%$) of the total Li flux (respectively F flux), corresponding to $f_{aHc}$ coefficients larger than 0.05 (respectively 0.10). The color bar encodes $\log_{10}(f_{aHc})$ from 2.0 (100) down to the minimal $f_{aHc}$ value shown; we recall that $\sum_{\forall a,b,c}f_{abc}>1$, so that the $f_{aHc}$ coefficients cannot be interpreted as percentage fractions. The top labels show the list of CR nuclei $a$ involved, while the right labels show the list of $c$ fragments produced from $a$ interacting on H. These fragments are GCR isotopes themselves but also short-lived, i.e.\ ghost, nuclei: for instance $^6$He in the left panel decays into $^6$Li (see Table~\ref{tab:ghosts}). The blue rectangles highlight the direct production of Li (left panel) and F (right panel); the reactions outside these rectangles involve intermediate steps in the production of the element under scrutiny.\label{fig:faHc_LiF}}
\end{figure*}

As stressed in Section~\ref{sec:coeffs_fabc}, the full information on the importance of the various production cross sections (involved in direct or multistep production) are encoded in the $f_{abc}$ coefficients. Tables of these ranked coefficients at 10~GeV/n are provided in Appendix~\ref{app:tablesxs_plot2DfaHc} for Li to Si species.
We consider below the $f_{aHc}$ coefficients, i.e.\ where the ISM target $b$ is H, discarding interactions on He (but keep in mind that some of the latter rank high, see Appendix~\ref{app:tablesxs_plot2DfaHc}). We show these coefficients in Fig.~\ref{fig:faHc_LiF} for the production of Li (respectively F) on the left (respectively right) panel. As in Fig.~\ref{fig:Zi_Zj}, we identify the dominant progenitors (here $^{56}$Fe, $^{28}$Si, $^{24}$Mg, etc.) for the production of Li via $^6$Li and $^7$Li (respectively F via $^{19}$F and the ghost nucleus $^{19}$Ne). At variance with Fig.~\ref{fig:Zi_Zj}, the $y$-axis on the right-hand side now includes all isotopes up to $^{55}$Fe. The latter can be intermediary steps to the production of Li (respectively F). As a result, we see a lot of smaller contributions associated to multistep fragmentation, which include fragmentation of $^{56}$Fe into Sc to Fe isotopes (respectively S to Fe isotopes). The sparse matrix of the $f_{aHc}$ coefficients directly tells us what reactions are involved to reach a given fraction of the total flux. We provide in Appendix~\ref{app:tablesxs_plot2DfaHc} similar figures for Be, B, N, Na, and Al.

We finally stress that for Li and F shown in Fig.~\ref{fig:faHc_LiF}, we limited ourselves to the $f_{abc}$ coefficients leading to $90\%$ and $85\%$ of the total flux respectively (in order to have legible and matching in size figures). For instance, to build up $97\%$ of the total flux, we need to consider 767 reaction for Li, 823 for Be, 550 for B, 85 for C, 452 for N, 33 for O, 1030 for F, 417 for Ne, 840 for Na, 189 for Mg, 651 for Al, and 139 for Si. Hence, even for mostly primary species, the number of reactions to consider can be large; the number of reactions needed to make up different fractions of the secondary or total flux are indicated in Tables~\ref{tab:sortedxs_Li} to~\ref{tab:sortedxs_Si}. As a result, reaching higher precision in the GCR flux modeling will always be at the price of increasing efforts to measure a growing number of reactions at high precision.
We can also stress that fragmentation of Fe has a growing importance when moving to heavier GCR elements.

\subsection{Desired reactions} \label{sec:res_errevol}
%%%%%%%%%%%%%%%%%%%%%%%%%%%%%%%%%%%%%%%%%%%%%%%%%%%%%%

We discussed in Section~\ref{sec:error_propag} different assumptions made on the status of current nuclear data uncertainties to propagate them to GCR fluxes predictions. These assumptions are fully uncorrelated (`uncorr.'), fully correlated (`corr'), or a mixture of both (`mix'), and are expected to bracket the reality of patchy data obtained over the years from very different facilities and measuring techniques. If future nuclear data are taken following the same patchiness, then we can use the above assumptions to make rough estimates on how new measurements will incrementally improve GCR modeling.

\begin{figure*}[t]
\includegraphics[width=0.32\textwidth]{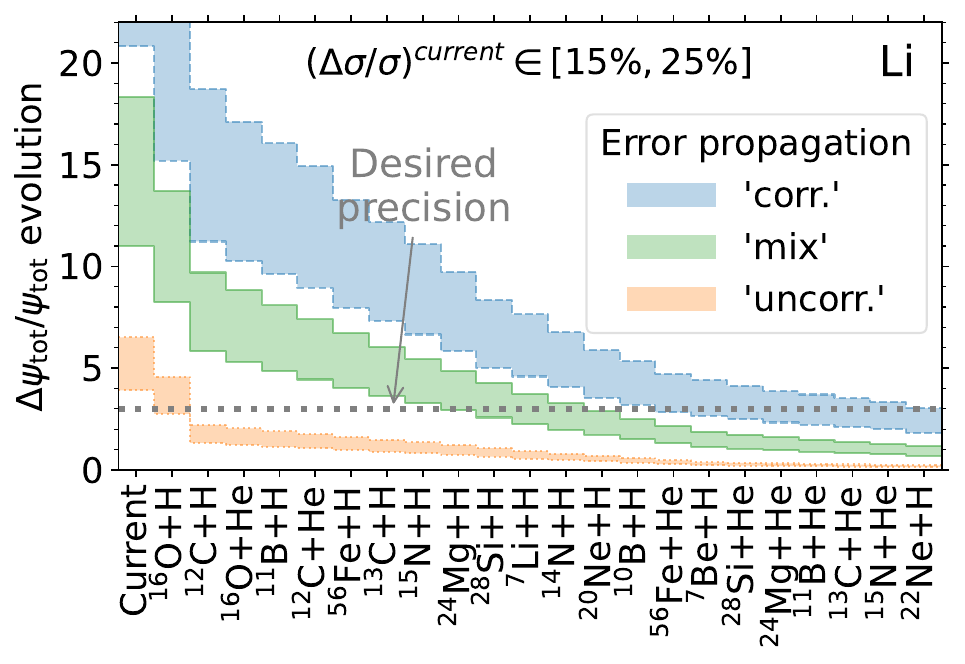}
\includegraphics[width=0.32\textwidth]{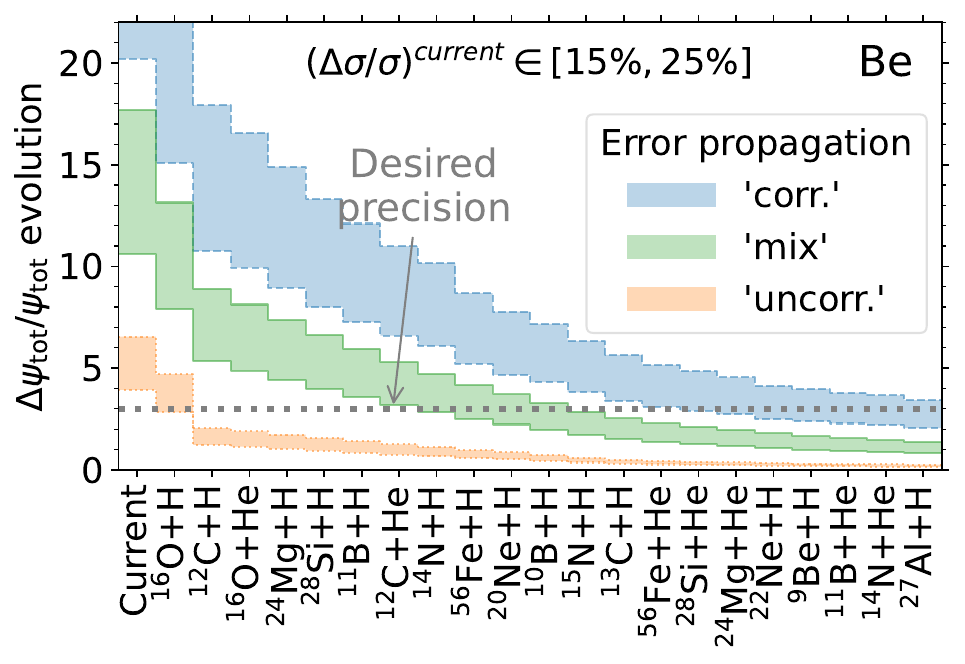}
\includegraphics[width=0.32\textwidth]{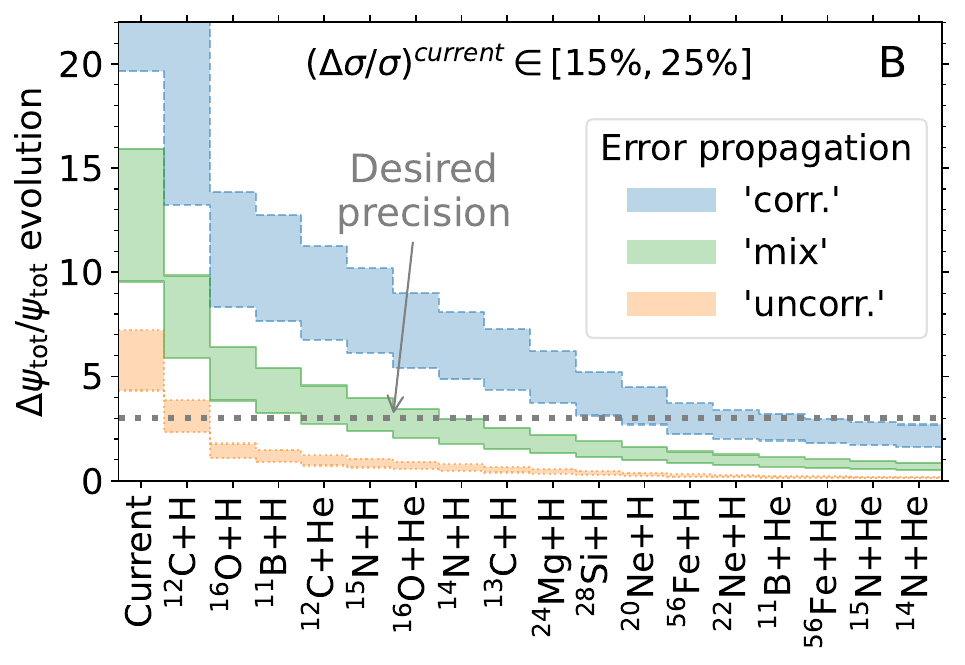}
\includegraphics[width=0.32\textwidth]{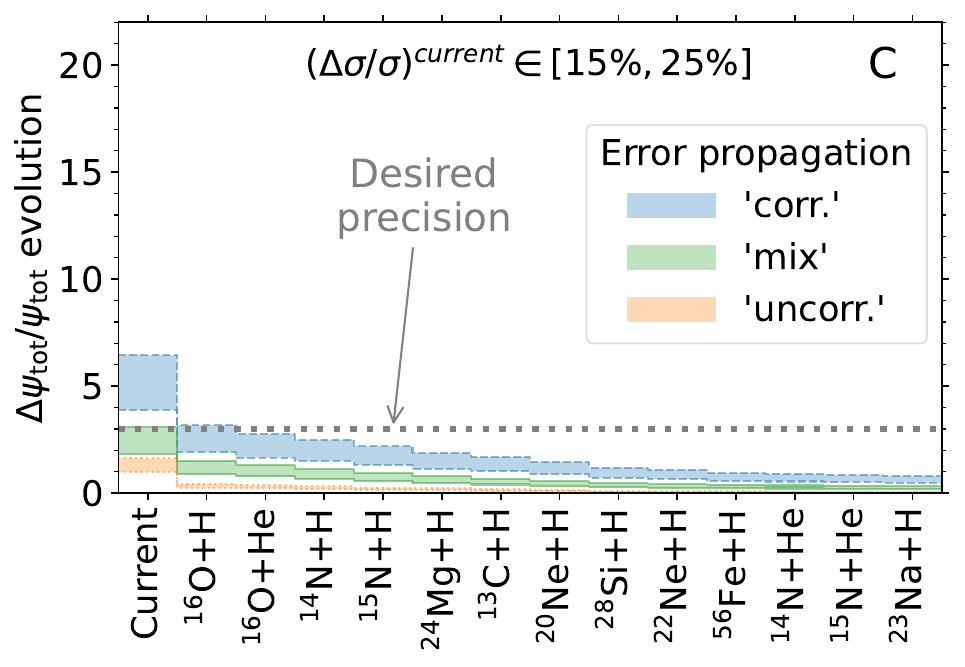}
\includegraphics[width=0.32\textwidth]{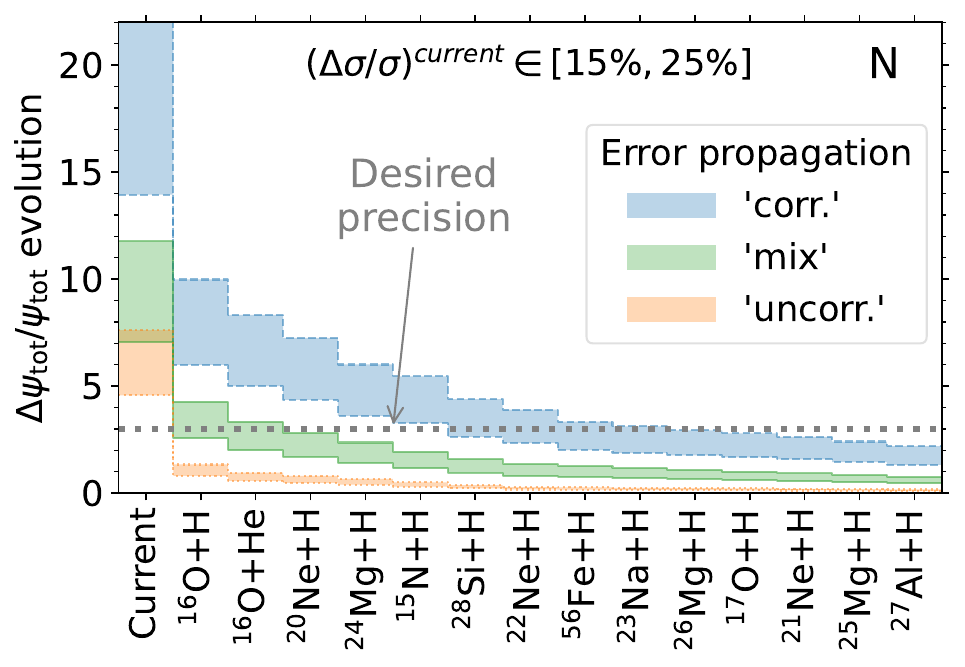}
\includegraphics[width=0.32\textwidth]{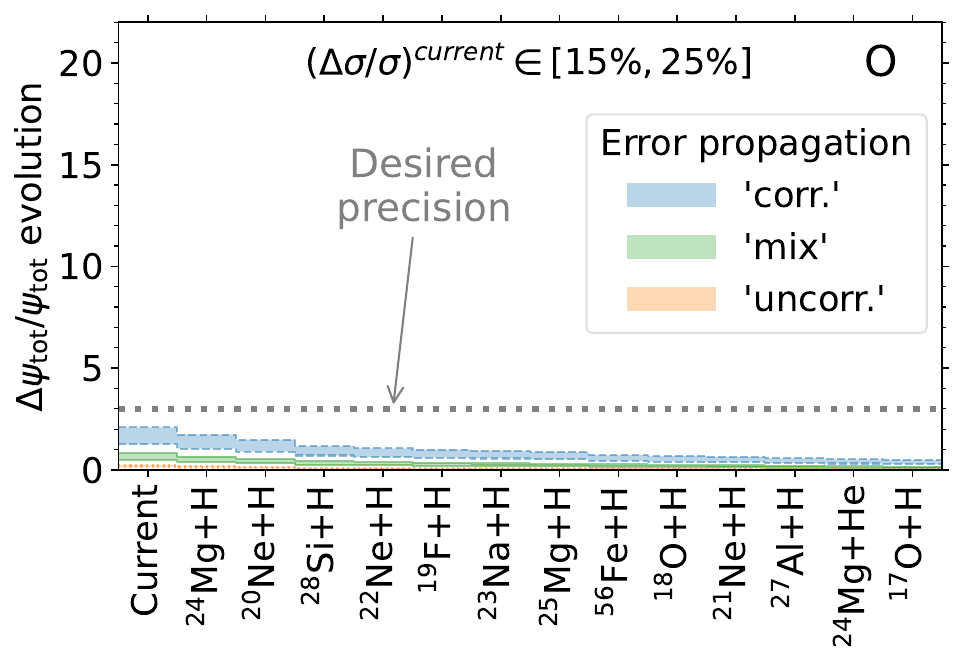}
\includegraphics[width=0.32\textwidth]{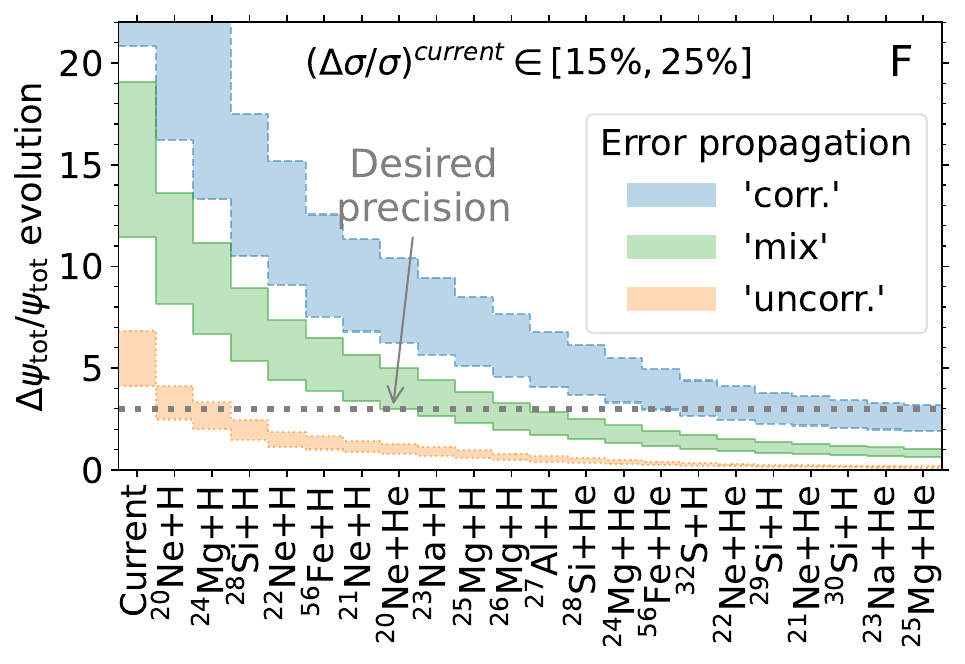}
\includegraphics[width=0.32\textwidth]{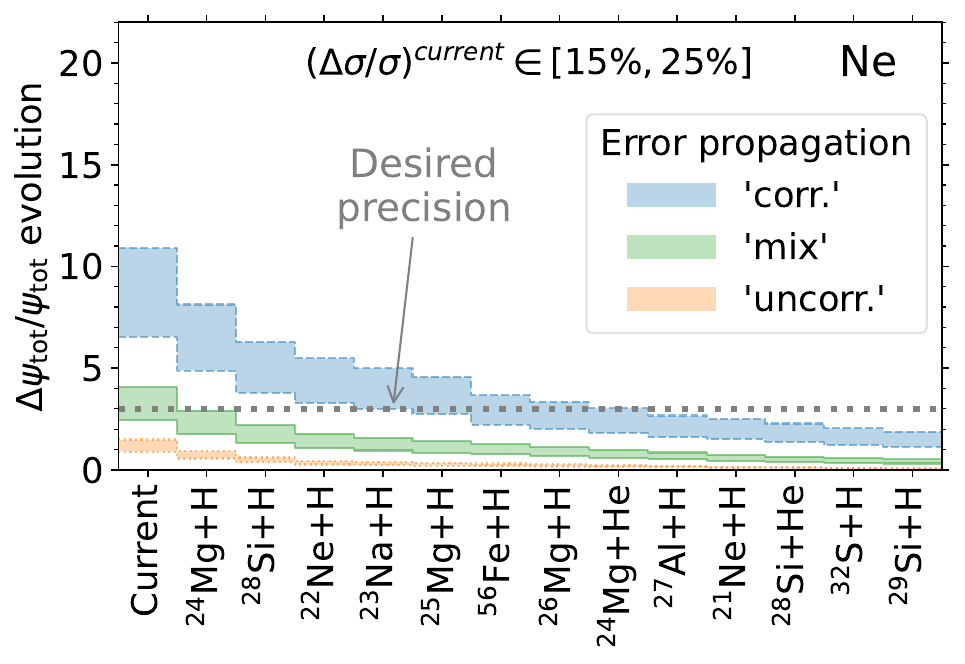}
\includegraphics[width=0.32\textwidth]{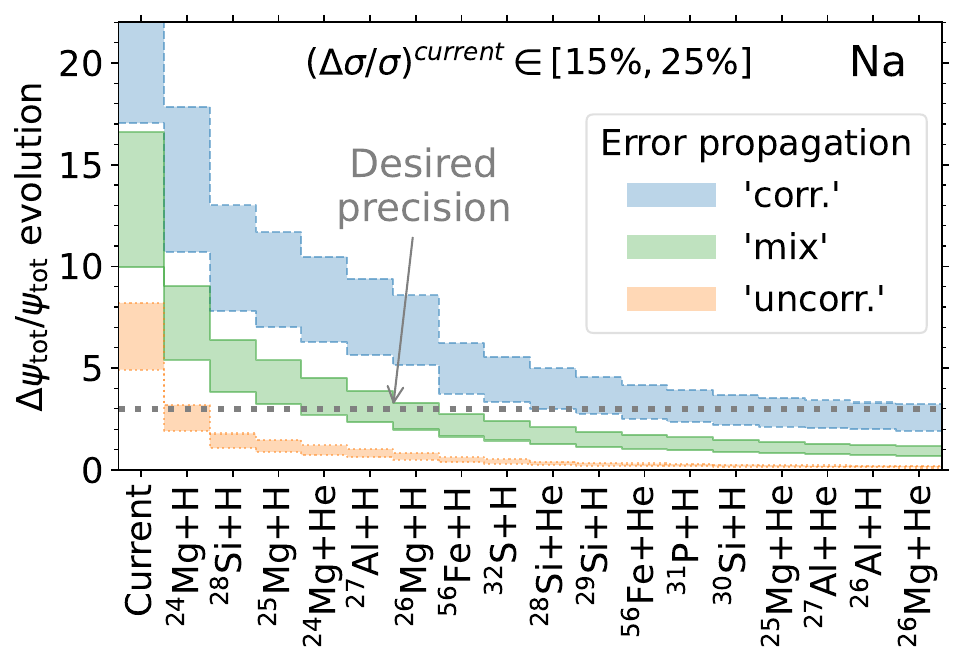}
\includegraphics[width=0.32\textwidth]{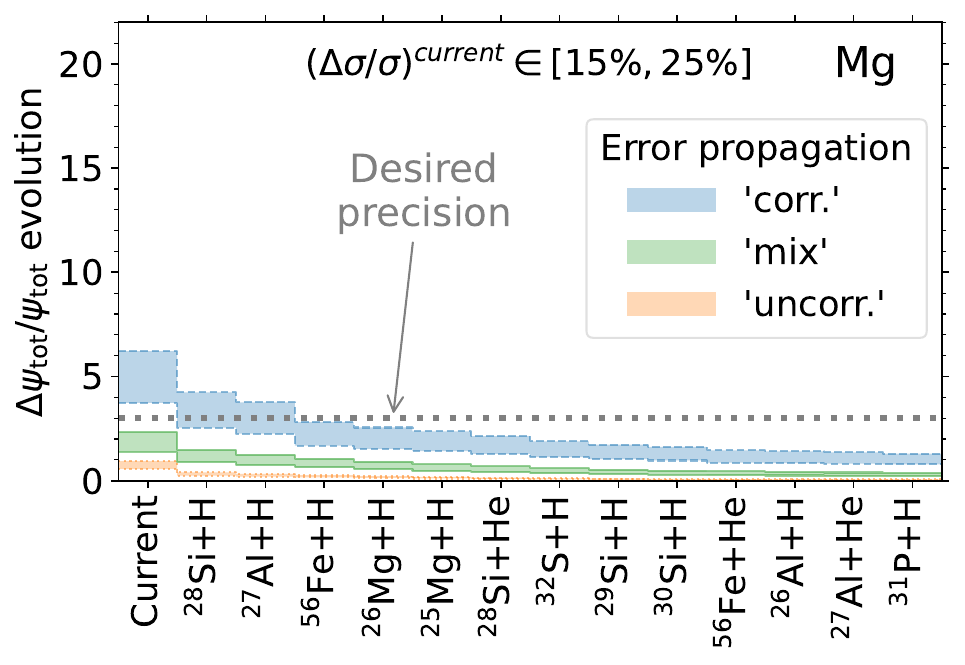}
\includegraphics[width=0.32\textwidth]{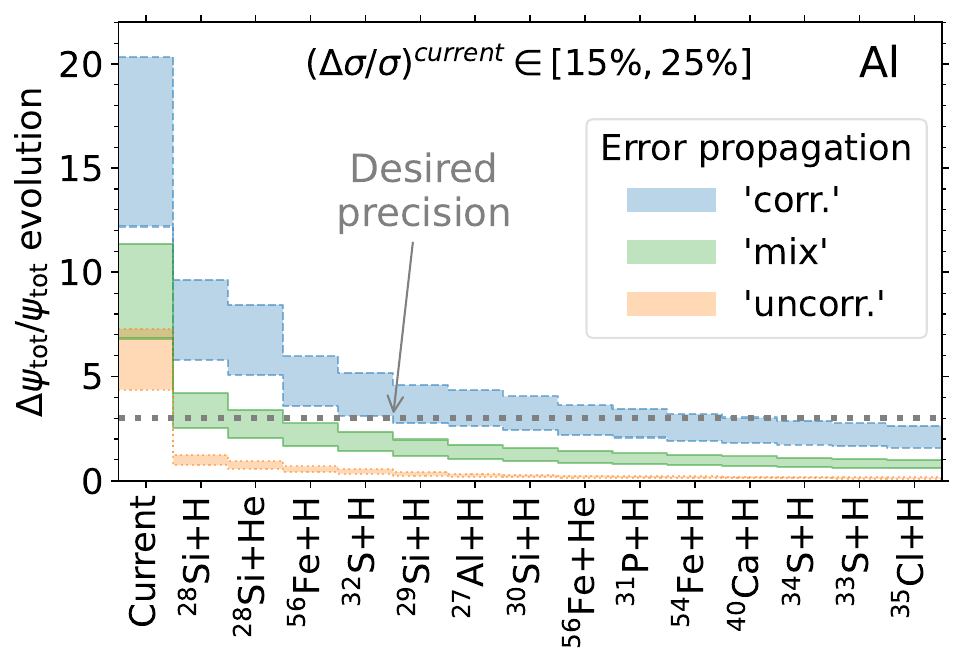}
\includegraphics[width=0.32\textwidth]{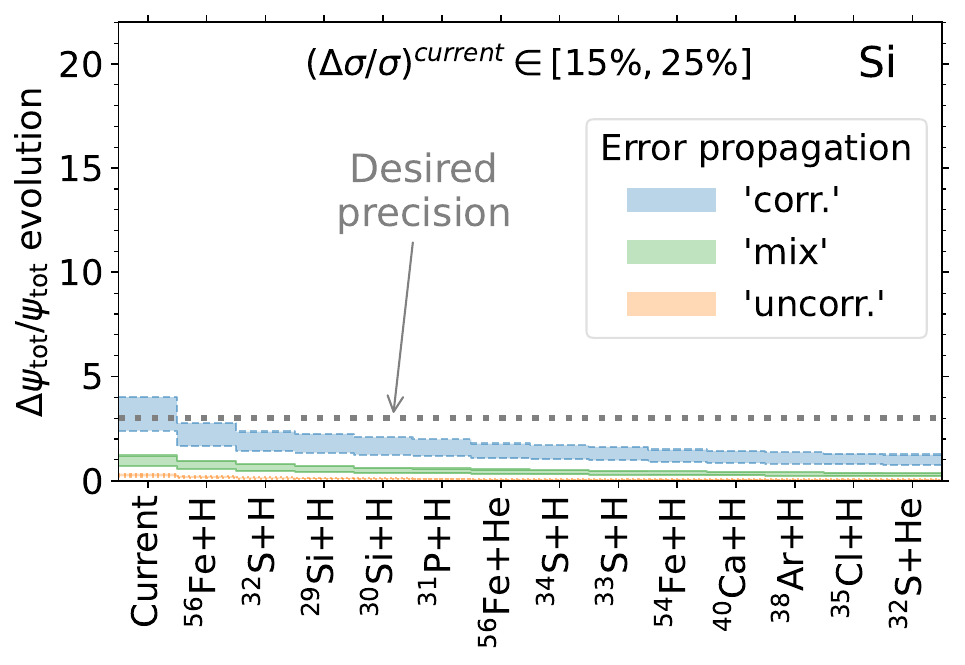}
\vspace{-0.25cm}
\caption{Improvement on the CR flux modeling precision when fragments of more and more reactions ($x$-axis labels) are ideally measured (i.e.\ no uncertainty). The first bin on the left-hand side, `Current', gives the error on the calculated elemental fluxes given the assumption made on the nuclear cross-section uncertainties without new cross-section measurements. From left to right, this error decreases when a growing number of reaction are measured. We consider three different hypotheses for the propagation of cross-section uncertainties (see Section~\ref{sec:error_propag}): correlated as given by Eq.~(\ref{eq:uncertainty_sumCorr}), uncorrelated given by Eq.~(\ref{eq:uncertainty_sumUncorr}), or a mixture of these two given by Eq.~(\ref{eq:uncertainty_sumMix}). The shaded areas encompasses the more or less optimistic assumption that current production cross-section uncertainties are in the $15-25\%$ range.\label{fig:err_evol_FSi}}
\end{figure*}

Following the approach used in \citetalias{2018PhRvC..98c4611G}, let us consider a scenario where we do incremental nuclear measurements over time, measuring first the fragments of the most highly ranked CR projectile (on H or He), then the second-highly ranked, etc. In an idealized situation where these measurements deliver near perfect nuclear data, Fig.~\ref{fig:err_evol_FSi} shows, in the above propagation uncertainties scenarios, which reactions to measure and how many of them need to be measured in order to get a better than $3\%$ uncertainty on the GCR flux prediction (on par with the GCR data uncertainties). We see that for primary species (O and Si) or mostly primary species (C and Mg), the current estimated $\approx 20\%$ uncertainty on production cross sections already ensures that the GCR fluxes of these elements already have a precision on par with the data. On the other end, purely secondary species (Li, Be, B, and F), and to a lesser extent mixed species (N, Ne, Na, and Al), need a significant number of reactions to be better measured. This number strongly depends on the propagation error scenario, moving from more up to 20 (`corr' scenario in blue) down to two (`uncorr' scenario in orange). We believe that the most realistic scenario is given by `mix' (green), which typically requires five to ten reactions to be measured.

Having a closer look at the reactions to measure, we recover the usual suspects (that depends on the CR element considered), i.e.\ mostly abundant GCR isotopes ($^{12}$C, $^{16}$O, $^{24}$Mg, $^{30}$Si, $^{56}$Fe) interacting on H. But we also have similar reactions on He that pop up for purely secondary GCR species.

\subsection{Desired number of events (beamtime)} \label{sec:res_beamtime}
%%%%%%%%%%%%%%%%%%%%%%%%%%%%%%%%%%%%%%%%%%%%%%%%%%%%%%

In the context of a recent pilot run at NA61/SHINE \cite{2019arXiv190907136U,2022icrc.confE.102A}, and possible future measurements at the Super Proton Synchrotron (SPS), it is interesting to consider the case in which every desired cross section is measured at once. By everything, we mean that, thanks to this facility, many progenitors and fragments of interest for CR physics can be measured in a `single' run. In such a case, we do not have to rely on approximations as above for error propagation, but we can directly calculate the desired number of events necessary for each reaction to reach the desired precision on GCR fluxes, as described in Section~\ref{sec:Cab_beamtime}.
\begingroup
%\squeezetable
\begin{table*}[t]
\caption{Table of sorted $C_{ab}$ coefficients for Li to Si, calculated from Eq.~(\ref{eq:Cab}). Only $C_{ab}>0.05$ are shown.\label{tab:c_ab}}
\begin{tabular}{rlcp{0.45cm}rlcp{0.45cm}rlcp{0.45cm}rlcp{0.45cm}rlcp{0.45cm}rlc}
\hline\hline
\multicolumn{3}{c}{Li}                     && \multicolumn{3}{c}{Be}                     && \multicolumn{3}{c}{B}                      && \multicolumn{3}{c}{C}                      && \multicolumn{3}{c}{N}                      && \multicolumn{3}{c}{O}                  \\
\cline{1-3}\cline{5-7}\cline{9-11}\cline{13-15}\cline{17-19}\cline{21-23}
\multicolumn{23}{c}{ }\\[-0.33cm]
\multicolumn{2}{c}{\;\;\;\,$a+b$} & $\C_{ab}$  &&  \multicolumn{2}{c}{\;\;\;\,$a+b$} & $\C_{ab}$ && \multicolumn{2}{c}{\;\;\;\,$a+b$} & $\C_{ab}$  && \multicolumn{2}{c}{\;\;\;\,$a+b$} & $\C_{ab}$  && \multicolumn{2}{c}{\;\;\;\,$a+b$} & $\C_{ab}$  && \multicolumn{2}{c}{\;\;\;\,$a+b$} & $\C_{ab}$  \\
\cline{1-3}\cline{5-7}\cline{9-11}\cline{13-15}\cline{17-19}\cline{21-23}
\multicolumn{23}{c}{ }\\[-0.33cm]
$^{16}\text{O}$  &\!\!$+\text{H}$  & 0.841 && $^{16}\text{O}$  &\!\!$+\text{H}$  & 1.079 && $^{12}\text{C}$  &\!\!$+\text{H}$  & 0.802 && $^{16}\text{O}$  &\!\!$+\text{H}$  & 1.054 && $^{16}\text{O}$  &\!\!$+\text{H}$  & 1.245 && $^{24}\text{Mg}$ &\!\!$+\text{H}$  & 0.499 \\
$^{12}\text{C}$  &\!\!$+\text{H}$  & 0.684 && $^{12}\text{C}$  &\!\!$+\text{H}$  & 0.928 && $^{16}\text{O}$  &\!\!$+\text{H}$  & 0.690 && $^{16}\text{O}$  &\!\!$+\text{He}$ & 0.185 && $^{16}\text{O}$  &\!\!$+\text{He}$ & 0.213 && $^{28}\text{Si}$ &\!\!$+\text{H}$  & 0.459 \\
$^{16}\text{O}$  &\!\!$+\text{He}$ & 0.176 && $^{16}\text{O}$  &\!\!$+\text{He}$ & 0.216 && $^{12}\text{C}$  &\!\!$+\text{He}$ & 0.147 && $^{24}\text{Mg}$ &\!\!$+\text{H}$  & 0.125 && $^{24}\text{Mg}$ &\!\!$+\text{H}$  & 0.138 && $^{20}\text{Ne}$ &\!\!$+\text{H}$  & 0.354 \\
$^{56}\text{Fe}$ &\!\!$+\text{H}$  & 0.165 && $^{28}\text{Si}$ &\!\!$+\text{H}$  & 0.200 && $^{16}\text{O}$  &\!\!$+\text{He}$ & 0.130 && $^{15}\text{N}$  &\!\!$+\text{H}$  & 0.116 && $^{28}\text{Si}$ &\!\!$+\text{H}$  & 0.125 && $^{56}\text{Fe}$ &\!\!$+\text{H}$  & 0.224 \\
$^{28}\text{Si}$ &\!\!$+\text{H}$  & 0.143 && $^{24}\text{Mg}$ &\!\!$+\text{H}$  & 0.200 && $^{11}\text{B}$  &\!\!$+\text{H}$  & 0.103 && $^{28}\text{Si}$ &\!\!$+\text{H}$  & 0.110 && $^{20}\text{Ne}$ &\!\!$+\text{H}$  & 0.113 && $^{22}\text{Ne}$ &\!\!$+\text{H}$  & 0.175 \\
$^{12}\text{C}$  &\!\!$+\text{He}$ & 0.140 && $^{56}\text{Fe}$ &\!\!$+\text{H}$  & 0.188 && $^{28}\text{Si}$ &\!\!$+\text{H}$  & 0.098 && $^{14}\text{N}$  &\!\!$+\text{H}$  & 0.104 && $^{15}\text{N}$  &\!\!$+\text{H}$  & 0.087 && $^{25}\text{Mg}$ &\!\!$+\text{H}$  & 0.092 \\
$^{24}\text{Mg}$ &\!\!$+\text{H}$  & 0.135 && $^{12}\text{C}$  &\!\!$+\text{He}$ & 0.181 && $^{24}\text{Mg}$ &\!\!$+\text{H}$  & 0.098 && $^{20}\text{Ne}$ &\!\!$+\text{H}$  & 0.092 && $^{56}\text{Fe}$ &\!\!$+\text{H}$  & 0.065 && $^{27}\text{Al}$ &\!\!$+\text{H}$  & 0.091 \\
$^{11}\text{B}$  &\!\!$+\text{H}$  & 0.114 && $^{11}\text{B}$  &\!\!$+\text{H}$  & 0.147 && $^{15}\text{N}$  &\!\!$+\text{H}$  & 0.093 && $^{13}\text{C}$  &\!\!$+\text{H}$  & 0.083 && $^{22}\text{Ne}$ &\!\!$+\text{H}$  & 0.054 && $^{23}\text{Na}$ &\!\!$+\text{H}$  & 0.089 \\
$^{15}\text{N}$  &\!\!$+\text{H}$  & 0.105 && $^{14}\text{N}$  &\!\!$+\text{H}$  & 0.120 && $^{56}\text{Fe}$ &\!\!$+\text{H}$  & 0.085 && $^{56}\text{Fe}$ &\!\!$+\text{H}$  & 0.068 &&                  &                 &       && $^{24}\text{Mg}$ &\!\!$+\text{He}$ & 0.086 \\
$^{13}\text{C}$  &\!\!$+\text{H}$  & 0.098 && $^{20}\text{Ne}$ &\!\!$+\text{H}$  & 0.117 && $^{14}\text{N}$  &\!\!$+\text{H}$  & 0.082 &&                  &                 &       &&                  &                 &       && $^{28}\text{Si}$ &\!\!$+\text{He}$ & 0.083 \\
$^{14}\text{N}$  &\!\!$+\text{H}$  & 0.091 && $^{15}\text{N}$  &\!\!$+\text{H}$  & 0.115 && $^{13}\text{C}$  &\!\!$+\text{H}$  & 0.074 &&                  &                 &       &&                  &                 &       && $^{19}\text{F}$  &\!\!$+\text{H}$  & 0.079 \\
$^{20}\text{Ne}$ &\!\!$+\text{H}$  & 0.086 && $^{13}\text{C}$  &\!\!$+\text{H}$  & 0.092 && $^{20}\text{Ne}$ &\!\!$+\text{H}$  & 0.066 &&                  &                 &       &&                  &                 &       && $^{26}\text{Mg}$ &\!\!$+\text{H}$  & 0.067 \\
$^{10}\text{B}$  &\!\!$+\text{H}$  & 0.057 && $^{10}\text{B}$  &\!\!$+\text{H}$  & 0.078 &&                  &                 &       &&                  &                 &       &&                  &                 &       && $^{21}\text{Ne}$ &\!\!$+\text{H}$  & 0.066 \\
$^{7}\text{Li}$  &\!\!$+\text{H}$  & 0.057 && $^{56}\text{Fe}$ &\!\!$+\text{He}$ & 0.058 &&                  &                 &       &&                  &                 &       &&                  &                 &       && $^{18}\text{O}$  &\!\!$+\text{H}$  & 0.062 \\
$^{56}\text{Fe}$ &\!\!$+\text{He}$ & 0.053 && $^{22}\text{Ne}$ &\!\!$+\text{H}$  & 0.054 &&                  &                 &       &&                  &                 &       &&                  &                 &       && $^{20}\text{Ne}$ &\!\!$+\text{He}$ & 0.060 \\
                 &                 &       &&                  &                 &       &&                  &                 &       &&                  &                 &       &&                  &                 &       && $^{56}\text{Fe}$ &\!\!$+\text{He}$ & 0.057 \\
                 &                 &       &&                  &                 &       &&                  &                 &       &&                  &                 &       &&                  &                 &       && $^{32}\text{S}$  &\!\!$+\text{H}$  & 0.056 \\
\multicolumn{23}{c}{\vspace*{-1.2mm}}\\
\hline\hline
\multicolumn{3}{c}{F}                      && \multicolumn{3}{c}{Ne}                     && \multicolumn{3}{c}{Na}                     && \multicolumn{3}{c}{Mg}                     && \multicolumn{3}{c}{Al}                      && \multicolumn{3}{c}{Si}                  \\
\cline{1-3}\cline{5-7}\cline{9-11}\cline{13-15}\cline{17-19}\cline{21-23}
\multicolumn{23}{c}{ }\\[-0.33cm]
\multicolumn{2}{c}{\;\;\;\,$a+b$} & $\C_{ab}$  &&  \multicolumn{2}{c}{\;\;\;\,$a+b$} & $\C_{ab}$ && \multicolumn{2}{c}{\;\;\;\,$a+b$} & $\C_{ab}$  && \multicolumn{2}{c}{\;\;\;\,$a+b$} & $\C_{ab}$  && \multicolumn{2}{c}{\;\;\;\,$a+b$} & $\C_{ab}$  && \multicolumn{2}{c}{\;\;\;\,$a+b$} & $\C_{ab}$  \\
\cline{1-3}\cline{5-7}\cline{9-11}\cline{13-15}\cline{17-19}\cline{21-23}
\multicolumn{23}{c}{ }\\[-0.33cm]
$^{20}\text{Ne}$  &\!\!$+\text{H}$  & 0.827 && $^{24}\text{Mg}$ &\!\!$+\text{H}$  & 0.662 && $^{24}\text{Mg}$ &\!\!$+\text{H}$  & 1.161 && $^{28}\text{Si}$ &\!\!$+\text{H}$  & 0.977 && $^{28}\text{Si}$ &\!\!$+\text{H}$  & 1.441 && $^{56}\text{Fe}$ &\!\!$+\text{H}$  & 1.112 \\
$^{24}\text{Mg}$  &\!\!$+\text{H}$  & 0.703 && $^{28}\text{Si}$ &\!\!$+\text{H}$  & 0.509 && $^{28}\text{Si}$ &\!\!$+\text{H}$  & 0.709 && $^{56}\text{Fe}$ &\!\!$+\text{H}$  & 0.484 && $^{56}\text{Fe}$ &\!\!$+\text{H}$  & 0.426 && $^{32}\text{S}$  &\!\!$+\text{H}$  & 0.336 \\
$^{28}\text{Si}$  &\!\!$+\text{H}$  & 0.702 && $^{56}\text{Fe}$ &\!\!$+\text{H}$  & 0.243 && $^{56}\text{Fe}$ &\!\!$+\text{H}$  & 0.320 && $^{27}\text{Al}$ &\!\!$+\text{H}$  & 0.225 && $^{28}\text{Si}$ &\!\!$+\text{He}$ & 0.229 && $^{56}\text{Fe}$ &\!\!$+\text{He}$ & 0.223 \\
$^{56}\text{Fe}$  &\!\!$+\text{H}$  & 0.433 && $^{22}\text{Ne}$ &\!\!$+\text{H}$  & 0.213 && $^{24}\text{Mg}$ &\!\!$+\text{He}$ & 0.185 && $^{28}\text{Si}$ &\!\!$+\text{He}$ & 0.156 && $^{32}\text{S}$  &\!\!$+\text{H}$  & 0.125 && $^{30}\text{Si}$ &\!\!$+\text{H}$  & 0.142 \\
$^{22}\text{Ne}$  &\!\!$+\text{H}$  & 0.306 && $^{23}\text{Na}$ &\!\!$+\text{H}$  & 0.119 && $^{25}\text{Mg}$ &\!\!$+\text{H}$  & 0.172 && $^{26}\text{Mg}$ &\!\!$+\text{H}$  & 0.147 && $^{56}\text{Fe}$ &\!\!$+\text{He}$ & 0.091 && $^{29}\text{Si}$ &\!\!$+\text{H}$  & 0.139 \\
$^{23}\text{Na}$  &\!\!$+\text{H}$  & 0.139 && $^{25}\text{Mg}$ &\!\!$+\text{H}$  & 0.112 && $^{27}\text{Al}$ &\!\!$+\text{H}$  & 0.155 && $^{25}\text{Mg}$ &\!\!$+\text{H}$  & 0.134 && $^{29}\text{Si}$ &\!\!$+\text{H}$  & 0.091 && $^{54}\text{Fe}$ &\!\!$+\text{H}$  & 0.112 \\
$^{20}\text{Ne}$  &\!\!$+\text{He}$ & 0.138 && $^{24}\text{Mg}$ &\!\!$+\text{He}$ & 0.108 && $^{26}\text{Mg}$ &\!\!$+\text{H}$  & 0.124 && $^{32}\text{S}$  &\!\!$+\text{H}$  & 0.121 && $^{27}\text{Al}$ &\!\!$+\text{H}$  & 0.070 && $^{31}\text{P}$  &\!\!$+\text{H}$  & 0.085 \\
$^{21}\text{Ne}$  &\!\!$+\text{H}$  & 0.137 && $^{27}\text{Al}$ &\!\!$+\text{H}$  & 0.100 && $^{28}\text{Si}$ &\!\!$+\text{He}$ & 0.116 && $^{56}\text{Fe}$ &\!\!$+\text{He}$ & 0.107 && $^{30}\text{Si}$ &\!\!$+\text{H}$  & 0.052 && $^{34}\text{S}$  &\!\!$+\text{H}$  & 0.069 \\
$^{25}\text{Mg}$  &\!\!$+\text{H}$  & 0.137 && $^{26}\text{Mg}$ &\!\!$+\text{H}$  & 0.088 && $^{32}\text{S}$  &\!\!$+\text{H}$  & 0.091 && $^{29}\text{Si}$ &\!\!$+\text{H}$  & 0.080 &&                  &                 &       && $^{33}\text{S}$  &\!\!$+\text{H}$  & 0.064 \\
$^{27}\text{Al}$  &\!\!$+\text{H}$  & 0.133 && $^{28}\text{Si}$ &\!\!$+\text{He}$ & 0.085 && $^{56}\text{Fe}$ &\!\!$+\text{He}$ & 0.074 && $^{30}\text{Si}$ &\!\!$+\text{H}$  & 0.054 &&                  &                 &       && $^{40}\text{Ca}$ &\!\!$+\text{H}$  & 0.063 \\
$^{26}\text{Mg}$  &\!\!$+\text{H}$  & 0.128 && $^{32}\text{S}$  &\!\!$+\text{H}$  & 0.064 && $^{29}\text{Si}$ &\!\!$+\text{H}$  & 0.058 &&                  &                 &       &&                  &                 &       && $^{38}\text{Ar}$ &\!\!$+\text{H}$  & 0.056 \\
$^{28}\text{Si}$  &\!\!$+\text{He}$ & 0.127 && $^{56}\text{Fe}$ &\!\!$+\text{He}$ & 0.058 &&                  &                 &       &&                  &                 &       &&                  &                 &       && $^{42}\text{Ca}$ &\!\!$+\text{H}$  & 0.054 \\
$^{24}\text{Mg}$  &\!\!$+\text{He}$ & 0.122 && $^{21}\text{Ne}$ &\!\!$+\text{H}$  & 0.056 &&                  &                 &       &&                  &                 &       &&                  &                 &       && $^{32}\text{S}$  &\!\!$+\text{He}$ & 0.053 \\
$^{56}\text{Fe}$  &\!\!$+\text{He}$ & 0.112 &&                  &                 &       &&                  &                 &       &&                  &                 &       &&                  &                 &       && $^{52}\text{Cr}$ &\!\!$+\text{H}$  & 0.051 \\
$^{32}\text{S}$   &\!\!$+\text{H}$  & 0.079 &&                  &                 &       &&                  &                 &       &&                  &                 &       &&                  &                 &       &&                  &                 &       \\
$^{22}\text{Ne}$  &\!\!$+\text{He}$ & 0.051 &&                  &                 &       &&                  &                 &       &&                  &                 &       &&                  &                 &       &&                  &                 &       \\
\hline\hline
\end{tabular}
\end{table*}
\endgroup

%%%%%%%%%%%%%%%%%%%%%

The constants $\C_{ab}$ are listed in Table~\ref{tab:c_ab} to
aid the optimization of the beam requests for future
measurements. These constants also allow to estimate the uncertainty
of a particular secondary flux due to the statistical uncertainty of
the cross section measurements in $a+b$ interactions to be predicted
for the given number of recorded interactions and they provide
guidelines on which combinations of projectile and target are the most
important ones to measure. For instance, it can be seen that the
dominating $\C_{ab}$ coefficients for Li are $\C_{\rm Cp}^{\rm Li}$ and
$\C_{\rm Op}^{\rm Li}$, but also Si and Fe interactions with
protons are important reactions. The constants derived here differ from
the values quoted in \citetalias{2018PhRvC..98c4611G} due to updated
cross-section values, in particular for the fragmentation of Fe (see Sect.~\ref{sec:setup_XS}).
Furthermore, the importance of a few reactions was
overestimated previously, due to an error in the decoding of the data
tables during calculation. In addition, \citetalias{2018PhRvC..98c4611G} did not
include the factor $f_{\rm sec}$ in Eq.~(\ref{eq:errProp}) leading to
a too large weight of reactions contributing to the flux of nuclei
that are mostly of primary origin. In summary, the $\C_{ab}$
coefficients and related calculations listed here supersede our previous results.

{We aim at the desired model uncertainty to be smaller than the uncertainty of the current and near future CR experiments. The AMS-02 experiment claims $\approx$3\% uncertainty for most of its data. Therefore, since the contribution from cross-section uncertainties should be a subdominant of the overall uncertainty, we
  investigate how keep this contribution at the $1\%$ level.
  If in addition an experimental systematic uncertainty of typically $0.5\%$
  can be achieved (e.g.~\citealt{NA61SHINE:2015bad}), then we arrive to the
required statistical accuracy of $\xi = \sqrt{0.01^2-0.005^2} = 0.0087$ as in \cite{Aduszkiewicz:2309890}. Adopting the optimal power-law exponent $\beta=1$ derived in Sec.~\ref{sec:error_propag}
results in the required number of interactions listed in
Table~\ref{tab:ninter}. It is worthwhile noting that a scaling with
$\beta=0$, as investigated in \citetalias{2018PhRvC..98c4611G}, would require about a
factor-of-two more interactions to be recorded to obtain the same
accuracy, but it involves fewer interaction channels.

\begin{table}[t]
    \caption{Required number of interactions to be recorded per reaction, as calculated from Eq.~(\ref{eq:nk}) with $\beta=1$. The reactions are given in three groups
      of increasing projectile mass (up to O, Si, or Fe). The cumulative number of required interactions is quoted at the end of each group.}
  \label{tab:ninter}
   \begin{tabular}{r@{+}lcc}
     \multicolumn{2}{c}{reaction} & $N_\text{int}$ \\
     \hline\hline
  $^{16}\text{O}$ & $\text{H}$  & 60k\\ %Be
  $^{12}\text{C}$ & $\text{H}$  & 50k\\ %Be
  $^{16}\text{O}$ & $\text{He}$  & 20k\\ %Be
  $^{11}\text{B}$ & $\text{H}$  & 10k\\ %Be
  $^{15}\text{N}$ & $\text{H}$  & 10k\\ %Be
  $^{14}\text{N}$ & $\text{H}$  & 10k\\ %Be
  $^{12}\text{C}$ & $\text{He}$  & 10k\\ %Be
  $^{10}\text{B}$ & $\text{H}$  & 5k\\ %Be
  $^{13}\text{C}$ & $\text{H}$  & 5k\\ %Be
  $^{7}\text{Li}$ & $\text{H}$  & 5k\\ %Li
  \multicolumn{2}{c}{} & $N(\leq\text{O}) = 1.9\times 10^5$ \vspace*{1mm}\\
  $^{28}\text{Si}$ & $\text{H}$  & 50k\\ %F
  $^{24}\text{Mg}$ & $\text{H}$  & 50k\\ %F
  $^{20}\text{Ne}$ & $\text{H}$  & 50k\\ %F
  $^{22}\text{Ne}$ & $\text{H}$  & 20k\\ %F
  $^{28}\text{Si}$ & $\text{He}$  & 10k\\ %F
  $^{27}\text{Al}$ & $\text{H}$  & 10k\\ %F
  $^{26}\text{Mg}$ & $\text{H}$  & 10k\\ %F
  $^{24}\text{Mg}$ & $\text{He}$  & 10k\\ %F
  $^{23}\text{Na}$ & $\text{H}$  & 10k\\ %F
  $^{25}\text{Mg}$ & $\text{H}$  & 10k\\ %F
  $^{21}\text{Ne}$ & $\text{H}$  & 10k\\ %F
  $^{20}\text{Ne}$ & $\text{He}$  & 10k\\ %F
  $^{32}\text{S}$ & $\text{H}$  & 5k\\ %F
  $^{29}\text{Si}$ & $\text{H}$  & 5k\\ %Na
  $^{22}\text{Ne}$ & $\text{He}$  & 5k\\ %F
  \multicolumn{2}{c}{} & $N(\leq\text{Si}) = 3.8\times 10^5$ \vspace*{1mm}\\
  $^{56}\text{Fe}$ & $\text{H}$  & 30k\\ %F
  $^{56}\text{Fe}$ & $\text{He}$  & 10k\\ %F
  \multicolumn{2}{c}{} & $N(\leq\text{Fe}) = 4.2\times 10^5$ \vspace*{1mm}\\
  \hline
  \end{tabular}
  \end{table}
  
%%%%%%%%%%%%%%%%%%%%%

\section{Forecasts from improved cross sections}
\label{sec:forecasts}
Here we investigate different scenarios for new nuclear cross-section measurements and the impact of these measurements on key questions addressed by the CR community, namely: the derivation of CR transport parameters and their impact on indirect DM searches. To assess the improvement brought by new nuclear cross-section measurements, we generate mock cross-section models which, within a given propagation scenario, are used to fit the transport parameters on the secondary to primary ratios. We start by explaining the framework used before coming to the results.

\subsection{Methodology} \label{sec:forecast_methodo}
%%%%%%%%%%%%%%%%%%%%%%%%%%%%%%%%%%%%%%%%%%%%%%%%%%%%%%

\subsubsection{Set-up of the mock cross sections} \label{sec:mockxs}
%%%%%%%%%%%%%%%%%%%%%%%%%%%%%%%%%%%%%%%%%%%%%%%%%%%%%%

We define several scenarios from which we draw a thousand of mock cross section models.

\paragraph*{Current uncertainty (Scenario I):}
In this scenario, for each cross-section reaction of our benchmark cross-section parameterization (see Section~\ref{sec:setup_XS}), we apply a default energy-independent random Gaussian bias with variance equal to 20\% of the cross-section value. Moreover, we apply another random Gaussian bias, chosen to be identical for reactions involving the same progenitor, with a variance again of 20\% of the cross-section value.
The first choice is primarily guided by the typical spread and uncertainty observed in the cross-section data. The second one is related to the presence of systematics between reactions from different progenitors, because their measurements were usually carried out at different facilities (see \citetalias{2018PhRvC..98c4611G} for a thorough discussion). This systematics, also expected in the cross-section parametrizations (deriving from these data), is particularly hard and a Sisyphean task to estimate. Indeed, nuclear data measurements spread over more than fifty years, with the respective publications providing more or less limited details on the different setups involved. For this reason, relying on rough estimates discussed in \citetalias{2018PhRvC..98c4611G}), we set the level of the systematics to that of the statistical ones.

\paragraph*{Updated cross sections up to O/Si/Fe (Scenario II):}
By default, the mock cross-section models of this kind are generated in the same way as in Scenario I.
However, they do include improvements of some of the cross-section uncertainties, according to our proposition (Table~\ref{tab:ninter}). For each reaction ($a+b\to c$) in this list, we estimate the Poisson parameter $\mu_{abc} = N_{a+b} \times \sigma_{a+b\to c}/\sigma_{a+b}$, with $N_{a+b}$ being the number of $a+b$ interactions proposed in Table~\ref{tab:ninter}. Here $\sigma_{a+b\to c}$ is the production cross section of $c$, and $\sigma_{a+b}$ is the $a+b$ total inelastic cross section. Hence $\mu_{abc}$ corresponds to the number of events including the multiplicity, and hence the following updates are given for a detector able to measure all fragments of a given fragmentation event. When the relative Poissonian error $1/\sqrt{\mu_{abc}}$ is smaller than 20\% for a given reaction, we draw a random value $Y$ from a Poissonian distribution of parameter $\mu_{abc}$, and apply the rescaling factor $Y/\mu_{abc}$ to the corresponding cross section. By drawing a thousand mock cross-section models, we mimic the improvement of the statistical uncertainty achieved with new fragmentation data. We consider three subscenarios, taking O, Si, and Fe consequently as the heaviest progenitors up to which our new experiment was able to take fragmentation data.

\subsubsection{Fit to secondary/primary ratios}
%%%%%%%%%%%%%%%%%%%%%%%%%%%%%%%%%%%%%%%%%%%%%%%%%%%%%%

For each scenario we fit alternatively the secondary/primary ratios: Li/C, Be/C, B/C and F/Si, to compare
improvements achieved for each observable. For the transport model, we use the minimalist, purely diffusive, \textit{SLIM} propagation benchmark presented above in the Section~\ref{sec:setup_propag}. For each mock cross-section model (a thousand for each scenario) we fit only the four parameters ($\delta_{\rm l}, R_{\rm l}, \delta, D_0$) which impact the low and intermediate rigidity dependence of the diffusion coefficient.

The results of the corresponding fits are shown as contour plots in Fig.~\ref{fig:delta_K0_bananas} with solid colored lines defining the $1\sigma$ contour in the $(\delta,D_0)$ relative error plane. This way we propagate the uncertainty of the cross section onto the propagation parameters values. To gauge if the improvements of the scenarios described in Section~\ref{sec:mockxs} are satisfactory, we have to compare the irreducible/intrinsic error on the propagation parameters coming from the uncertainty of the secondary/primary data. We do so by fitting a thousand mock AMS-02 datasets generated from a given known model (i.e.\ with fixed $\delta_{\rm l}, R_{\rm l}, \delta, D_0$) from which each data point is scattered according to a Gaussian law, whose variance is $\sigma=\sqrt{\sigma_{stat}^2+ \sigma_{\rm syst}^2}$, with $\sigma_{stat}$ and $\sigma_{\rm syst}$ being the statistical and systematic uncertainties provided by the AMS-02 collaboration. Note that the fragmentation cross sections are fixed to their benchmark value (see Section~\ref{sec:setup_XS}). The results of these fits are shown in Fig.~\ref{fig:delta_K0_bananas} with a solid black line defining the $1\sigma$ contour in the $(\delta,D_0)$ relative error plane. The central point of that region is highlighted in red.

\subsection{Improvements on propagation parameters} \label{subsec:improvements_on_pp}
%%%%%%%%%%%%%%%%%%%%%%%%%%%%%%%%%%%%%%%%%%%%%%%%%%%%%%

\begin{figure*}[t]
   \includegraphics[width=0.49\textwidth]{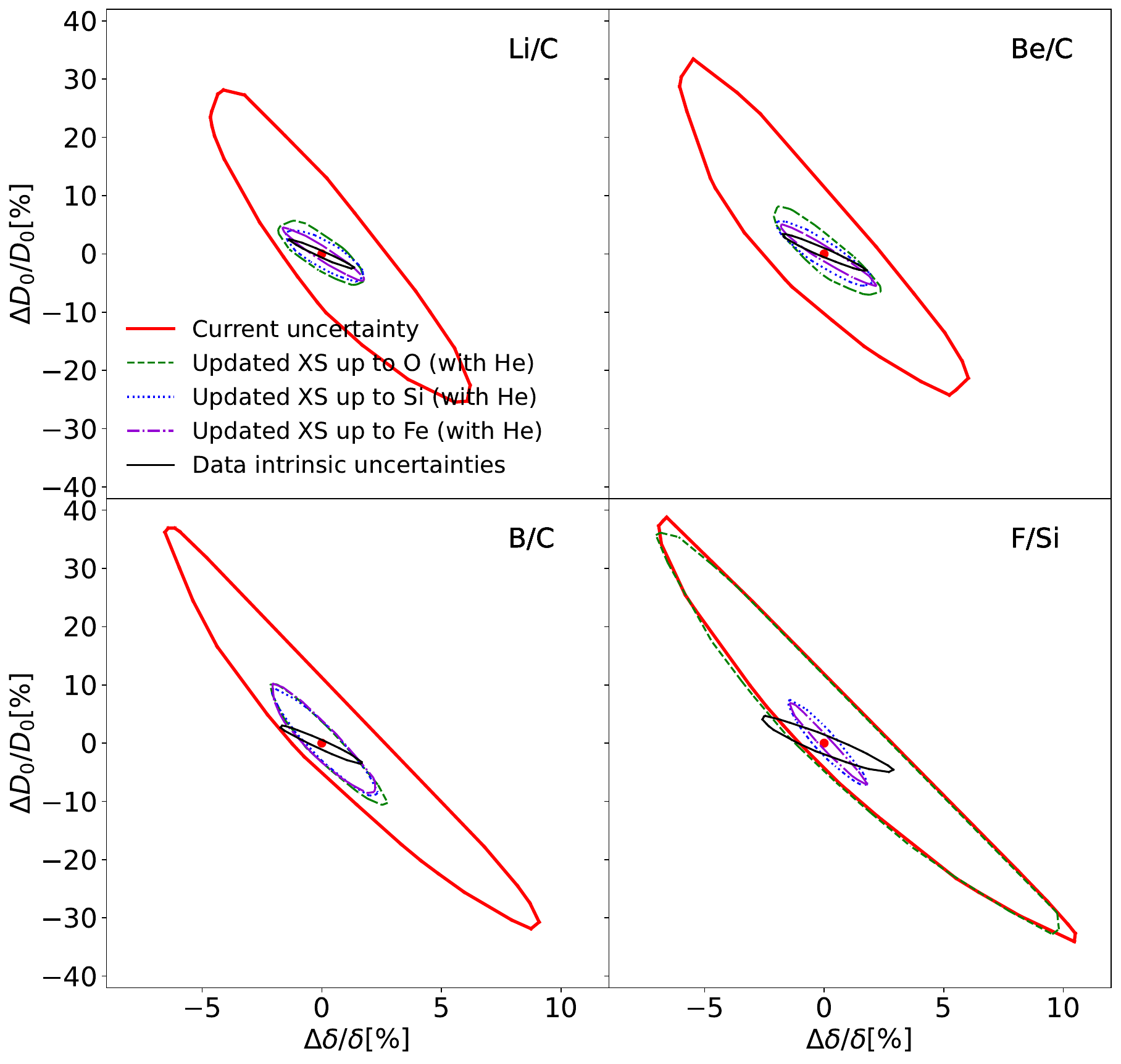}
   \includegraphics[width=0.49\textwidth]{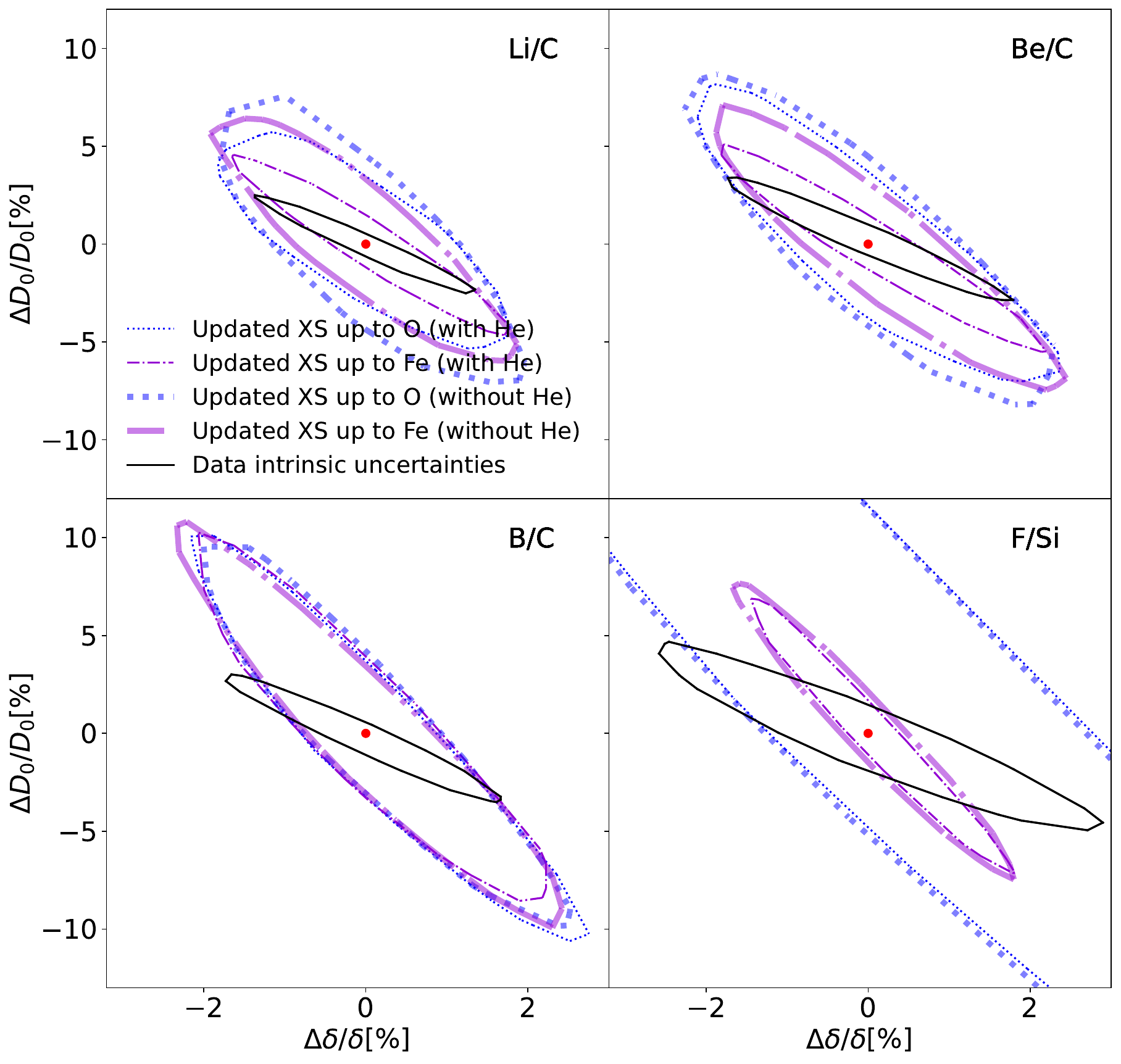}
   \caption{Forecast of transport parameters determination from new cross-section measurement campaigns. Each figure shows $1\sigma$ contours in the ($D_0$, $\delta$) relative error plane in different scenarios. The \textit{left panel} shows the estimated current uncertainty (solid red line) and three cases were a subsets of cross sections have been updated according to our proposition Table~\ref{tab:ninter}, increasing the mass of the heavier progenitor from O to Fe. Finally, for comparison, we show the irreducible/intrinsic data uncertainty (solid black line). The \textit{right panel} is a zoom of the left one, and compare subcases where we would not measure the fragmentations of Table~\ref{tab:ninter} on Helium target. More details on how these bounds were computed can be found in the text.}\label{fig:delta_K0_bananas}
\end{figure*}

Several important remarks can be drawn from the left panel of Fig.~\ref{fig:delta_K0_bananas}: (i) For each secondary/primary ratio considered, the current uncertainty on $(D_0,\delta)$ (solid red line) is much larger than the data would allow if the uncertainties on the fragmentation cross sections (solid black line) were vanishing. Note the difference in correlation in the ($D_0$,$\delta$) plane, where the relative range of $\delta$ is much smaller than that of $D_0$. This implies that consequent improvements in the accuracy of the cross sections would result in tighter relative constrains for $\delta$ than for $D_0$. On the other hand, such measurements will bring down the uncertainty in $D_0$ four-fold from 40\% to better than 10\%.
(ii) Our proposition (Table~\ref{tab:ninter}) leads to significant improvements in the accuracy of ($D_0$,$\delta$). The three subscenarios, taking different heaviest progenitors for the new fragmentation data (O, Si, or Fe), give similar improvements derived from the Li/C, Be/C, and B/C ratios, with the slightly reduced uncertainties as we move from O to Si and to Fe. (iii) For the F/Si ratio, significant improvements in the accuracy of ($D_0$,$\delta$) are observed when the new fragmentation data involve progenitors heavier than F, as anticipated. We anticipate that this will also be the case for all heavier ratios, hence the importance to get a new cross-section dataset including heavy nuclei fragmentation.

While the left panel of Fig.~\ref{fig:delta_K0_bananas} includes the fragmentations on H and He targets proposed in Table~\ref{tab:ninter}, the right panel shows a comparison when fragmentation on He target is not measured.
We remark that including the He fragmentation always lead to reduced uncertainties for ($D_0$,$\delta$), however the improvement is mild for the heavier nuclei ratios (B/C and F/Si) while it is significant for the lightest ones (Li/C and Be/C), decreasing the uncertainty in $D_0$ by a few percentages.

\subsection{Reduced uncertainty for secondary $\bar{p}$}
%%%%%%%%%%%%%%%%%%%%%%%%%%%%%%%%%%%%%%%%%%%%%%%%%%%%%%

GCR antiprotons is one of the most powerful probe for the indirect detection of dark matter \cite{Heisig:2020jvs,2004PhRvD..69f3501D,Giesen:2015ufa,Jin:2015sqa,Evoli:2015vaa,Cuoco:2016eej,Cui:2016ppb,Huang:2016tfo,Feng:2017tnz,Cuoco:2017rxb,2018JCAP...01..055R,Cuoco:2017iax,2019PhRvD..99j3026C,DiMauro:2021qcf,2020PhRvR...2b3022B}. However, as highlighted in many studies \cite{Heisig:2020jvs,Giesen:2015ufa,Evoli:2015vaa,Cuoco:2017rxb,2018JCAP...01..055R,DiMauro:2021qcf} and in particular in \cite{2020PhRvR...2b3022B}, the modeling uncertainties of the $\bar{p}$ secondary astrophysical background are significantly larger than AMS-02 data uncertainties \cite{2016PhRvL.117i1103A}. The two most important modeling uncertainties are (i) the nuclear production cross sections of $\bar{p}$ and $\bar{n}$ from GCRs and (ii) the transport parameter uncertainties. As underlined just above, the latter are related to the nuclear production uncertainties.

\begin{figure}[t]
   \includegraphics[width=0.49\textwidth]{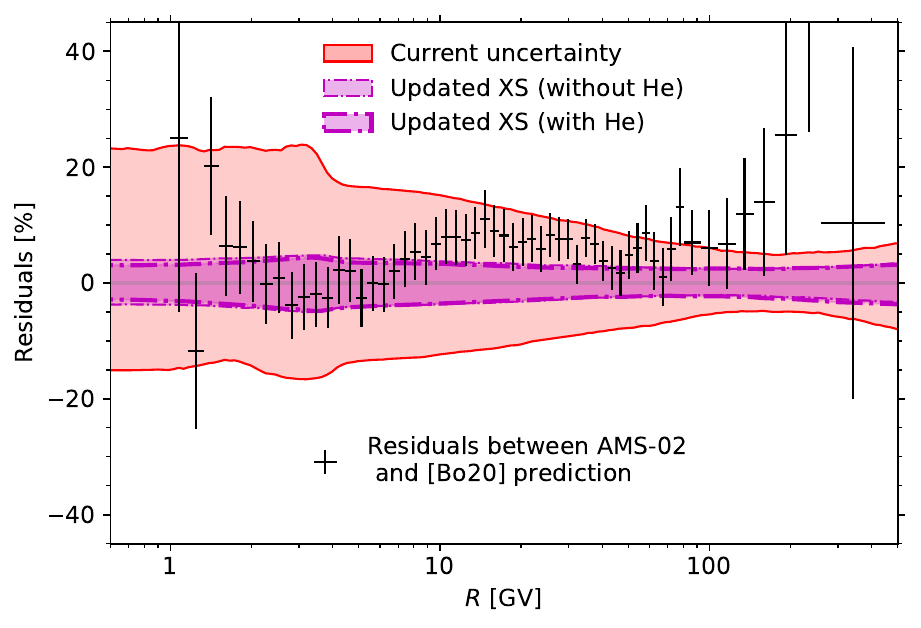}
   \caption{Reduction of the uncertainties on secondary $\bar{p}$ flux brought by new nuclear data. The red envelope shows the current `transport' uncertainty, i.e.\ uncertainty related to the transport parameter uncertainties (linked to the nuclear cross-section uncertainties). The magenta envelopes show the envelopes with new cross section measurements (including reactions up to Fe), for the case of new measurements both on H and He as targets (thick lines) or H target only (thin lines). With this improvement, the uncertainty is smaller than the current AMS-02 data uncertainties \cite{2016PhRvL.117i1103A}; the symbols show the residuals of the data with regards to the best-fit $\bar{p}$ prediction of \cite{2020PhRvR...2b3022B}, [Bo20] in the legend. See text for discussion.}
   \label{fig:forecast_pbar}
\end{figure}

Considering the same scenarios as above, we calculate from the best-fit coefficients the associated $\bar{p}$ flux (for each mock cross-section model). Doing so, we can calculate at each energy the distribution of $\bar{p}$ values and its quantiles, forming an envelope of uncertainties over all energies, as shown in Fig.~\ref{fig:forecast_pbar}. In this plot, the red envelope shows the current `transport' uncertainties\footnote{This band does not exactly match the one shown in \cite{2020PhRvR...2b3022B} because we do not use exactly the same approach. In particular, here we fix the solar modulation instead of propagating its uncertainty as in \cite{2020PhRvR...2b3022B}, where we were also making use of the \textit{BIG} benchmark scenario instead of \textit{SLIM} here. Both these effects enlarge the uncertainty band at low rigidities.}, which is much larger than the $\bar{p}$ data uncertainties. The black symbols show the residuals between the AMS-02 data and the best-fit model of \cite{2020PhRvR...2b3022B}, featuring a `bump' seen at $\approx 10$~GV (also from other studies) which has triggered a lot of excitement in the community, possibly hinting at a new physics signal. This feature can actually be related to some correlations in the data, as advocated in \cite{2020PhRvR...2b3022B}. In any case, the current precision on the transport parameters, related to our current knowledge of nuclear cross sections, is not on par with the AMS-02 data precision.

The magenta envelopes show the configurations where we fit the transport parameters using the B/C ratio only (see Section~\ref{subsec:improvements_on_pp}), in a scenario where all nuclear cross sections have been improved for reactions both on H and He targets (thick lines) or only on H targets (thin lines). We do not consider transport parameters extracted from the fits of  Li/C, B/C or F/Si, because they only need to be calibrated with the best-measured ratio: B/C, or ideally, a combined fit of these ratios. Actually, using Li/C, Be/C or B/C only leads to minor differences as these ratios all lead to similar transport parameter uncertainties. Using transport parameter uncertainties from F/Si would lead to minor or similar improvements depending on the new nuclear data taken, as discussed above. The improved nuclear data, and so the improved transport parameter uncertainties, would be the game changers for indirect dark matter searches, as AMS-02 data could be fully exploited then. Actually, as advocated in Fig.~2 of \cite{2020PhRvR...2b3022B}, both the nuclear data (highlighted in this paper) and $\bar{p}$ and $\bar{n}$ production (see also \cite{2018PhRvD..97j3011K}) should be improved to overcome this challenge.

\subsection{Reduced uncertainty for $L$ and primary $\bar{p}$}
%%%%%%%%%%%%%%%%%%%%%%%%%%%%%%%%%%%%%%%%%%%%%%%%%%%%%%

Primary dark matter signals also suffer from transport parameter uncertainties. However, as highlighted in several studies \cite{2008PhRvD..78d3506D,2021PhRvD.104h3005G}, these uncertainties are dominated by the uncertainty on the halo size of the Galaxy, $L$ (the primary flux is actually roughly proportional to $L$).

\begin{figure}[t]
   \includegraphics[width=0.49\textwidth]{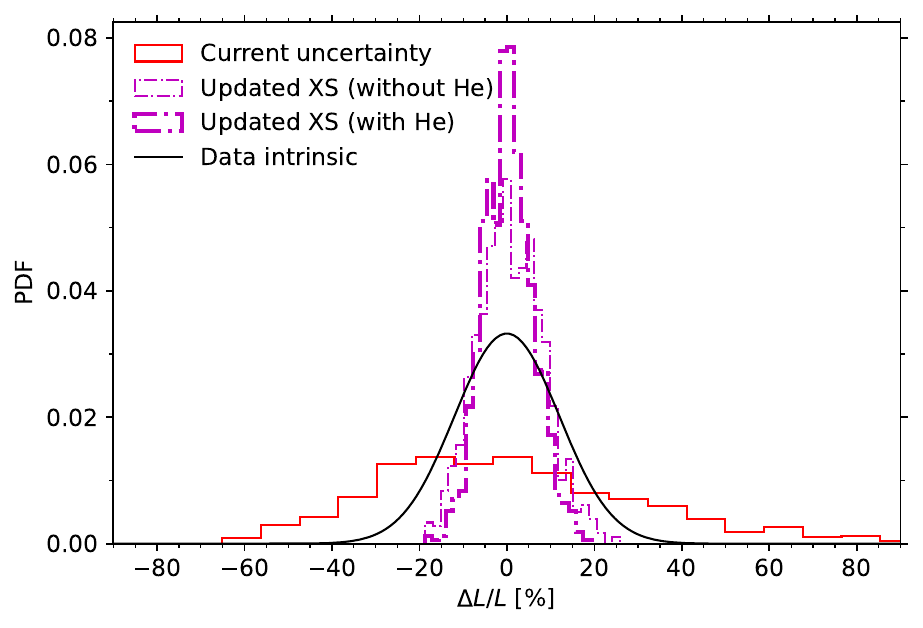}
   \caption{Forecast for the relative precision of the halo size $L$ of the Galaxy. The red envelope shows the current uncertainty on $L$, and the magenta envelopes shows the improvement achieved with by new cross section measurements (including reactions up to Fe), for the case of measurements both on H and He as targets (thick line) or H target alone (thin line). With this improvement, the uncertainty becomes smaller than the uncertainties from current data (as estimated in \cite{2022A&A...667A..25M}). See text for discussion.}
   \label{fig:forecast_L}{}
\end{figure}

This halo size cannot be directly extracted from the above secondary-to-primary nuclei ratios because it largely constraints the $L/D_0$ ratio, but this degeneracy can be broken using the radioactive clocks measurements, such as $^{10}$Be (e.g., \cite{1998ApJ...509..212S, 1998A&A...337..859P, 1998ApJ...506..335W, 2002A&A...381..539D,2010A&A...516A..66P}). Recent studies emphasized the role of the accuracy of the nuclear cross sections in the determination of $L$ via the $^{10}$Be/$^9$Be ratio \cite{2020A&A...639A..74W,2020PhRvD.101b3013E,2021JCAP...03..099D,2022A&A...667A..25M,PhysRevD.101.023013}. We take advantage of a simplified calculation of this ratio proposed in \cite{2022A&A...667A..25M} to fit the halo size $L$ from $^{10}$Be/$^9$Be ratio for each mock cross-section scenarios discussed above (using the best-fit transport parameters derived for each of these mocks). This allows us to estimate the uncertainties on the determination of $L$, as shown in Fig.~\ref{fig:forecast_L}. The red histogram shows the probability distribution function of $L$ for current nuclear data. The standard deviation of this distribution ($\approx$30\%) is much larger than that of the black histogram $\approx$12\%, estimated from the data uncertainties only assuming perfectly known nuclear data \cite{2022A&A...667A..25M}. New nuclear data (magenta histogram) allow to decrease this number down to $\approx$5\%, with slightly better precision if these nuclear data are taken on both H and He targets (thick line) compared to the case of  H target alone (thin line).

The black histogram is actually estimated from pre-AMS-02 data, with $^{10}$Be/$^9$Be data precision at the level of ten to twenty percent. Preliminary data from the AMS-02 and the HELIX project \cite{2019ICRC...36..121P} are expected to reach similar precision $\approx$15\%), but to cover a much larger energy range. {The black line estimate is thus expected to improve (see e.g. \cite{2023arXiv230510337J}), although these challenging forthcoming measurements may still not go as far to high energies as desired to provide significant improvement \cite{2022A&A...667A..25M}. In any case, the message is that the new cross section data will allow the uncertainty in the halo size $L$ to be reduced by more than a factor of two. This is already a huge step forwards as it means that, for instance, dark matter exclusion plots based on $\bar{p}$ studies \cite{2022ScPP...12..163C} will benefit from a similar gain. Additionally, such an improvement will help constraining the underlying physics at the origin of the diffusive halo.

\section{Conclusions} \label{sec:conclusions}
%%%%%%%%%%%%%%%%%%%%%%%%%%%%%%%%%%%%%%%%%%%%%%%%%%%%%%
%%%%%%%%%%%%%%%%%%%%%%%%%%%%%%%%%%%%%%%%%%%%%%%%%%%%%%

This paper follows up on our previous work (\citetalias{2018PhRvC..98c4611G}) to prioritize the list of cross sections of interest for Galactic CR studies that have to be measured with a higher precision. The current generation of CR experiments (AMS-02, CALET, DAMPE, Fermi-LAT) has been bringing a revolution in astrophysics of CRs, and to fully exploit these data, we need a combined effort of the astrophysics, nuclear, and particle physics communities and their facilities to meet the demand for high-precision nuclear fragmentation data.

In this paper, we have reviewed the production of F to Si and also updated our previous predictions for Li to O GCR elements. We have provided several plots to highlight the importance of direct production of Li to Si elements, but also to highlight the respective importance of the primary progenitors. The ranking of these quantities is mostly of interest for the CR community.

As in \citetalias{2018PhRvC..98c4611G}, the main result of our paper is the calculation of the so-called $f_{abc}$ coefficients, which directly rank the most important reactions involved in the production of GCRs. This allows us to select the most critical reactions for production of the so-called secondary Li, Be, B, Fe, and Na elements, used to calibrate the interstellar transport of GCRs. Their accurate prediction significantly impacts our ability to make progress and fully benefit from recent GCR data on a variety of important questions, such as understanding the origin of CRs, properties of the interstellar matter, our local galactic environment, indirect dark matter searches, and many others.

From these $f_{abc}$ coefficients, we have highlighted the way towards actionable measurements, by providing the exact reactions to measure and the desired number of reactions necessary to render GCR model accuracy on par with existing GCR data precision.
In particular, the cross-section plots (parameterizations and available data) shown in the Supplemental Material \cite{supp..mat..arxiv} (including references [155-253]) illustrate that many relevant reactions still have no data, especially those for high-$Z$ progenitors (up to Fe) or at energies above a few GeV/n; this strongly stresses the need for new measurements.
From our numbers, the beam time can be easily estimated thus facilitating the experiment planning.
We have shown that new nuclear data would dramatically improve the precision on the derived transport parameters. They would (i) reduce the uncertainties of some transport parameters by a factor of ten, allowing many disputed interpretations of the observed features in spectra of CR species to be addressed; (ii) reduce the uncertainties in production of secondary antiprotons by a factor of five, allowing the dark matter searches and exploration of discrepancies between the data and the models with acute precision; and (iii) reduce the uncertainty on the halo size of the Galaxy by a factor of a few, allowing the constraints on dark matter candidates to be automatically improved by the same factor. While we focused in this paper on these three examples, the benefit of these new nuclear data would not end here, as multimessengers studies of the Galaxy (radio- to $\gamma$-rays, neutrinos, and CRs) are to some level all calibrated on the transport coefficients that can only be improved with these new nuclear data.

In this second paper of the series, we have covered GCR elements from Li to Si, and similar studies for heavier species are highly desirable. AMS-02 has already provided measurements of the Fe flux \cite{2021PhRvL.126d1104A} and will soon fill the gap of the elements between Si and Fe. Of particular importance are the so-called sub-Fe elements: Sc, Ti, and V. Because of their mostly secondary origin, they can be used to probe GCR transport of heavy species, and comparing them with light secondary nuclei results, will provide new insights on GCR transport. It is already known that fragmentation of Fe is a key ingredient for this modeling. Our follow-up effort should quantify how much so.
In addition, AMS-02 has also published data on $^3$He/$^4$He \cite{2019PhRvL.123r1102A} and those on $^2$H/$^4$He should follow soon. The production of these light nuclei should also be revisited because these ratios provide complementary constraints on CR propagation. Indeed, $^2$H and $^3$He are secondary species and their predictions also suffer from nuclear reaction uncertainties \cite{2012A&A...539A..88C}. A future work will also quantify what are the critical reactions to be measured in that case.

%%%%%%%%%%%%%%%%%%%%%%%%%%%%%%%%%%%%%%%%%%%%%%%%%%%%%%
%%%%%%%%%%%%%%%%%%%%%%%%%%%%%%%%%%%%%%%%%%%%%%%%%%%%%%{}
%%%%%%%%%%%%%%%%%%%%%%%%%%%%%%%%%%%%%%%%%%%%%%%%%%%%%%

\acknowledgments

This work has been supported by the ``Investissements d'avenir, Labex ENIGMASS'' and by the Programme National des Hautes Energies of CNRS/INSU with INP and IN2P3, co-funded by CEA and CNES. IVM acknowledges partial support through NASA Grants No.\ 80NSSC23K0169 and No.\ 80NSSC22K0718. M.U. acknowledges the support of the German Research Foundation DFG (Project No.\ 426579465).
We thank the NA61/SHINE collaboration for their very useful feedback during the {\em NA61++/SHINE: NA61++/SHINE: Physics Opportunities from Ions to Pions} meeting, which motivated the calculations made for Section~V.

%%%%%%%%%%%%%%%%%%%%%%%%%%%%%%%%%%%%%%%%%%%%%%%%%%%%%%
%%%%%%%%%%%%%%%%%%%%%%%%%%%%%%%%%%%%%%%%%%%%%%%%%%%%%%
%\eject
%\clearpage
\appendix

\section{Energy-dependent view of the fragmentation contributions, channels, and progenitors}
\label{app:plot_Edep_rankings}
%%%%%%%%%%%%%%%%%%%%%%%%%%%%%%%%%%%%%%%%%%%%%%%%%%%%%%
%%%%%%%%%%%%%%%%%%%%%%%%%%%%%%%%%%%%%%%%%%%%%%%%%%%%%%

%\clearpage
%\onecolumngrid
In this Appendix, we present an energy-dependent view of the various fragmentation contributions (shown at 10~GeV/n only in Table~\ref{tab:origin}), alongside an energy-dependent view of the rankings of the progenitors and channels.

\begin{figure*}[!t]
\includegraphics[width=\columnwidth]{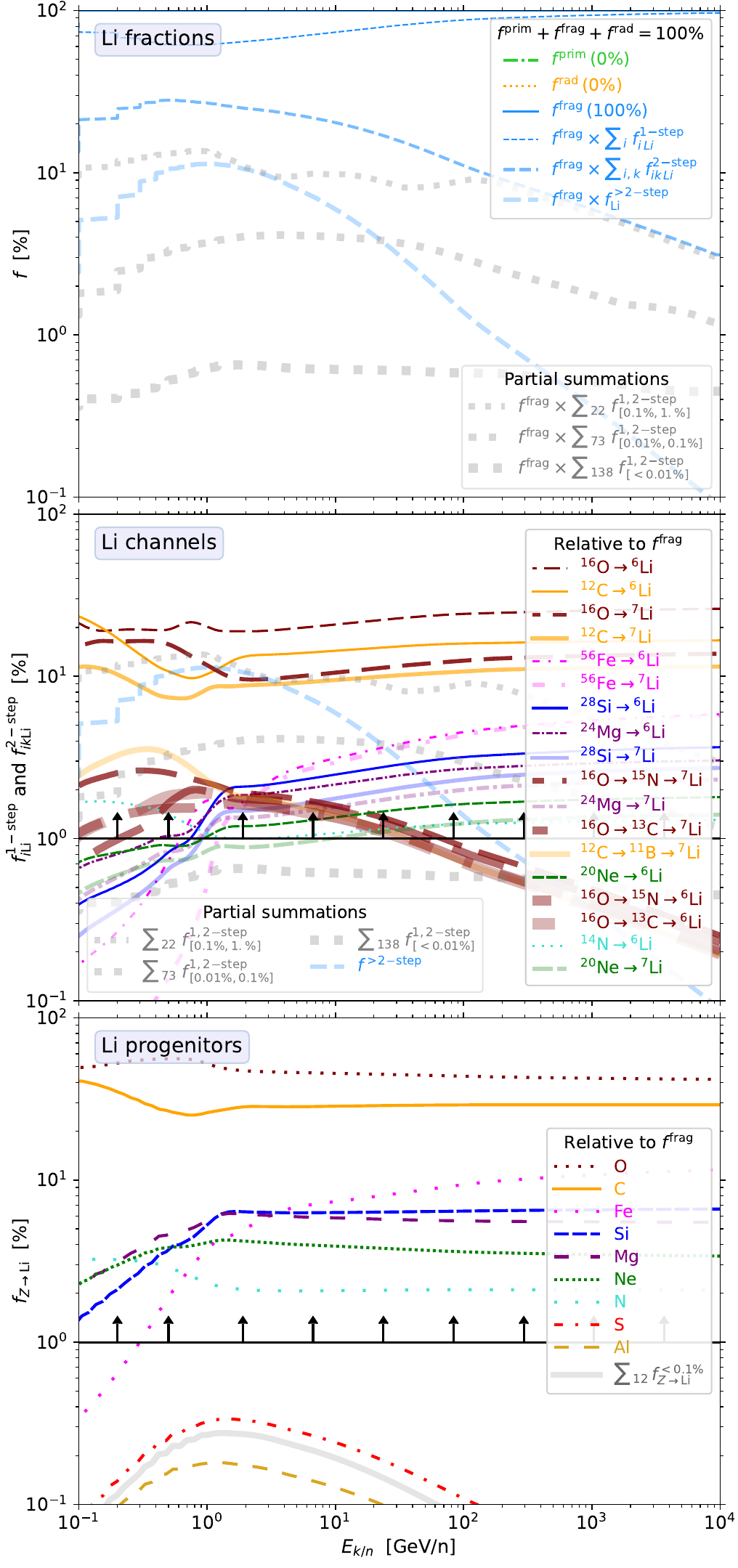}
\includegraphics[width=\columnwidth]{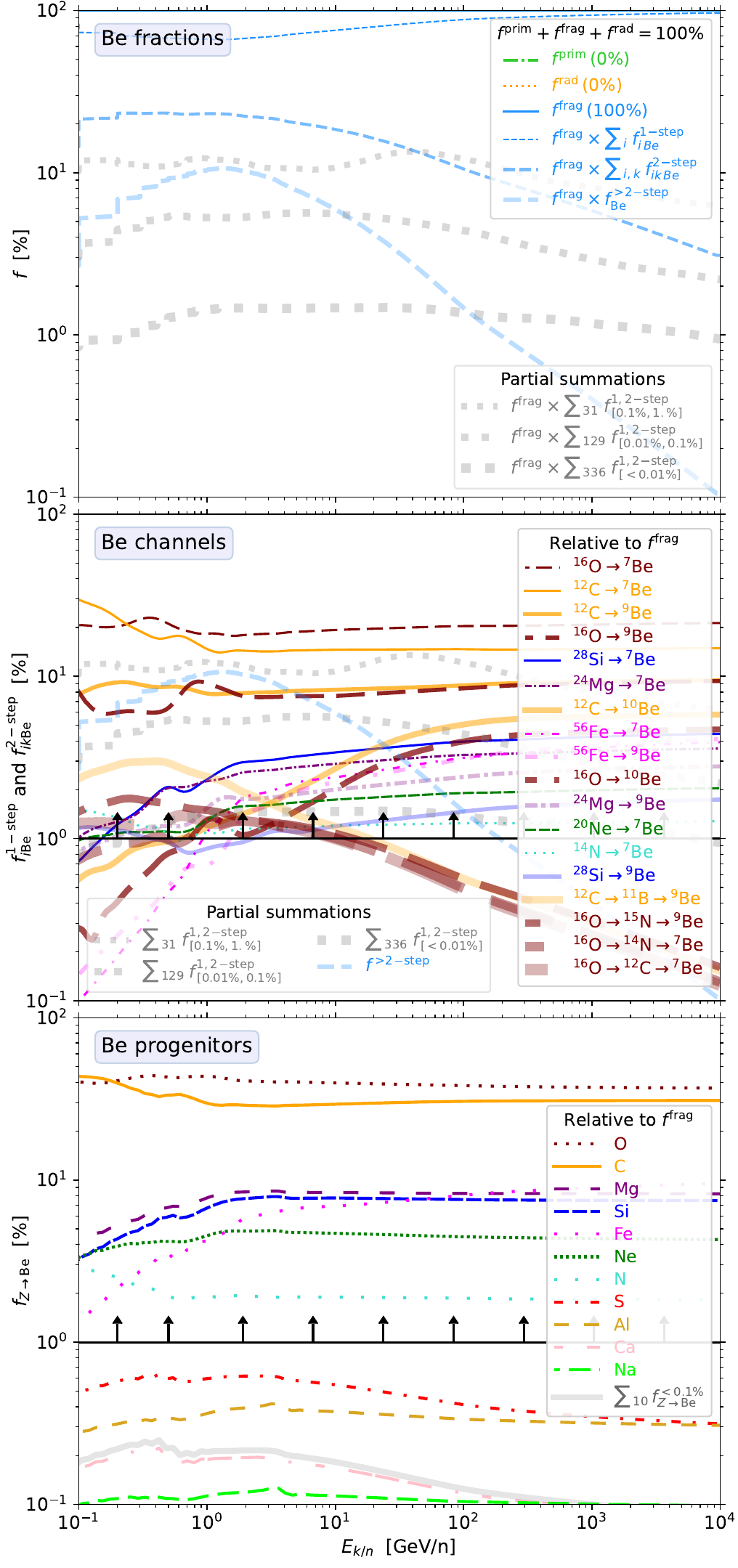}
\vspace{-0.2cm}
\caption{
Contributions and rankings for Li (left) and Be (right) fluxes at $\phi_{\rm FF}=700$~MV.
{\bf Top panel}: primary, radioactive, and fragmentation fractions also broken down into the sum of all 1-step, 2-step, or $>2$-step contributions, see Eqs.~(\ref{eq:sum_Pij_Pijk}-\ref{eq:sum_sup2step}); gray dotted lines highlight the importance of subpercentage 1-step and 2-step contributions, when grouped according to their value at 10~GeV/n (in $\sum_N$, $N$ is their number).
{\bf Middle panel}: relative to $f_{\rm sec}$, 1-step and 2-step channels (see Eq.~\ref{eq:sum_Pij_Pijk}) shown individually when larger than $1\%$ at 10~GeV/n (right legend), or grouped by contributing fractions (gray-dotted lines in bottom legend). The sum of all these curves, plus all $>\!2$-step channels, $f^{>\!2\rm-step}$ (blue-dashed line in bottom legend), makes up $100\%$ of $f_{\rm sec}$. The black solid line (with vertical arrows) separates contributions larger or smaller than $1\%$ of the total flux (nontrivial shape if $f_{\rm sec}<100\%$).
{\bf Bottom panel}: relative to $f_{\rm sec}$, CR progenitors whose contribution $f_{Z \to \rm Li}$ (see Eq.~\ref{eq:coeffs_Pij}) is larger than $0.1\%$ at 10~GeV/n. Their sum, plus all smaller contributions (thick gray solid line) makes up $100\%$ of $f_{\rm sec}$. As in the middle panel, the black solid line separates contributions above or below $1\%$ of the total flux.
\label{fig:ranked_LiBe}}
\end{figure*}
\begin{figure*}[t]
\includegraphics[width=\columnwidth]{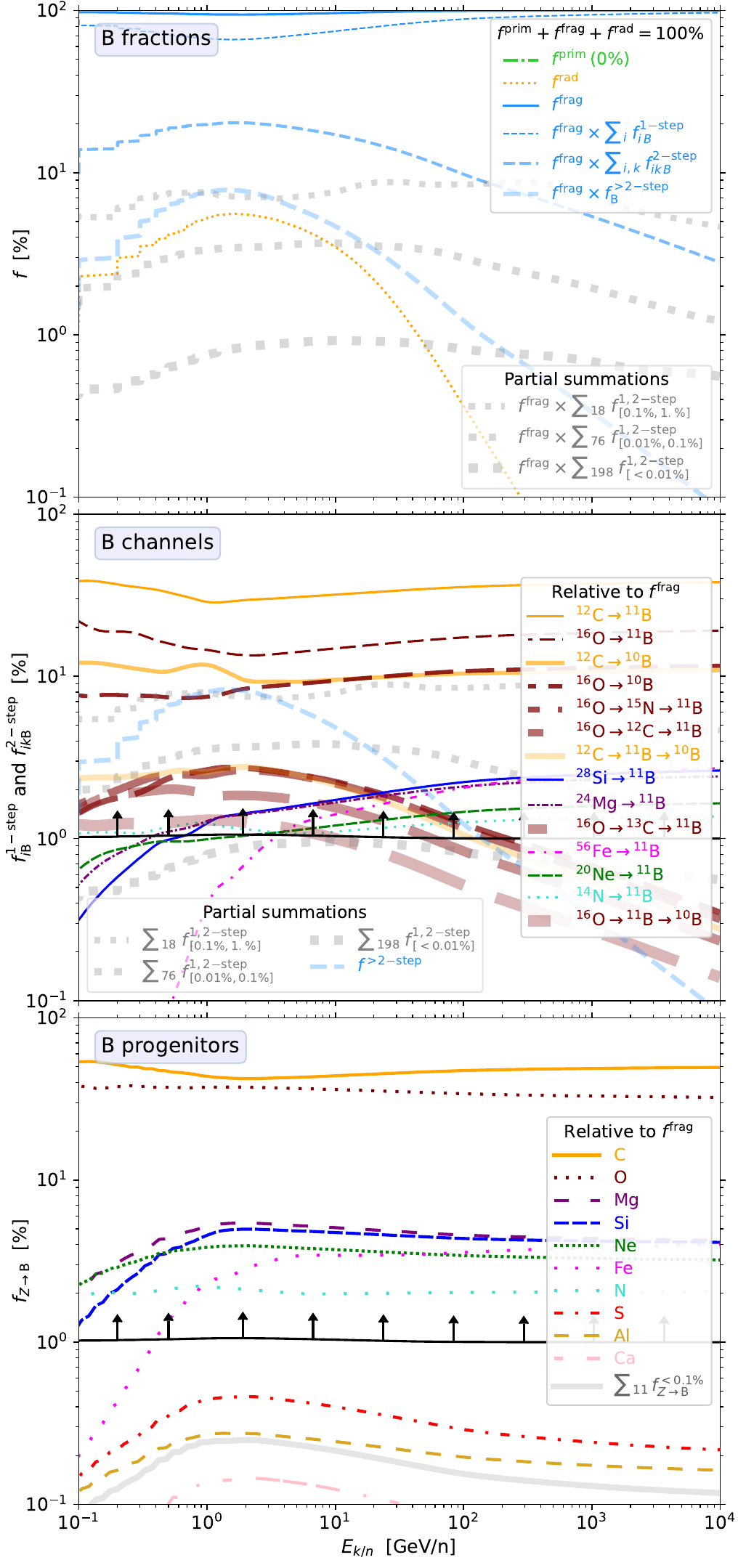}
\includegraphics[width=\columnwidth]{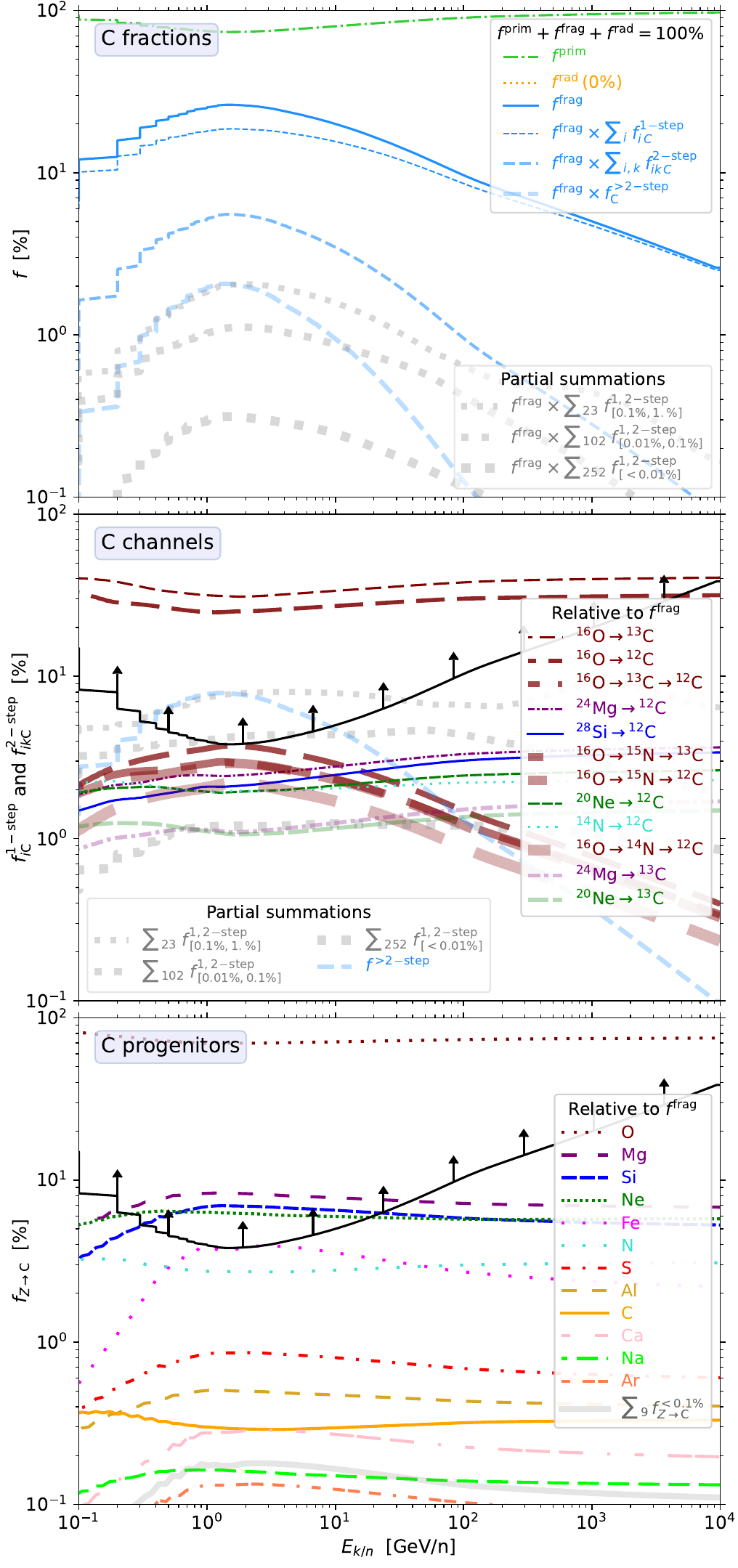}
\caption{Same as in Fig.~\ref{fig:ranked_LiBe}, but for B (left) and C (right).
\label{fig:ranked_BC}}
\end{figure*}
\begin{figure*}[t]
\includegraphics[width=\columnwidth]{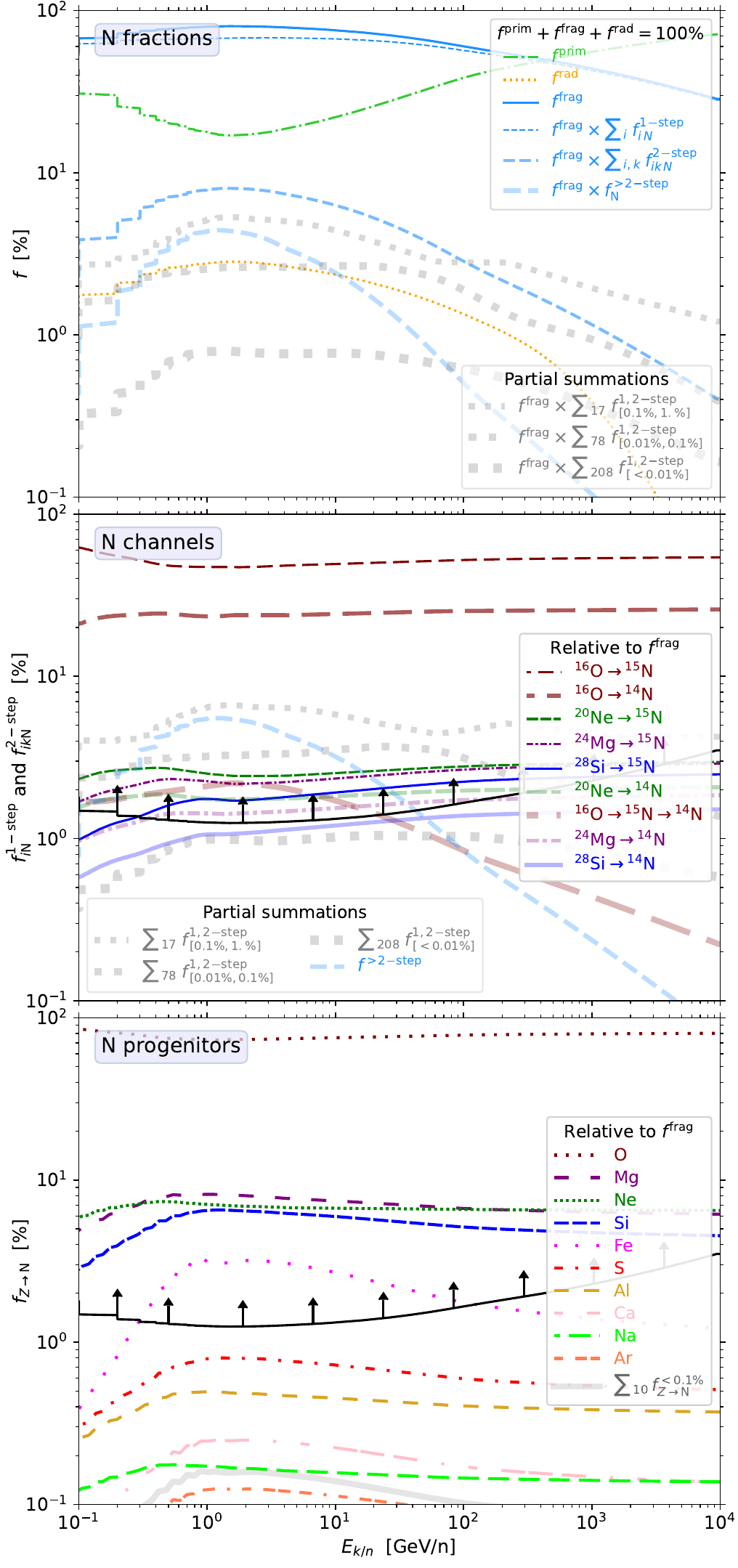}
\includegraphics[width=\columnwidth]{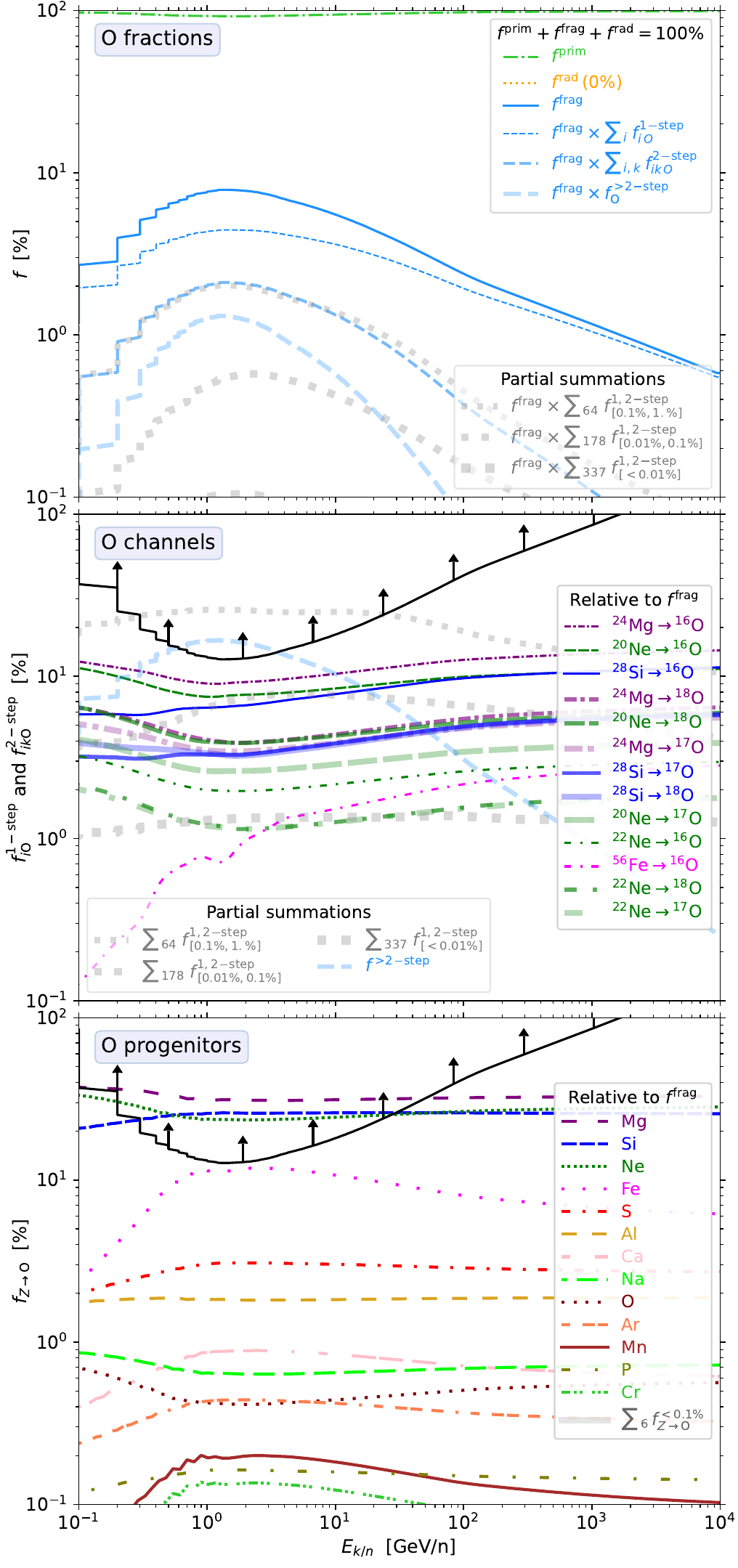}
\caption{Same as in Fig.~\ref{fig:ranked_LiBe}, but for N (left) and O (right).
\label{fig:ranked_NO}}
\end{figure*}
\begin{figure*}[t]
\includegraphics[width=\columnwidth]{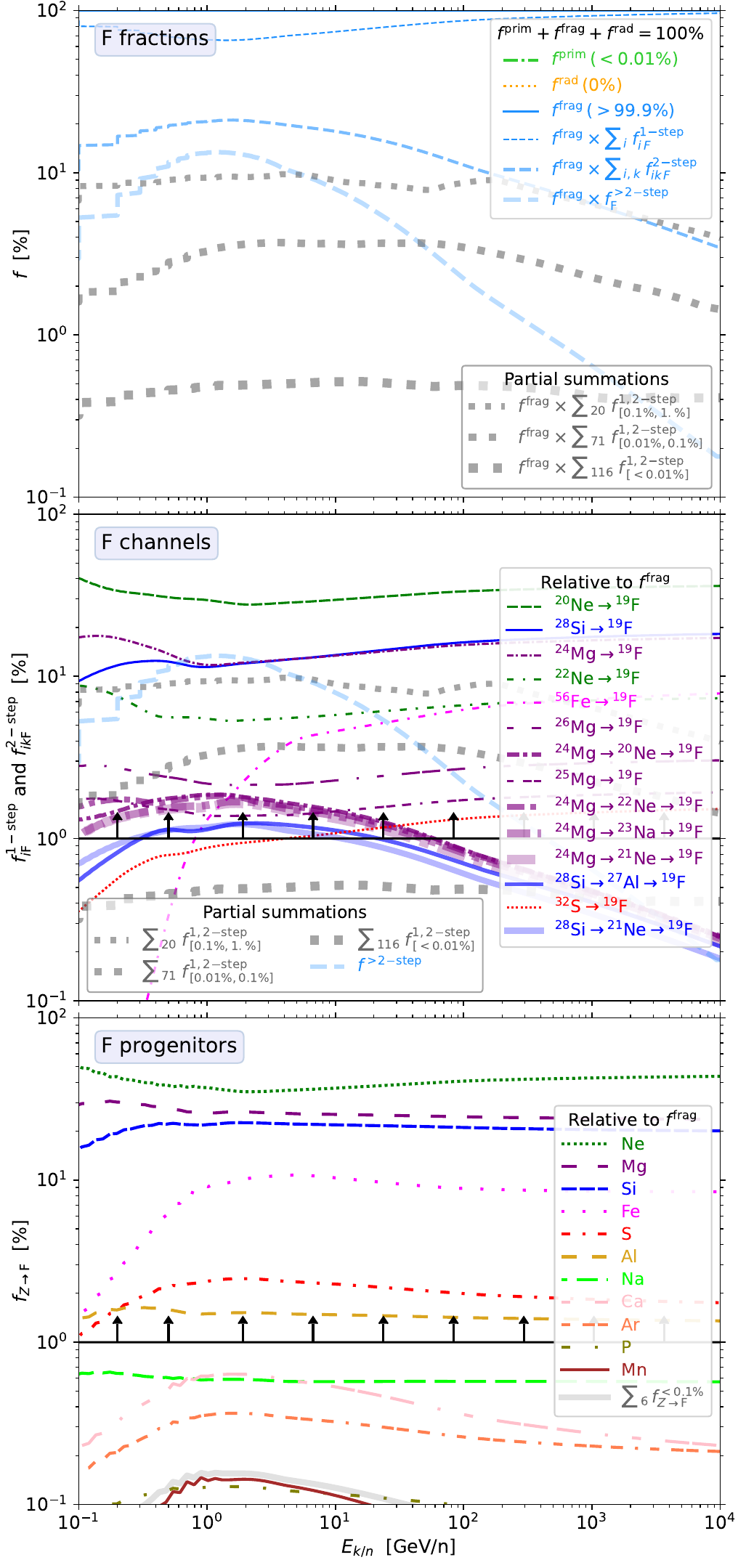}
\includegraphics[width=\columnwidth]{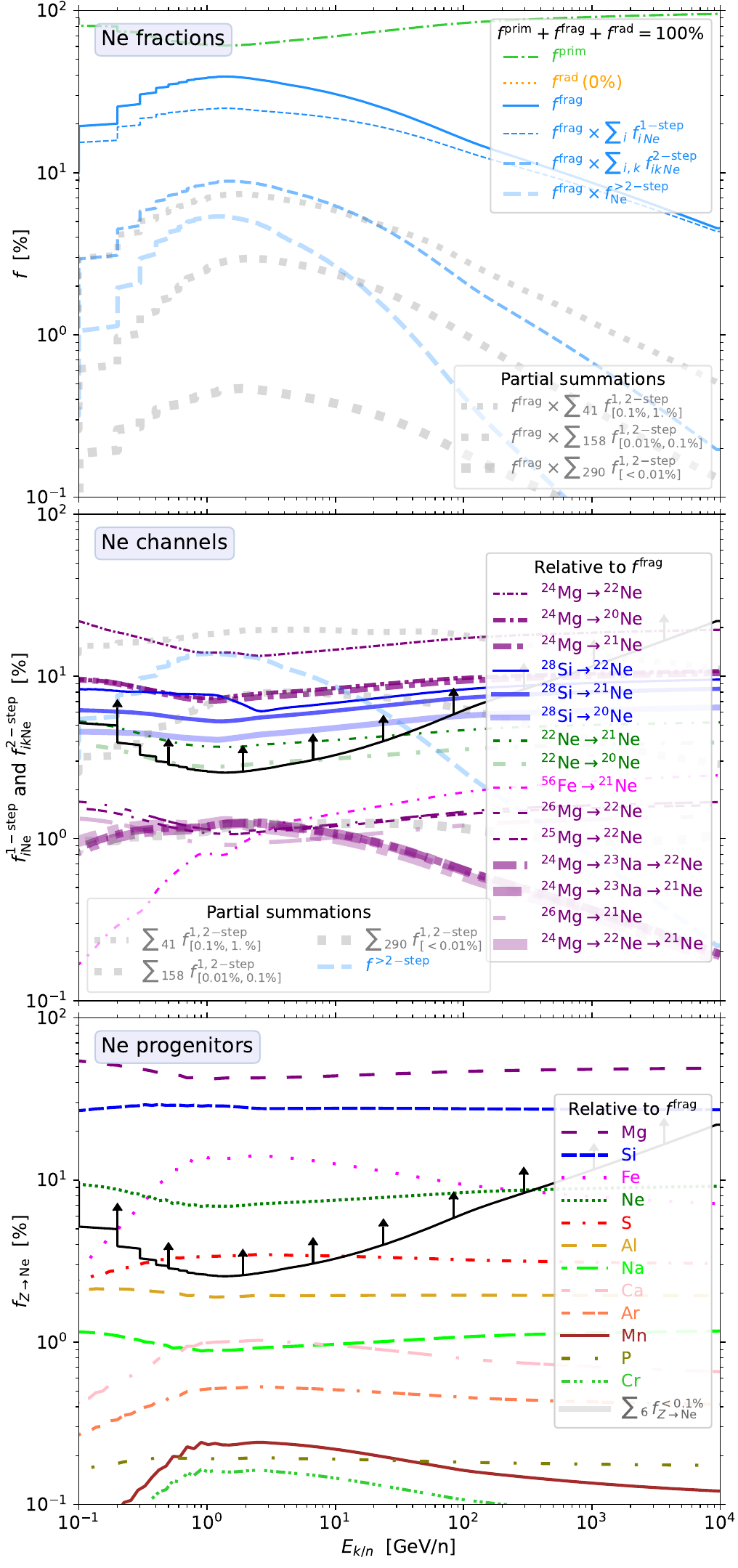}
\caption{Same as in Fig.~\ref{fig:ranked_LiBe}, but for F (left) and Ne (right).
\label{fig:ranked_FNe}}
\end{figure*}
\begin{figure*}[t]
\includegraphics[width=\columnwidth]{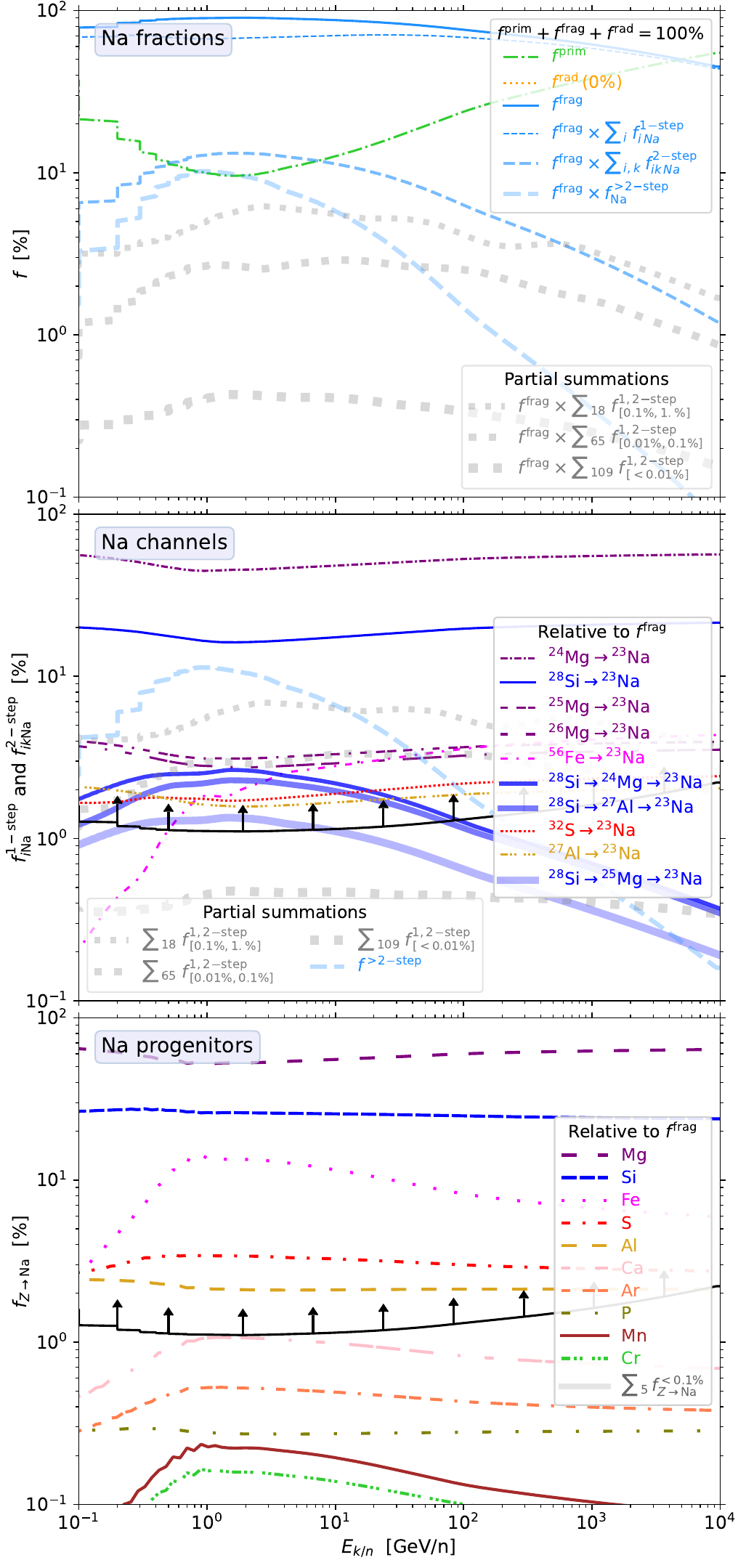}
\includegraphics[width=\columnwidth]{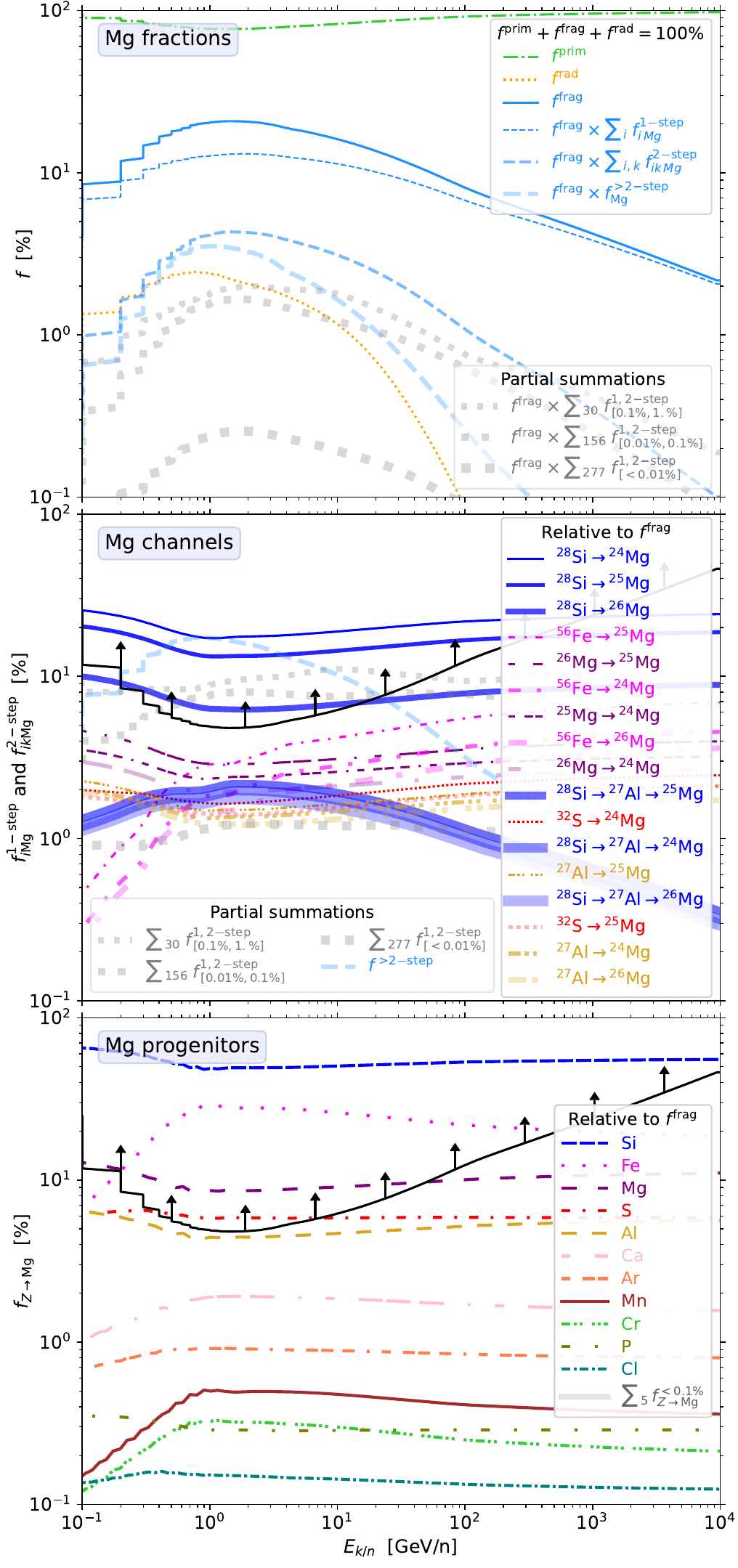}
\caption{Same as in Fig.~\ref{fig:ranked_LiBe}, but for Na (left) and Mg (right).
\label{fig:ranked_NaMg}}
\end{figure*}
\begin{figure*}[t]
\includegraphics[width=\columnwidth]{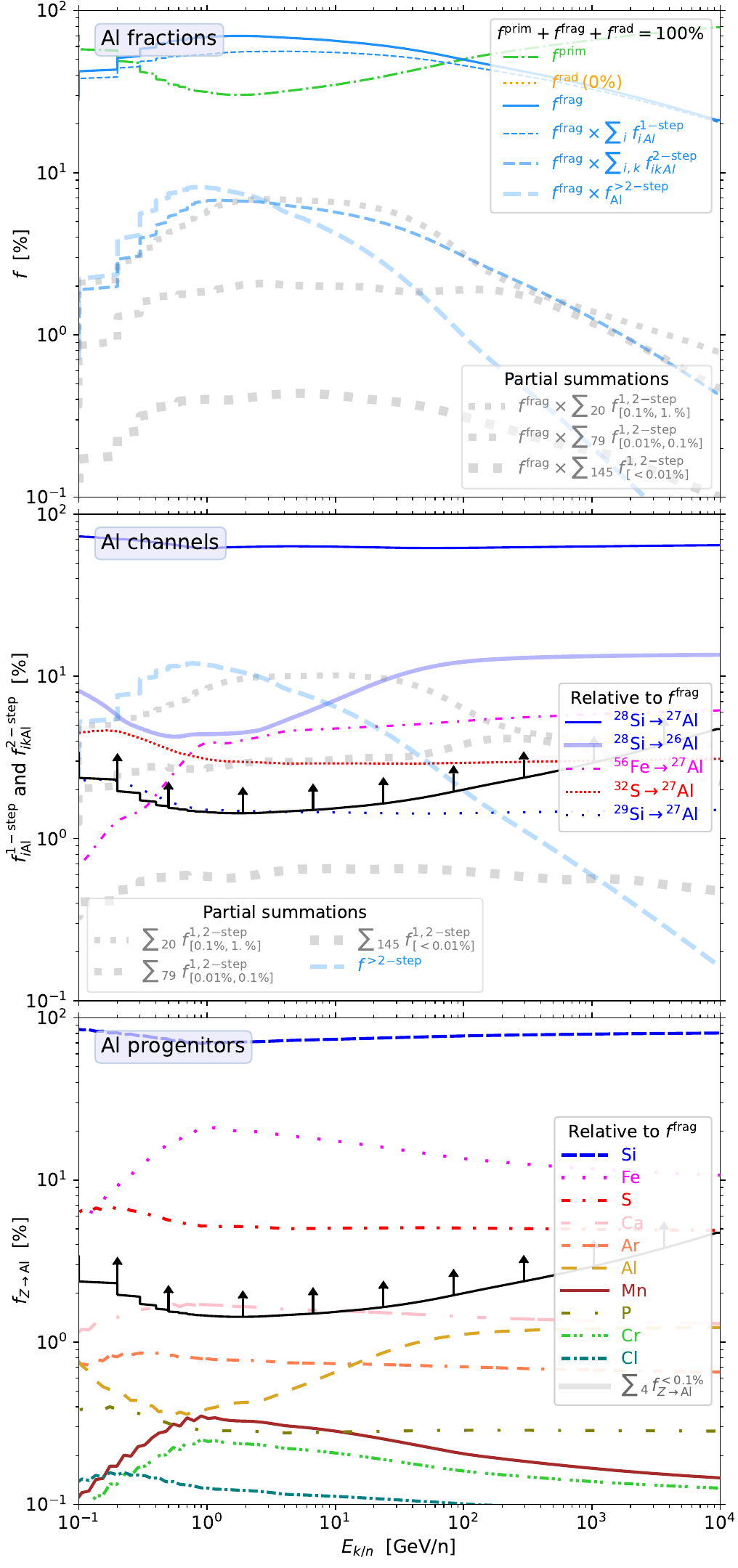}
\includegraphics[width=\columnwidth]{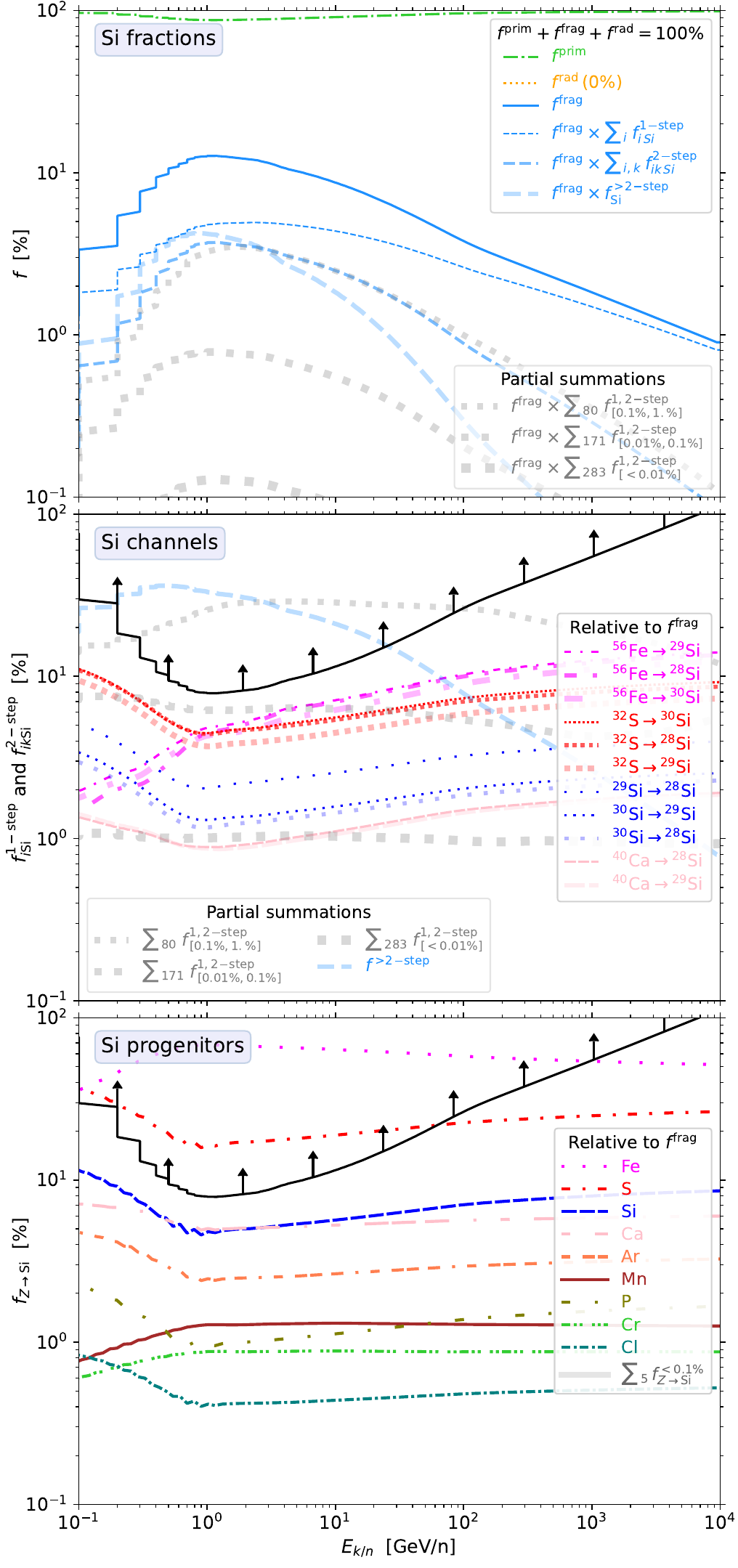}
\caption{Same as in Fig.~\ref{fig:ranked_LiBe}, but for Al (left) and Si (right).
\label{fig:ranked_AlSi}}
\end{figure*}

Figures \ref{fig:ranked_LiBe} to \ref{fig:ranked_AlSi} provide, as a function of $E_{k/n}$, a graphical view of the CR origin, and the ranking of channels and progenitors (in three different panels, from top to bottom respectively). We choose to show a TOA energy domain ranging from hundreds of MeV/n to hundreds of TeV/n, which corresponds to the range covered by recent and current CR experiments, Voyager 1, ACE-CRIS, AMS-02, and TRACER, ATIC, CREAM, NUCLEON, CALET, DAMPE. We consider here all CR elements from Li to Si, updating the results shown in \citetalias{2018PhRvC..98c4611G} for Li to N.

\subsection{Origin of contributions}
The top panels of Figs.~\ref{fig:ranked_LiBe}-\ref{fig:ranked_AlSi} extend the conclusions already drawn (at one energy) from Table~\ref{tab:origin}, relative to the primary ($f_{\rm prim}$), fragmentation ($f_{\rm sec}$), and radioactive ($f_{\rm rad}$) nature of the various CR elements. Concerning $f_{\rm rad}$, only B (top-left panel of Fig.~\ref{fig:ranked_BC}) and Mg (top-right panel of Fig.~\ref{fig:ranked_NaMg}) have non-zero contributions (orange dotted lines), which quickly drop to zero as energy increases owing to a boosted lifetime of the associated $\beta$-unstable parent CR. Concerning the energy dependence of $f_{\rm sec}$ with respect to $f_{\rm prim}$ (for mixed species), it peaks at a few GeV/n, as already discussed in Section~\ref{sec:origin}: the interaction times is constant with energy (cross sections are assumed to be constant above a few GeV/n) while the escape time (from the Galaxy) decreases. As a result, as illustrated for instance in the N plot in the top-left panel of Fig.~\ref{fig:ranked_NO}, the fragmentation contribution (blue solid line) decreases with energy while the primary component grows (green dash-dotted line). A similar energy dependence is observed when moving from 1-step to 2-step reactions, and then to $>$2-step contributions with respect to the total flux (thin to thicker blue dashed lines) reflecting the decrease in the probability of multiple interactions. For pure secondary species like F (top left panel of Fig.~\ref{fig:ranked_FNe}), the steepening of the energy dependence of such a contribution with the increase of the number of steps involved in its production is seen even more clearly.

In addition, these plots show how a larger number of individually negligible contributions adds up to a significant amount. To do so, we focus on individual 1-step and 2-step contributions whose $P^{\rm 1-step}_{ij}$ and $P^{\rm 2-step}_{ikj}$ coefficients (see Eq.~\ref{eq:coeffs_Pij_Pikj}) fall in the range $[x\%,y\%]$ with $(x,y)<0.1$: there are $N$ such coefficients, and we sum over all of them weighted by the secondary fraction $f^{\rm frac}$ to get the desired result. The thin to thick gray dotted lines and their caption show, for instance for the pure secondary F (top-left panel of Fig.~\ref{fig:ranked_FNe}), that when moving from $[0.1\%,1\%]$ to $[0.01\%,0.1\%]$, and then to $[0.001\%,0.01\%]$, a growing number of channels is involved (respectively $N=20$, 75, and 119), adding up to a sizable amount of the total flux (respectively $\approx$10\%, 3\%, 0.4\%); because these contributions combine 1- and 2-step channels, the different grouping intervals share the same energy dependence. Roughly, to get a $3\%$ percent precision on the modeled fluxes, all reactions making up at least $97\%$ of the total flux must be accounted for, which can implicate up to hundreds of sub-percent reactions.

\subsection{Ranking of channels}
%%%%%%%%%%%%%%%%%%%%%%%%%%%%%%%%%%%%%%%%%%%%%%%%%%%%%%

The middle panels of Figs.~\ref{fig:ranked_LiBe}-\ref{fig:ranked_AlSi} are very busy and difficult to read at a single glance. Nevertheless, we decided to keep them this way, because they give a comprehensive view of all channels (see definitions in Section~\ref{sec:coeffs_Pij_steps}), in the sense that the sum of all the curves shown corresponds to $100\%$ of the fragmentation flux. They are separated into 1-step or 2-step channels whose individual contributions are larger than $1\%$ at 10 GeV/n (color-coded legend), and those below this level which are further grouped by range (gray dotted lines). We also group together all $>\!2$-step contributions (dashed-blue curve). These groups were already shown in the top panels with respect to the total flux, while they are shown here with respect to the secondary component\footnote{For pure secondary species for which $f_{\rm sec}=100\%$ (Li, Be, B, and F), the curves are by construction the same.} $f_{\rm sec}$. We do not repeat the discussion of the energy dependence of these contributions, as they were amply commented in the previous paragraph.

As expected and as already stressed in \citetalias{2018PhRvC..98c4611G}, most of the LiBeB (Figs.~\ref{fig:ranked_LiBe} and \ref{fig:ranked_BC}) comes from the direct and 2-step fragmentation of $^{12}$C and $^{16}$O. However, the ranking of the reactions at 10~GeV/n is slightly different from those shown in \citetalias{2018PhRvC..98c4611G} (as gathered in Tables~V,~VI, and~VII there): reactions with $^{56}$Fe progenitor are ranked significantly higher for Li and Be here. As another illustration of the impact of some cross-section updates, the $^{16}$O$\to^{6}$Li reaction is now contributing about twice the $^{12}$C$\to^{6}$Li reaction (left panel of Fig.~\ref{fig:ranked_LiBe}), whereas it was reported as only a few percentages larger in \citetalias{2018PhRvC..98c4611G} (see Table~V there): this is related to the renormalization of the of $^{16}$O+p$\to^{6}$Li cross section to a higher value as measured in \cite{2005JETPL..81..140B}.

Looking globally at the dominant channels for the various elements from F up to Si, we see that: F production is dominated by fragmentation of $^{20,22}$Ne, $^{24}$Mg, $^{28}$Si, and $^{56}$Fe; Ne and Na by $^{24}$Mg and $^{28}$Si; Mg and Al by $^{28}$Si; Si by $^{56}$Fe and $^{32}$S. However the detail of the number of contributions larger than $1\%$ at 10~GeV/n, or whether one single channel dominates vs several ones contribute equally, is strongly dependent on the CR species under scrutiny. Interestingly, in almost all elements shown, the sum of all $>$2-step contributions (blue-dashed line) ranks second at 10~GeV/n, though it rapidly decreases below and above this energy; we will come back to this in the next paragraph. For the primary species (O, Ne, and Si), the dominant contribution appears to be the sum of $\approx$50 1-step and 2-step channels in the $[0.1\%-1\%]$ range (thinnest dotted line), rather than a certain reaction.

It is important to question the relevance of these various channels (or groups of channels) with respect to the total flux, especially for fluxes with a significant primary fraction. To do so, the solid black lines (with upward arrows) separate the region above which a channel (or group of channels) accounts for more than $1\%$ of the total flux; this black line is a straight line at $1\%$ for secondary species (Li, Be, B, and F), but it has a more complicated shape for mixed (combining primary and secondary components) species. For instance, for Mg (right-middle panel of Fig.~\ref{fig:ranked_NaMg}), only the direct production from $^{28}$Si is above this line for the individual channels. The grouped 1-step and 2-step channels in the $[0.1\%-1\%]$ range (thinnest dotted line) and the sum of all $>\!2$-step contributions (blue-dashed line) also are. However, as the energy increases and the element becomes predominantly primary, the number of relevant channels (and groups of channels) further decreases.

\subsection{Ranking of progenitors}
%%%%%%%%%%%%%%%%%%%%%%%%%%%%%%%%%%%%%%%%%%%%%%%%%%%%%%

The bottom panels of Figs.~\ref{fig:ranked_LiBe}-\ref{fig:ranked_AlSi} show yet another complementary perspective on the reactions to consider. Compared to the channel ranking above, the progenitor ranking is based on the sum of the direct and all multistep fragmentation channels starting from this progenitor (see Section~\ref{sec:coeffs_Pij}).
The colored curves show CR progenitors whose contribution is larger than $0.1\%$ at 10~GeV/n, and the thick gray line--the sum of all remaining progenitors contributions. As discussed in \cite{2022arXiv220801337F}, the different energy dependencies between the contributions of various elements reflects the different energy dependencies of the elemental fluxes themselves. Indeed, the ratios of heavier-to-lighter primary CR fluxes decrease at low energy \cite{2011A&A...526A.101P,2021PhRvL.126d1104A} owing to a larger fragmentation cross section of heavier species and their faster ionization energy losses. For instance, for the production of secondary Si (right panel of Fig.~\ref{fig:err_evol_FSi}), Fe is the dominant progenitor (pink dotted line). Similarly, elements with similar atomic numbers share the same energy dependence, i.e.\ contribution of heavier elements, e.g., Mn and Cr, decreases while contribution of lighter elements, e.g., S, Si, Ca, etc., increases at low energy.

One can also compare contributions of various progenitors to the black solid line (with upward arrows) corresponding to 1\% of the total flux. We account for small individual contributions of different channels tagging them to the original element they came from. Interestingly, in the case of Mg (right panel of Fig.~\ref{fig:ranked_NaMg}), Fe progenitor appears above the 1\% line in the bottom panel, whereas none of the individual channels originating from $^{56}$Fe is contributing above the black solid line in the middle panel.

\section{Figures for $D_{i\to j}$ and $P_{i\to j}$ at 10~GeV/n}
\label{app:fig_Dij_Pij}
%%%%%%%%%%%%%%%%%%%%%%%%%%%%%%%%%%%%%%%%%%%%%%%%%%%%%%
%%%%%%%%%%%%%%%%%%%%%%%%%%%%%%%%%%%%%%%%%%%%%%%%%%%%%%

In this Appendix, we show the $D_{i\to j}$ and $P_{i\to j}$ coefficients at 10~GeV/n (Fig.~\ref{fig:Dij_Pij}). These coefficients are introduced and defined in Section~\ref{sec:coeffs_Dij} and Eq.~(\ref{eq:coeffs_Dij}), and Section~\ref{sec:coeffs_Pij} and Eq.~(\ref{eq:coeffs_Pij}) respectively. They describe the contributing fraction of a progenitor $i$ into the isotope $j$ for elements Li through Si studied in this paper.

%%%%%%%%%%%%%%%%%%%%%
%%%%%%%%%%%%%%%%%%%%%

Similarly to the discussion of Section~\ref{sec:res_rankings} for the $D$ and $P$ coefficients for elements, the $D_{i\to j}$ coefficients in the left panel of Fig.~\ref{fig:Dij_Pij} show the direct production of the various CR isotopes of Li to Si from the GCR isotopic fluxes as measured. The $P_{i \to j}$ coefficients on the right panel correspond to the cumulated production from GCR primary fluxes only. In that respect, the values for species of pure secondary origin (LiBeB isotopes), mostly secondary origin (Sc, Ti, V isotopes), and subdominant isotopes in primary species (for instance $^{41-44}$Ca) are null or below $0.1\%$.

\section{Tables of ranked $f_{abc}$ and plots at 10~GeV/n}
\label{app:tablesxs_plot2DfaHc}
%%%%%%%%%%%%%%%%%%%%%%%%%%%%%%%%%%%%%%%%%%%%%%%%%%%%%%
%%%%%%%%%%%%%%%%%%%%%%%%%%%%%%%%%%%%%%%%%%%%%%%%%%%%%%

In this Appendix, we provide the ranking of the most impacting production cross sections based on the ranking of the $f_{abc}$ coefficients introduced in Section~\ref{sec:coeffs_fabc}. In order to mitigate possible cross-section uncertainties for those channels where no data is available, our calculations are based on the average $f_{abc}$ coefficients over the three cross-section models discussed in Section~\ref{sec:setup_XS}. The full list of $f_{abc}$ coefficients can be obtained from the original ASCII files, available to download from ZENODO \cite{genolini_yoann_2023_8143305}.

%%%%%%%%%%%%%%%%%%%%%

In the first column, Tables \ref{tab:sortedxs_Be} to \ref{tab:sortedxs_Si} list the reactions ranked at 10~GeV/n for the production of secondary Li through Si species in CRs. The next 3-column block shows the flux impact (minimum, mean, and maximum value calculated from our three cross-section parameterizations), i.e.\ how much this reaction contributes, in fraction, to the calculated GCR fluxes. As discussed in Section~\ref{sec:coeffs_fabc} and contrarily to the $D$ and $P$ coefficients, we recall that the sum of the $f_{abc}$ coefficients (for a given $c$) is larger than one. The fifth column presents the minimal and maximal  cross-section values for each reaction. The next-to-last column (check symbol) indicates when at least one data point exists for the listed cross section. While there can be several data points available for some reactions, for some others only 1--2 points may exist, and yet in a limited energy range. This is for instance the case for the reactions involved in the production of F (see Table~\ref{tab:sortedxs_F}): at the first glance, the cross sections are well-constrained as the most important reactions have at least one data point, but as discussed in \cite{2022arXiv220801337F}, it appears that all these data points are located in the rising portion of the cross sections, leaving the exact asymptotic values at high energy quite uncertain.

The last column shows the ratio of the reaction cross section to the cumulative one, $\sigma/\sigma^{\rm cumul}$ (see Section~\ref{sec:ghosts}); in \citetalias{2018PhRvC..98c4611G}, we showed instead $\sigma^{\rm cumul}/\sigma$, but we feel the inverse of this ratio is a natural choice. Reactions with $\sigma/\sigma^{\rm cumul}<1$ indicate that the isotope produced comes from its direct production and the production of ghosts (highlighted in boldface); see also Section~\ref{sec:ghosts} and Eq.~\ref{eq:cumul}). For instance, for F (see Table~\ref{tab:sortedxs_F}), the contribution from the short-lived nucleus $^{19}$Ne (see Table~\ref{tab:ghosts}) can reach half the cumulative production at most.
We stress that these coefficients and tables are relative to the secondary production. As shown in Table~\ref{tab:origin}, some of these species are mostly primary (e.g., O and Si). Hence, the relative importance of the reactions for the total flux is reduced. To illustrate this, we insert lines (in the tables) highlighting that coefficients above these lines make up some given fractions ($25\%$, $50\%$, $75\%$; etc.) of the secondary or total fluxes. We also provide graphical views in Figs.~\ref{fig:faHc_BeAl} and \ref{fig:faHc_BNNa} of the $f_{aHc}$ coefficients not shown in Section~\ref{sec:results}; we refer the reader to this section for detailed information on how to read and interpret these plots.

To conclude this Appendix, we underline that tables for Li to N supersede those presented in \citetalias{2018PhRvC..98c4611G}, mostly differing from the ranking position of reactions involving $^{56}$Fe (as already discussed in n Section~\ref{sec:coeffs_fabc}).

\begin{figure*}[p]%[t]
\includegraphics[width=1.1\textwidth, angle=90]{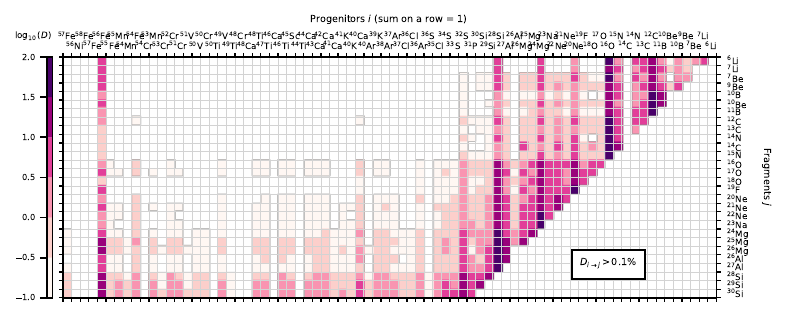}
\includegraphics[width=1.1\textwidth, angle=90]{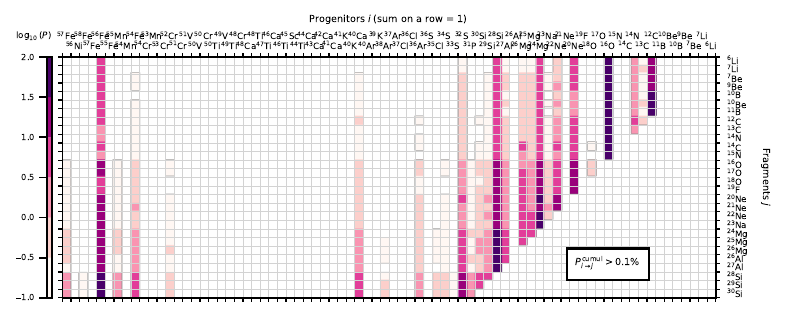}
\caption{Same as Fig.~\ref{fig:Zi_Zj} for isotopes instead of elements. The top panel corresponds to the $D_{i\to j}$ coefficients (\ref{sec:coeffs_Dij}) and the bottom panel to the $P_{i\to j}$ ones (\ref{sec:coeffs_Pij}). For both cases, the sum of coefficients on one row is 100\%, i.e.\ all the secondary production. We nevertheless recall that the latter ($f_{\rm sec}$) is sometimes only a small fraction of the total flux (see Table~\ref{tab:origin}). See text and Section~\ref{sec:res_rankings} for discussion.
\label{fig:Dij_Pij}}
\end{figure*}

\begingroup
\begin{table}[!ht]
\caption{Reactions sorted according to their flux impact $f_{abc}$ (with $\sum f_{abc}=1.39$) on Li at 10 GeV/n ($f_{\rm sec}=100\%$, see Table~\ref{tab:origin}).  The first 50 reactions shown provide 76.4\% of the total flux (767 reactions to reach 97\%); reactions in {\bf bold} highlight short-lived fragments (see Sect.~\ref{sec:ghosts} and  Table~\ref{tab:ghosts}).\label{tab:sortedxs_Li}}
\begin{tabular}{lrclc@{\hskip 0.4cm}c@{\hskip 0.2cm}c}
\hline\hline
Reaction &  \multicolumn{3}{c}{\!\!\!Flux impact $f_{abc}$} & $\sigma$ [mb]& \!\!\!\!\!Data\! & $\displaystyle\frac{\sigma}{\sigma^{\rm cumul}}$\\
\cline{2-4}
  &  Min & \!{\bf Mean}\! & Max &  [Min,Max] & & \\
\hline
$\rm ^{16}O\!+\!H\!\!\to^{6}$$\rm Li$ & 11.4 & {\bf15.2} & 17.1 &  [13.0, 18.4] & \cmark & 1.0\\[-1.30mm]
$\rm ^{12}C\!+\!H\!\!\to^{6}$$\rm Li$ & 11.3 & {\bf12.5} & 13.1 &  [14.0, 15.4] & \cmark & 0.9\\[-1.30mm]
$\rm ^{12}C\!+\!H\!\!\to^{7}$$\rm Li$ & 9.91 & {\bf9.93} & 9.97 &  [11.9, 12.6] & \cmark & 1.0\\[-1.30mm]
\multicolumn{7}{c}{{\em ($>$25\% of $f_{\rm sec,tot}$)}}\\[-1.30mm]
$\rm ^{16}O\!+\!H\!\!\to^{7}$$\rm Li$ & 9.67 & {\bf9.74} & 9.77 &  [10.7, 11.2] & \cmark & 1.0\\[-1.30mm]
$\rm ^{11}B\!+\!H\!\!\to^{7}$$\rm Li$ & 2.89 & {\bf2.92} & 2.99 &  [21.5, 21.5] & \cmark & 1.0\\
\hline\hline
 \end{tabular}
\end{table}
\endgroup
\begingroup
\begin{table}[!ht]
\caption{Reactions sorted according to their flux impact $f_{abc}$ (with $\sum f_{abc}=1.35$) on Be at 10 GeV/n ($f_{\rm sec}=100\%$, see Table~\ref{tab:origin}).  The first 50 reactions shown provide 74.7\% of the total flux (823 reactions to reach 97\%); reactions in {\bf bold} highlight short-lived fragments (see Sect.~\ref{sec:ghosts} and  Table~\ref{tab:ghosts}).\label{tab:sortedxs_Be}}
\begin{tabular}{lrclc@{\hskip 0.4cm}c@{\hskip 0.2cm}c}
\hline\hline
Reaction &  \multicolumn{3}{c}{\!\!\!Flux impact $f_{abc}$} & $\sigma$ [mb]& \!\!\!\!\!Data\! & $\displaystyle\frac{\sigma}{\sigma^{\rm cumul}}$\\
\cline{2-4}
  &  Min & \!{\bf Mean}\! & Max &  [Min,Max] & & \\
\hline
$\rm ^{16}O\!+\!H\!\!\to^{7}$$\rm Be$ & 16.5 & {\bf16.8} & 16.9 &  [9.7, 10.0] & \cmark & 1.0\\[-1.30mm]
$\rm ^{12}C\!+\!H\!\!\to^{7}$$\rm Be$ & 14.4 & {\bf14.5} & 14.6 &  [9.0, 9.7] & \cmark & 1.0\\[-1.30mm]
$\rm ^{12}C\!+\!H\!\!\to^{9}$$\rm Be$ & 8.12 & {\bf8.25} & 8.51 &  [6.1, 6.8] & \cmark & 1.0\\[-1.30mm]
\multicolumn{7}{c}{{\em ($>$25\% of $f_{\rm sec,tot}$)}}\\[-1.30mm]
$\rm ^{16}O\!+\!H\!\!\to^{9}$$\rm Be$ & 4.89 & {\bf6.09} & 6.68 &  [3.6, 4.6] & \cmark & 1.0\\[-1.30mm]
$\rm ^{16}O\!+\!He\!\!\to^{7}$$\rm Be$ & 2.68 & {\bf2.87} & 2.97 &  [14.7, 15.4] &  & 1.0\\
\hline\hline
 \end{tabular}
\end{table}
\endgroup
\begingroup
\begin{table}[!ht]
\caption{Reactions sorted according to their flux impact $f_{abc}$ (with $\sum f_{abc}=1.36$) on B at 10 GeV/n ($f_{\rm sec}=96\%$, see Table~\ref{tab:origin}).  The first 50 reactions shown provide 81.9\% of the total flux (550 reactions to reach 97\%); reactions in {\bf bold} highlight short-lived fragments (see Sect.~\ref{sec:ghosts} and  Table~\ref{tab:ghosts}).\label{tab:sortedxs_B}}
\begin{tabular}{lrclc@{\hskip 0.4cm}c@{\hskip 0.2cm}c}
\hline\hline
Reaction &  \multicolumn{3}{c}{\!\!\!Flux impact $f_{abc}$} & $\sigma$ [mb]& \!\!\!\!\!Data\! & $\displaystyle\frac{\sigma}{\sigma^{\rm cumul}}$\\
\cline{2-4}
  &  Min & \!{\bf Mean}\! & Max &  [Min,Max] & & \\
\hline
$\rm ^{12}C\!+\!H\!\!\to^{11}$$\rm B$ & 17.1 & {\bf17.2} & 17.4 &  [27.7, 30.0] & \cmark & 0.5\\[-1.30mm]
$\rm \mathbf{^{12}C\!+\!H\!\!\to^{11}}$${\rm\mathbf C}$ & 15.6 & {\bf16.6} & 17.2 &  [26.9, 27.9] & \cmark & 0.5\\[-1.30mm]
$\rm ^{16}O\!+\!H\!\!\to^{11}$$\rm B$ & 7.5 & {\bf10.8} & 17.3 &  [11.2, 27.3] & \cmark & 0.5\\[-1.30mm]
\multicolumn{7}{c}{{\em ($>$25\% of $f_{\rm sec}$ $\Rightarrow$ 28\% of $f_{\rm tot}$)}}\\[-1.30mm]
$\rm ^{12}C\!+\!H\!\!\to^{10}$$\rm B$ & 7.15 & {\bf7.85} & 8.19 &  [12.3, 13.3] & \cmark & 0.9\\[-1.30mm]
$\rm ^{16}O\!+\!H\!\!\to^{10}$$\rm B$ & 6.98 & {\bf7.05} & 7.09 &  [10.5, 11.0] & \cmark & 0.9\\
\hline\hline
 \end{tabular}
\end{table}
\endgroup
\begingroup
\begin{table}[!ht]
\caption{Reactions sorted according to their flux impact $f_{abc}$ (with $\sum f_{abc}=1.29$) on C at 10 GeV/n ($f_{\rm sec}=20\%$, see Table~\ref{tab:origin}).  The first 50 reactions shown provide 95.9\% of the total flux (85 reactions to reach 97\%); reactions in {\bf bold} highlight short-lived fragments (see Sect.~\ref{sec:ghosts} and  Table~\ref{tab:ghosts}).\label{tab:sortedxs_C}}
\begin{tabular}{lrclc@{\hskip 0.4cm}c@{\hskip 0.2cm}c}
\hline\hline
Reaction &  \multicolumn{3}{c}{\!\!\!Flux impact $f_{abc}$} & $\sigma$ [mb]& \!\!\!\!\!Data\! & $\displaystyle\frac{\sigma}{\sigma^{\rm cumul}}$\\
\cline{2-4}
  &  Min & \!{\bf Mean}\! & Max &  [Min,Max] & & \\
\hline
$\rm ^{16}O\!+\!H\!\!\to^{12}$$\rm C$ & 21.2 & {\bf21.5} & 22.2 & 32.3 & \cmark & 0.9\\[-1.30mm]
\multicolumn{7}{c}{{\em ($>$80\% of $f_{\rm tot}$ $\Rightarrow$ 2\% of $f_{\rm sec}$)}}\\[-1.30mm]
$\rm \mathbf{^{16}O\!+\!H\!\!\to^{13}}$${\rm\mathbf O}$ & 18.2 & {\bf18.4} & 18.8 & 30.5 & \cmark & 0.5\\[-1.30mm]
\multicolumn{7}{c}{{\em ($>$25\% of $f_{\rm sec}$ $\Rightarrow$ 85\% of $f_{\rm tot}$)}}\\[-1.30mm]
\multicolumn{7}{c}{{\em ($>$85\% of $f_{\rm tot}$ $\Rightarrow$ 26\% of $f_{\rm sec}$)}}\\[-1.30mm]
$\rm ^{16}O\!+\!H\!\!\to^{13}$$\rm C$ & 11.7 & {\bf11.9} & 12.1 & 17.5 & \cmark & 0.4\\
\hline\hline
 \end{tabular}
\end{table}
\endgroup
\begingroup
\begin{table}[!ht]
\caption{Reactions sorted according to their flux impact $f_{abc}$ (with $\sum f_{abc}=1.23$) on N at 10 GeV/n ($f_{\rm sec}=75\%$, see Table~\ref{tab:origin}).  The first 50 reactions shown provide 87.6\% of the total flux (452 reactions to reach 97\%); reactions in {\bf bold} highlight short-lived fragments (see Sect.~\ref{sec:ghosts} and  Table~\ref{tab:ghosts}).\label{tab:sortedxs_N}}
\begin{tabular}{lrclc@{\hskip 0.4cm}c@{\hskip 0.2cm}c}
\hline\hline
Reaction &  \multicolumn{3}{c}{\!\!\!Flux impact $f_{abc}$} & $\sigma$ [mb]& \!\!\!\!\!Data\! & $\displaystyle\frac{\sigma}{\sigma^{\rm cumul}}$\\
\cline{2-4}
  &  Min & \!{\bf Mean}\! & Max &  [Min,Max] & & \\
\hline
$\rm ^{16}O\!+\!H\!\!\to^{15}$$\rm N$ & 25.1 & {\bf25.4} & 26.0 & 34.3 & \cmark & 0.5\\[-1.30mm]
\multicolumn{7}{c}{{\em ($>$25\% of $f_{\rm tot}$ $\Rightarrow$ 0\% of $f_{\rm sec}$)}}\\[-1.30mm]
$\rm \mathbf{^{16}O\!+\!H\!\!\to^{15}}$${\rm\mathbf O}$ & 22.3 & {\bf22.6} & 23.1 & 30.5 & \cmark & 0.5\\[-1.30mm]
\multicolumn{7}{c}{{\em ($>$25\% of $f_{\rm sec}$ $\Rightarrow$ 44\% of $f_{\rm tot}$)}}\\[-1.30mm]
\multicolumn{7}{c}{{\em ($>$50\% of $f_{\rm tot}$ $\Rightarrow$ 34\% of $f_{\rm sec}$)}}\\[-1.30mm]
$\rm ^{16}O\!+\!H\!\!\to^{14}$$\rm N$ & 17.5 & {\bf20.2} & 21.6 &  [22.9, 29.2] & \cmark & 1.0\\
\hline\hline
 \end{tabular}
\end{table}
\endgroup
\begingroup
\begin{table}[!ht]
\caption{Reactions sorted according to their flux impact $f_{abc}$ (with $\sum f_{abc}=1.50$) on O at 10 GeV/n ($f_{\rm sec}=6\%$, see Table~\ref{tab:origin}).  The first 50 reactions shown provide 97.5\% of the total flux (33 reactions to reach 97\%); reactions in {\bf bold} highlight short-lived fragments (see Sect.~\ref{sec:ghosts} and  Table~\ref{tab:ghosts}).\label{tab:sortedxs_O}}
\begin{tabular}{lrclc@{\hskip 0.4cm}c@{\hskip 0.2cm}c}
\hline\hline
Reaction &  \multicolumn{3}{c}{\!\!\!Flux impact $f_{abc}$} & $\sigma$ [mb]& \!\!\!\!\!Data\! & $\displaystyle\frac{\sigma}{\sigma^{\rm cumul}}$\\
\cline{2-4}
  &  Min & \!{\bf Mean}\! & Max &  [Min,Max] & & \\
\hline
$\rm ^{24}Mg\!+\!H\!\!\to^{16}$$\rm O$ & 9.19 & {\bf9.49} & 9.64 &  [30.2, 30.4] & \cmark & 0.9\\[-1.30mm]
\multicolumn{7}{c}{{\em ($>$90\% of $f_{\rm tot}$)}}\\[-1.30mm]
$\rm ^{20}Ne\!+\!H\!\!\to^{16}$$\rm O$ & 6.33 & {\bf7.96} & 8.77 &  [33.4, 39.0] & \cmark & 1.0\\[-1.30mm]
$\rm ^{28}Si\!+\!H\!\!\to^{16}$$\rm O$ & 5.51 & {\bf6.12} & 6.43 &  [18.2, 20.3] & \cmark & 0.9\\[-1.30mm]
\multicolumn{7}{c}{{\em ($>$95\% of $f_{\rm tot}$ $\Rightarrow$ 14\% of $f_{\rm sec}$)}}\\[-1.30mm]
$\rm \mathbf{^{20}Ne\!+\!H\!\!\to^{18}}$${\rm\mathbf F}$ & 2.70 & {\bf3.83} & 4.40 &  [14.2, 18.3] & \cmark & 0.8\\
\hline\hline
 \end{tabular}
\end{table}
\endgroup
\begingroup
\begin{table}[!ht]
\caption{Reactions sorted according to their flux impact $f_{abc}$ (with $\sum f_{abc}=1.39$) on F at 10 GeV/n ($f_{\rm sec}=100\%$, see Table~\ref{tab:origin}).  The first 50 reactions shown provide 75.0\% of the total flux (1030 reactions to reach 97\%); reactions in {\bf bold} highlight short-lived fragments (see Sect.~\ref{sec:ghosts} and  Table~\ref{tab:ghosts}).\label{tab:sortedxs_F}}
\begin{tabular}{lrclc@{\hskip 0.4cm}c@{\hskip 0.2cm}c}
\hline\hline
Reaction &  \multicolumn{3}{c}{\!\!\!Flux impact $f_{abc}$} & $\sigma$ [mb]& \!\!\!\!\!Data\! & $\displaystyle\frac{\sigma}{\sigma^{\rm cumul}}$\\
\cline{2-4}
  &  Min & \!{\bf Mean}\! & Max &  [Min,Max] & & \\
\hline
$\rm \mathbf{^{20}Ne\!+\!H\!\!\to^{19}}$${\rm\mathbf Ne}$ & 15.5 & {\bf16.8} & 17.4 &  [25.8, 27.2] & \cmark & 0.5\\[-1.30mm]
$\rm ^{20}Ne\!+\!H\!\!\to^{19}$$\rm F$ & 12.5 & {\bf13.9} & 14.6 &  [21.6, 21.9] & \cmark & 0.5\\[-1.30mm]
$\rm ^{24}Mg\!+\!H\!\!\to^{19}$$\rm F$ & 8.59 & {\bf8.80} & 9.22 &  [9.0, 10.1] & \cmark & 0.7\\[-1.30mm]
\multicolumn{7}{c}{{\em ($>$25\% of $f_{\rm sec,tot}$)}}\\[-1.30mm]
$\rm ^{28}Si\!+\!H\!\!\to^{19}$$\rm F$ & 7.89 & {\bf8.12} & 8.23 &  [8.6, 8.7] & \cmark & 0.7\\[-1.30mm]
$\rm ^{22}Ne\!+\!H\!\!\to^{19}$$\rm F$ & 5.33 & {\bf6.13} & 6.53 &  [15.1, 21.9] & \cmark & 0.8\\
\hline\hline
 \end{tabular}
\end{table}
\endgroup
\begingroup
\begin{table}[!ht]
\caption{Reactions sorted according to their flux impact $f_{abc}$ (with $\sum f_{abc}=1.42$) on Ne at 10 GeV/n ($f_{\rm sec}=32\%$, see Table~\ref{tab:origin}).  The first 50 reactions shown provide 89.0\% of the total flux (417 reactions to reach 97\%); reactions in {\bf bold} highlight short-lived fragments (see Sect.~\ref{sec:ghosts} and  Table~\ref{tab:ghosts}).\label{tab:sortedxs_Ne}}
\begin{tabular}{lrclc@{\hskip 0.4cm}c@{\hskip 0.2cm}c}
\hline\hline
Reaction &  \multicolumn{3}{c}{\!\!\!Flux impact $f_{abc}$} & $\sigma$ [mb]& \!\!\!\!\!Data\! & $\displaystyle\frac{\sigma}{\sigma^{\rm cumul}}$\\
\cline{2-4}
  &  Min & \!{\bf Mean}\! & Max &  [Min,Max] & & \\
\hline
$\rm \mathbf{^{24}Mg\!+\!H\!\!\to^{22}}$${\rm\mathbf Na}$ & 7.48 & {\bf7.96} & 8.19 & 23.6 & \cmark & 0.5\\[-1.30mm]
\multicolumn{7}{c}{{\em ($>$50\% of $f_{\rm tot}$)}}\\[-1.30mm]
$\rm ^{24}Mg\!+\!H\!\!\to^{20}$$\rm Ne$ & 7.01 & {\bf7.42} & 7.63 & 23.0 & \cmark & 0.9\\[-1.30mm]
$\rm ^{24}Mg\!+\!H\!\!\to^{22}$$\rm Ne$ & 6.97 & {\bf7.41} & 7.64 & 22.0 & \cmark & 0.5\\[-1.30mm]
$\rm ^{24}Mg\!+\!H\!\!\to^{21}$$\rm Ne$ & 6.66 & {\bf7.06} & 7.26 & 21.2 & \cmark & 0.8\\[-1.30mm]
$\rm ^{28}Si\!+\!H\!\!\to^{21}$$\rm Ne$ & 5.29 & {\bf5.74} & 5.97 &  [16.9, 17.4] & \cmark & 1.0\\
\hline\hline
 \end{tabular}
\end{table}
\endgroup
\begingroup
\begin{table}[!ht]
\caption{Reactions sorted according to their flux impact $f_{abc}$ (with $\sum f_{abc}=1.29$) on Na at 10 GeV/n ($f_{\rm sec}=89\%$, see Table~\ref{tab:origin}).  The first 50 reactions shown provide 81.1\% of the total flux (840 reactions to reach 97\%); reactions in {\bf bold} highlight short-lived fragments (see Sect.~\ref{sec:ghosts} and  Table~\ref{tab:ghosts}).\label{tab:sortedxs_Na}}
\begin{tabular}{lrclc@{\hskip 0.4cm}c@{\hskip 0.2cm}c}
\hline\hline
Reaction &  \multicolumn{3}{c}{\!\!\!Flux impact $f_{abc}$} & $\sigma$ [mb]& \!\!\!\!\!Data\! & $\displaystyle\frac{\sigma}{\sigma^{\rm cumul}}$\\
\cline{2-4}
  &  Min & \!{\bf Mean}\! & Max &  [Min,Max] & & \\
\hline
$\rm ^{24}Mg\!+\!H\!\!\to^{23}$$\rm Na$ & 27.0 & {\bf28.0} & 28.5 & 39.2 & \cmark & 0.6\\[-1.30mm]
\multicolumn{7}{c}{{\em ($>$25\% of $f_{\rm tot}$ $\Rightarrow$ 16\% of $f_{\rm sec}$)}}\\[-1.30mm]
$\rm \mathbf{^{24}Mg\!+\!H\!\!\to^{23}}$${\rm\mathbf Mg}$ & 19.9 & {\bf20.0} & 20.1 &  [27.6, 29.0] & \cmark & 0.4\\[-1.30mm]
\multicolumn{7}{c}{{\em ($>$25\% of $f_{\rm sec}$ $\Rightarrow$ 33\% of $f_{\rm tot}$)}}\\[-1.30mm]
$\rm ^{28}Si\!+\!H\!\!\to^{23}$$\rm Na$ & 10.7 & {\bf11.2} & 11.4 &  [15.7, 15.7] & \cmark & 0.7\\[-1.30mm]
\multicolumn{7}{c}{{\em ($>$50\% of $f_{\rm tot}$ $\Rightarrow$ 44\% of $f_{\rm sec}$)}}\\
\hline\hline
 \end{tabular}
\end{table}
\endgroup
\begingroup
\begin{table}[!ht]
\caption{Reactions sorted according to their flux impact $f_{abc}$ (with $\sum f_{abc}=1.55$) on Mg at 10 GeV/n ($f_{\rm sec}=16\%$, see Table~\ref{tab:origin}).  The first 50 reactions shown provide 94.2\% of the total flux (189 reactions to reach 97\%); reactions in {\bf bold} highlight short-lived fragments (see Sect.~\ref{sec:ghosts} and  Table~\ref{tab:ghosts}).\label{tab:sortedxs_Mg}}
\begin{tabular}{lrclc@{\hskip 0.4cm}c@{\hskip 0.2cm}c}
\hline\hline
Reaction &  \multicolumn{3}{c}{\!\!\!Flux impact $f_{abc}$} & $\sigma$ [mb]& \!\!\!\!\!Data\! & $\displaystyle\frac{\sigma}{\sigma^{\rm cumul}}$\\
\cline{2-4}
  &  Min & \!{\bf Mean}\! & Max &  [Min,Max] & & \\
\hline
$\rm ^{28}Si\!+\!H\!\!\to^{24}$$\rm Mg$ & 14.3 & {\bf14.5} & 14.6 &  [28.0, 28.8] & \cmark & 0.8\\[-1.30mm]
\multicolumn{7}{c}{{\em ($>$85\% of $f_{\rm tot}$ $\Rightarrow$ 7\% of $f_{\rm sec}$)}}\\[-1.30mm]
$\rm ^{28}Si\!+\!H\!\!\to^{25}$$\rm Mg$ & 11.2 & {\bf11.7} & 11.9 &  [21.3, 22.8] & \cmark & 0.8\\[-1.30mm]
$\rm ^{28}Si\!+\!H\!\!\to^{26}$$\rm Mg$ & 6.88 & {\bf7.00} & 7.26 &  [13.1, 13.8] & \cmark & 1.0\\[-1.30mm]
$\rm ^{28}Si\!+\!H\!\!\to^{26}$$\rm Al$ & 4.79 & {\bf4.81} & 4.83 & 16.2 & \cmark & 1.0\\[-1.30mm]
$\rm ^{28}Si\!+\!H\!\!\to^{27}$$\rm Al$ & 3.39 & {\bf3.61} & 3.72 & 50.4 & \cmark & 0.6\\
\hline\hline
 \end{tabular}
\end{table}
\endgroup
\begingroup
\begin{table}[!ht]
\caption{Reactions sorted according to their flux impact $f_{abc}$ (with $\sum f_{abc}=1.25$) on Al at 10 GeV/n ($f_{\rm sec}=65\%$, see Table~\ref{tab:origin}).  The first 50 reactions shown provide 86.0\% of the total flux (651 reactions to reach 97\%); reactions in {\bf bold} highlight short-lived fragments (see Sect.~\ref{sec:ghosts} and  Table~\ref{tab:ghosts}).\label{tab:sortedxs_Al}}
\begin{tabular}{lrclc@{\hskip 0.4cm}c@{\hskip 0.2cm}c}
\hline\hline
Reaction &  \multicolumn{3}{c}{\!\!\!Flux impact $f_{abc}$} & $\sigma$ [mb]& \!\!\!\!\!Data\! & $\displaystyle\frac{\sigma}{\sigma^{\rm cumul}}$\\
\cline{2-4}
  &  Min & \!{\bf Mean}\! & Max &  [Min,Max] & & \\
\hline
$\rm ^{28}Si\!+\!H\!\!\to^{27}$$\rm Al$ & 37.3 & {\bf38.3} & 38.8 & 50.4 & \cmark & 0.6\\[-1.30mm]
\multicolumn{7}{c}{{\em ($>$25\% of $f_{\rm sec}$ $\Rightarrow$ 51\% of $f_{\rm tot}$)}}\\[-1.30mm]
\multicolumn{7}{c}{{\em ($>$50\% of $f_{\rm tot}$ $\Rightarrow$ 23\% of $f_{\rm sec}$)}}\\[-1.30mm]
$\rm \mathbf{^{28}Si\!+\!H\!\!\to^{27}}$${\rm\mathbf Si}$ & 21.1 & {\bf21.4} & 21.9 &  [27.4, 29.6] & \cmark & 0.4\\[-1.30mm]
$\rm ^{28}Si\!+\!H\!\!\to^{26}$$\rm Al$ & 6.13 & {\bf6.30} & 6.38 & 16.2 & \cmark & 1.0\\[-1.30mm]
\multicolumn{7}{c}{{\em ($>$50\% of $f_{\rm sec}$ $\Rightarrow$ 67\% of $f_{\rm tot}$)}}\\
\hline\hline
 \end{tabular}
\end{table}
\endgroup
\begingroup
\begin{table}[!ht]
\caption{Reactions sorted according to their flux impact $f_{abc}$ (with $\sum f_{abc}=1.80$) on Si at 10 GeV/n ($f_{\rm sec}=9\%$, see Table~\ref{tab:origin}).  The first 50 reactions shown provide 95.3\% of the total flux (139 reactions to reach 97\%); reactions in {\bf bold} highlight short-lived fragments (see Sect.~\ref{sec:ghosts} and  Table~\ref{tab:ghosts}).\label{tab:sortedxs_Si}}
\begin{tabular}{lrclc@{\hskip 0.4cm}c@{\hskip 0.2cm}c}
\hline\hline
Reaction &  \multicolumn{3}{c}{\!\!\!Flux impact $f_{abc}$} & $\sigma$ [mb]& \!\!\!\!\!Data\! & $\displaystyle\frac{\sigma}{\sigma^{\rm cumul}}$\\
\cline{2-4}
  &  Min & \!{\bf Mean}\! & Max &  [Min,Max] & & \\
\hline
$\rm ^{56}Fe\!+\!H\!\!\to^{30}$$\rm Si$ & 4.93 & {\bf5.56} & 6.82 &  [6.3, 9.1] & \cmark & 0.8\\[-1.30mm]
\multicolumn{7}{c}{{\em ($>$90\% of $f_{\rm tot}$)}}\\[-1.30mm]
$\rm ^{56}Fe\!+\!H\!\!\to^{29}$$\rm Si$ & 5.05 & {\bf5.32} & 5.46 &  [6.8, 7.4] & \cmark & 0.8\\[-1.30mm]
$\rm ^{32}S\!+\!H\!\!\to^{28}$$\rm Si$ & 4.40 & {\bf4.54} & 4.61 &  [33.6, 33.9] & \cmark & 0.8\\[-1.30mm]
$\rm \mathbf{^{56}Fe\!+\!H\!\!\to^{28}}$${\rm\mathbf Al}$ & 4.10 & {\bf4.36} & 4.87 &  [5.8, 6.9] & \cmark & 0.7\\[-1.30mm]
$\rm \mathbf{^{32}S\!+\!H\!\!\to^{30}}$${\rm\mathbf P}$ & 3.41 & {\bf4.11} & 4.46 &  [24.9, 29.6] & \cmark & 0.7\\
\hline\hline
 \end{tabular}
\end{table}
\endgroup

\begin{figure*}[t]
\includegraphics[width=0.47\textwidth]{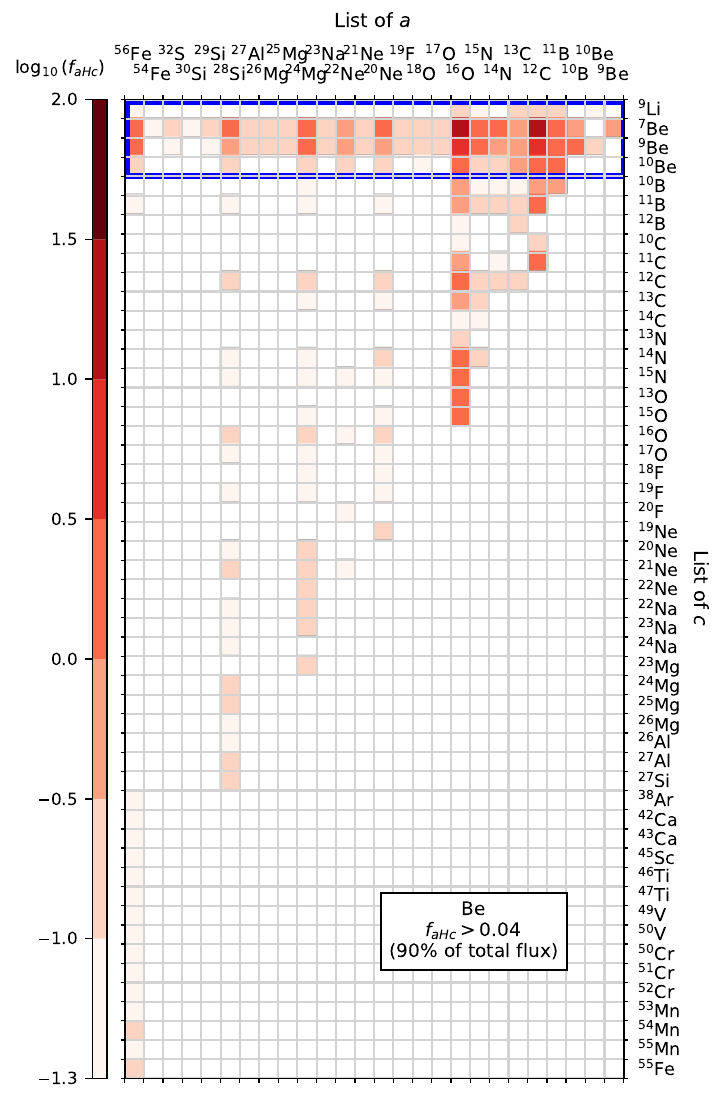}
\includegraphics[width=0.52\textwidth]{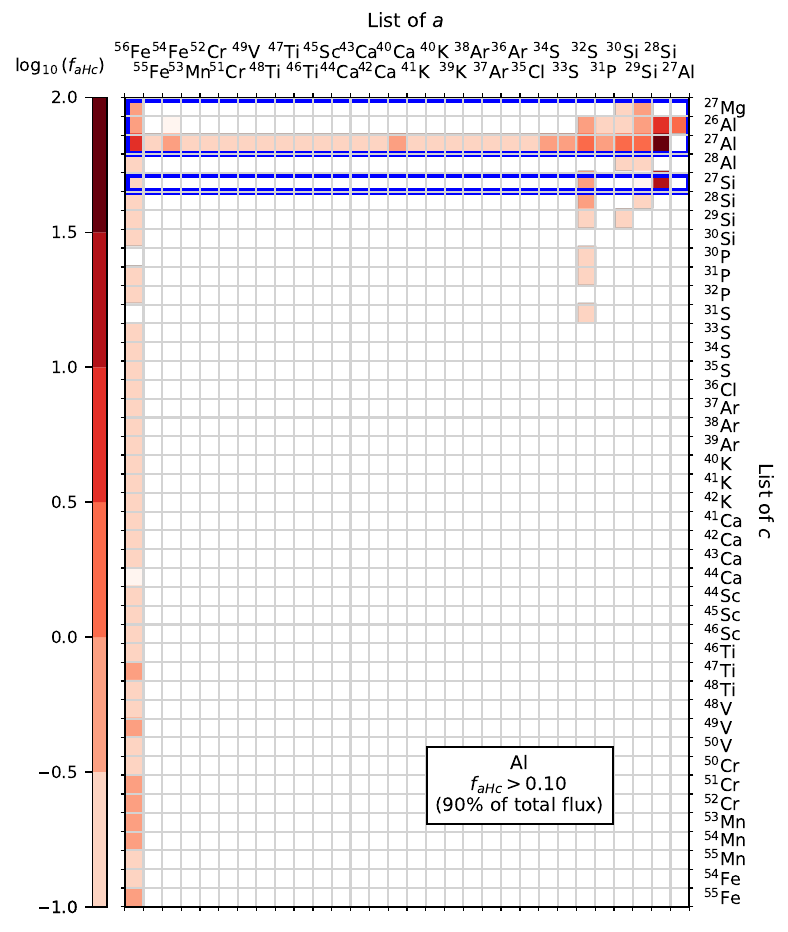}
\label{fig:faHc_BeAl}
\caption{Same as Fig.~\ref{fig:faHc_LiF} but for Be (left) and Al (right).}
\end{figure*}

\begin{figure*}[t]
\includegraphics[width=0.4\textwidth]{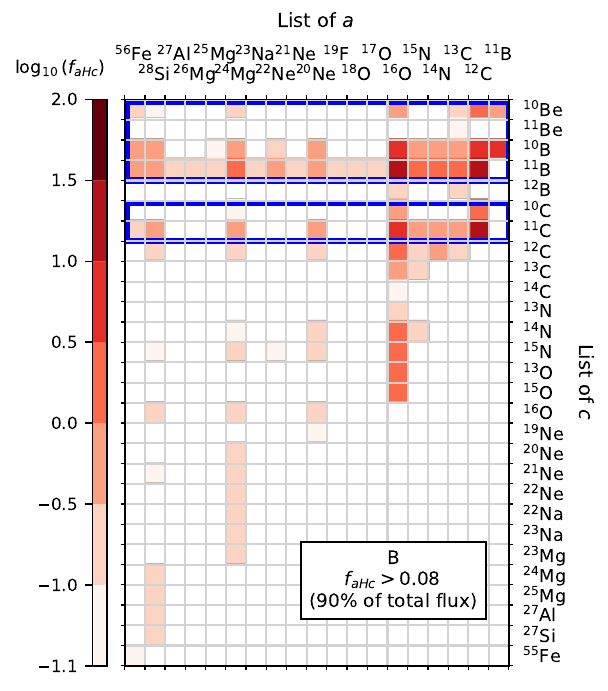}
\includegraphics[width=0.37\textwidth]{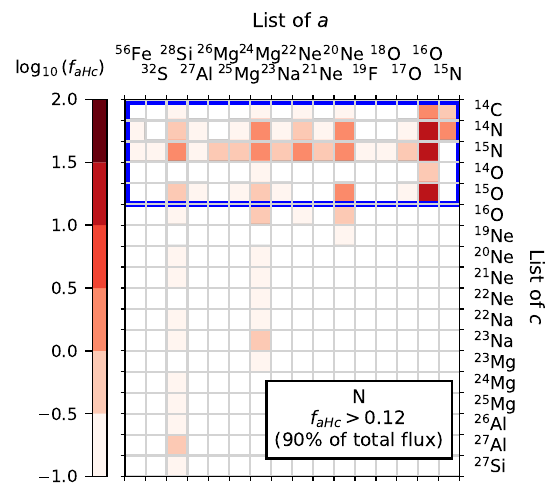}
\includegraphics[width=0.601\textwidth]{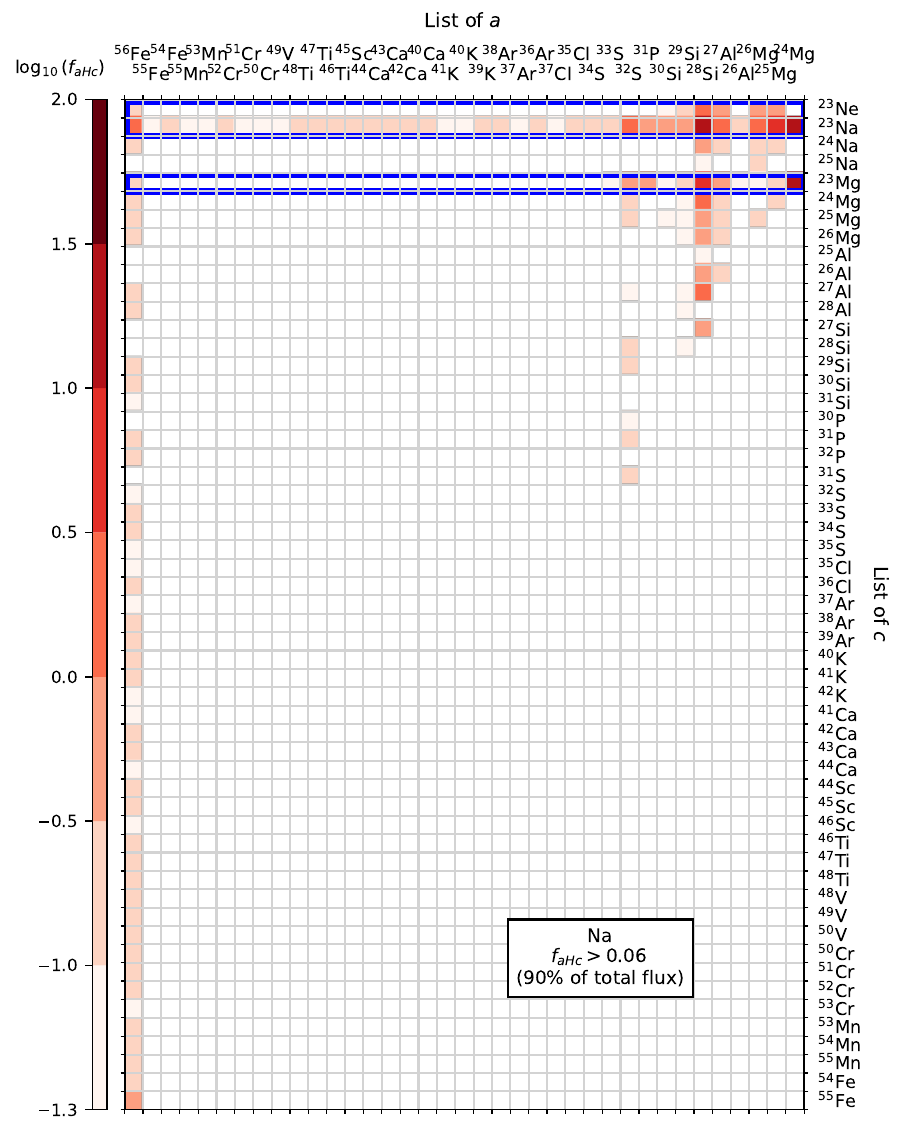}
\caption{Same as Fig.~\ref{fig:faHc_LiF} but for B (top left) and N (top right) and Na (bottom).}
\label{fig:faHc_BNNa}
\end{figure*}

% %%%%%%%%%%%%%%%%%%%%%%%%%%%%%%%%%%%%%%%%%%%%%%%%%%%%%
%%%%%REFERENCES SM.tex%%%%%%%%%%%%%%%%%%%%%%%%%%%%%%
\nocite{1997ApJ...488..730R}
\nocite{epherre1969comparison}   \nocite{fontes1977b}         \nocite{1979PhRvC..20..787R} \nocite{1983PhRvC..28.1602O}   \nocite{1998ApJ...501..911S} \nocite{1999ICRC....4..267K} \nocite{korejwo2002isotopic} \nocite{1984ADNDT..31..359R} \nocite{NUCLEX} \nocite{1995NIMPB.103..183M} \nocite{1996NIMPB.114...91S} \nocite{doi:10.1063/1.39166} \nocite{1990PhRvC..41..547W} \nocite{1990PhRvC..41..547W} \nocite{1997ApJ...479..504C} \nocite{1997PhRvC..56..398K} \nocite{1997PhRvC..56.1536C} \nocite{1998ApJ...508..949W} \nocite{1998PhRvC..58.3539W}
\nocite{2019NIMPB.454...50B}     \nocite{2011PAN....74..523T}
\nocite{2008PhRvC..78c4615T}  \nocite{2006NIMPA.562..729H} \nocite{2005JETPL..81..140B} \nocite{2004PhRvC..70e4607N} \nocite{KETTERN2004939} \nocite{2004NIMPB.226..243Y} \nocite{2002NIMPB.196..239K}
\nocite{1999ICRC....4..267K} \nocite{1998ApJ...508..949W,1998PhRvC..58.3539W} \nocite{Fa98ref} \nocite{1997NIMPB.123..324S} \nocite{1997NIMPB.129..153M} \nocite{1997M&PSA..32Q..78L} \nocite{fassbender1997activation} \nocite{1997NIMPB.123..324S} \nocite{1996NIMPB.114...91S}  \nocite{1995NIMPB.103..183M} \nocite{1993PhRvC..48.2617S}  \nocite{1992LPI....23.1305S} \nocite{1990PhRvC..41..566W} \nocite{1990NIMPA.291..662K} \nocite{1990NIMPB..52..588D,NSR1990DI06} \nocite{aleksandrov1990production}
\nocite{1989Ana...114..287M}  \nocite{1986NIMPB..16...61M} \nocite{1984ADNDT..31..359R} \nocite{1983PhRvC..28.1602O} \nocite{1981E&PSL..53..203R}
\nocite{1979PhRvC..20..787R}     \nocite{1977ICRC....2..203R} \nocite{1977PhRvC..15.2159F} \nocite{1976PhRvC..13..253K} \nocite{INOUE19761425} \nocite{1976PhRvC..14..753H} \nocite{1976PhRvC..14.1506H} \nocite{1975CRASB.280..513R} \nocite{1975PhRvC..12..915R} \nocite{1975PhLB...57..186R} \nocite{1974PhRvC...9.1385R}  \nocite{1972PhRvC...6..685R} \nocite{1972PhRvC...6.1153B} \nocite{1972NuPhA.195..311A} \nocite{1971NuPhA.175..124S} \nocite{1971PhRvL..27..875R} \nocite{1971NuPhA.165..405F} \nocite{bimbot1971spallation} \nocite{BARBIER19712720} 
\nocite{1968NuPhB...5..188S} \nocite{RAYUDU19682311} \nocite{1968PhRv..169..836D} \nocite{ANDREWS1968689} \nocite{PhysRev.154.1005} \nocite{1966NucPh..78..476M}  \nocite{NSR1966GA15} \nocite{1965NucPh..62...81V} \nocite{REEDER19651879} \nocite{1965PhRv..139.1513D} \nocite{1965PhL....15..147B} \nocite{warshaw1954cross} \nocite{doi:10.1139/v64-178} \nocite{1964PhRv..133.1507P}  \nocite{1964NucPh..51..363H} \nocite{1963PhL.....7..163V} \nocite{1963JETP...16....1L}  \nocite{1962NucPh..39..447G} \nocite{1962NucPh..31...43F} \nocite{1962PhRv..127..950C} \nocite{brun1962determination} \nocite{1961NucPh..25..216L} \nocite{1961PPS....78..681C} \nocite{1960NucPh..18..646P} \nocite{1960PhRv..118.1618H} \nocite{1960PhRv..119..316B}
\nocite{PhysRev.112.1319} \nocite{Symonds_1957} \nocite{prokoshkin1957investigation} \nocite{Burcham_1955} \nocite{Dickson_1951}
%%%%%%%%%%%%%%%%%%%%%%%%%%%%%%%%%%%%
%%%%% Ajouts Yoann %%%%%%%%%%%%%%%%%
\nocite{2007PhRvC..75d4603V} % Napolitani 2007
\nocite{1998NIMPB.145..449L} % Leya 1998 
\nocite{1995NIMPB.103..183M} % mchel 1995
\nocite{sah1960n15} %Sa60
\nocite{1970PhRvC...1..193K} %Ko70
\nocite{1951PhRv...83...47M} %Me51
\nocite{1961NucPh..24...28G} %Go61
\nocite{1952PhRv...88...19H} %Hi52
\nocite{1979PhRvC..19..962K} %Ka79
\nocite{1952PhRv...86..405M} %Ma52
\nocite{moskaleva1971yields} %Mo70
\nocite{1993ICRC....2..163T} % Tu93
\nocite{1997PhRvC..55.2458V}

\clearpage
\bibliography{nuc_channels}
%\bibliography{nuc_channels,biblio_SigFrag}

\end{document}